\documentclass[twoside]{article}

\usepackage[accepted]{aistats2025}
\usepackage[round]{natbib}

\usepackage[utf8]{inputenc} 
\usepackage[T1]{fontenc}    
\usepackage{hyperref}       
\usepackage{url}            
\usepackage{booktabs}       
\usepackage{amsfonts}       
\usepackage{nicefrac}       
\usepackage{microtype}      
\usepackage{xcolor}         

\usepackage{algorithm}
\usepackage{algcompatible}
\usepackage{subfigure}
\usepackage{graphicx}
\usepackage{multirow}
\usepackage{amsmath}
\usepackage{amssymb}
\usepackage{mathtools}
\usepackage{amsthm}

\usepackage[capitalize,noabbrev]{cleveref}

\theoremstyle{plain}
\newtheorem{theorem}{Theorem}[section]
\newtheorem{proposition}[theorem]{Proposition}

\theoremstyle{definition}

\newtheorem{assumption}[theorem]{Assumption}
\theoremstyle{remark}

\begin{document}

%

%

\twocolumn[

\aistatstitle{Differentiable Calibration of Inexact Stochastic Simulation Models via Kernel Score Minimization}

\aistatsauthor{ Ziwei Su \And Diego Klabjan }

\aistatsaddress{Northwestern University \And  Northwestern University } ]

\begin{abstract}
Stochastic simulation models are generative models that mimic complex systems to help with decision-making. The reliability of these models heavily depends on well-calibrated input model parameters. However, in many practical scenarios, only output-level data are available to learn the input model parameters, which is challenging due to the often intractable likelihood of the stochastic simulation model. Moreover, stochastic simulation models are frequently inexact, with discrepancies between the model and the target system. No existing methods can effectively learn and quantify the uncertainties of input parameters using only output-level data.
In this paper, we propose to learn differentiable input parameters of stochastic simulation models using output-level data via kernel score minimization with stochastic gradient descent. We quantify the uncertainties of the learned input parameters using a frequentist confidence set procedure based on a new asymptotic normality result that accounts for model inexactness. The proposed method is evaluated on exact and inexact G/G/1 queueing models.
\end{abstract}

\section{INTRODUCTION}
\label{sec:intro}
Stochastic simulation models play a crucial role in diverse domains, including manufacturing, supply chain management, cloud computing, and epidemiology. 
They serve as powerful tools for mimicking complex systems that are either costly or impractical to study directly.
Examples of stochastic simulation models include queueing models \citep{pan2021simulation} (Section~\ref{sec:queue} of the Supplementary Material introduces queues), digital twins \citep{biller2022simulation}, stochastic inventory models \citep{zhang2021analysis}, and compartmental models \citep{frazier2022modeling}.
These generative models generate random samples to help with decision-making.
For instance, queueing models for the waiting lines of cloud computing systems help develop strategies to reduce customer waiting times.
 
The output distribution of a stochastic simulation model is determined by a set of unknown input values.
We term this set of input values the \emph{simulation parameter}.
Calibrating the simulation parameter to align the simulation to the target system is key to the reliability of the simulation model. 
Typically, this calibration is performed using input-level data or domain knowledge about the simulation parameter, termed input modeling \citep{barton_input_2022}.
In practical scenarios where only output-level data are available,  the process of learning the input simulation parameter is termed \emph{stochastic simulation calibration} \citep{kleijnen1995verification}.

Existing methods for stochastic simulation calibration, such as likelihood-free inference techniques \citep{cranmer2020frontier}, often assume that the simulation model is exact, i.e., there exists a simulation parameter that perfectly aligns the model output with the target system. 
However, stochastic simulation models are frequently inexact in reality due to the complexity of real-world target systems.
No existing frequentist methods can effectively learn and quantify the uncertainties of input parameters using only output-level data under model inexactness.
We address this challenge by exhibiting a frequentist learning and confidence set procedure based on kernel score \citep{dawid2007geometry} minimization.

\paragraph{Notation and Setting}
Stochastic simulation calibration leverages a set of multi-dimensional output-level data $X_1, \ldots, X_m$ from the target system, presumed drawn independently from the target system output $P_{\star}$, i.e. $X_1, \ldots, X_m \stackrel{\mathrm{i.i.d.}}{\sim} P_{\star}$.
The simulation parameter $\theta$ in some parameter space $\Theta \subset \mathbb{R}^p$ determines the output distribution of the stochastic simulation model denoted $P_{\theta}$.
When the probability density function $f_\theta$ exists, the maximum likelihood estimation (MLE) is the standard approach to estimate $\theta$: $\widehat{\theta}^{\operatorname{MLE}}_m \in \arg \max_{\theta \in \Theta} \frac{1}{m} \sum_{i=1}^{m} \log f_\theta(X_i).$
MLE is an instance of the \emph{optimum score estimation} \citep{gneiting2007strictly}:
$\widehat{\theta}_{m} \in \arg \min_{\theta \in \Theta} L_m(\theta) =  \frac{1}{m} \sum_{i=1}^{m} S\left(P_{\theta}, X_{i}\right),$
where $S(\cdot, \cdot)$ is a \emph{strictly proper scoring rule} \citep{gneiting2007strictly}, $L_m(\theta)$ is the \emph{optimum score}, and $\widehat{\theta}_{m}$ is the \emph{optimum score estimator}. 
The negative log-likelihood $-\log f_\theta(x)$ is an example of a strictly proper scoring rule (\emph{logarithmic score} \citep{good1952rational}) that corresponds to MLE. Minimizing the optimum score $L_m(\theta)$ amounts to minimizing a statistical distance induced by $S$ between $P_{\theta}$ and the empirical measure $\widehat{P}_\star^m = \frac{1}{m} \sum_{i=1}^{m} \delta_{X_i}$, where $\delta$ denotes the Dirac delta function. Under model inexactness, minimizing the optimum score aims to find the best possible $\theta$ that minimizes the statistical distance between $P_{\theta}$ and $\widehat{P}_\star^m$. The optimum score estimator is hence also referred to as the minimum distance estimator \citep{basu2011statistical}.

Despite the asymptotic properties of optimum score estimators, such as MLE, the likelihood-free nature of stochastic simulation models hinders their direct application for simulation parameter estimation. 
While it is feasible to simulate realizations from $P_\theta$, evaluating $P_\theta$ is usually impractical due to the complex probabilistic mapping between $\theta$ and $P_\theta$. 
We denote the simulated sample from $P_\theta$ by $Y_1(\theta), \ldots, Y_n(\theta) \stackrel{\mathrm{i.i.d.}}{\sim} P_{\theta}$.
Consider a queueing model where the service time distributions for servers are parameterized by unknown service rates, forming the simulation parameter $\theta$.
It is feasible to simulate from the queueing model to obtain a random sample for any $\theta$, such as generating a sample of the average waiting times for customers. 
However, deriving an expression for $P_\theta$ is analytically challenging.
The likelihood of the model output $P_\theta$ is hence inaccessible.
Stochastic simulation models can thus be viewed as intractable generative models analogous to deep generative models such as generative adversarial networks (GANs) \citep{goodfellow2014generative}, diffusion models \citep{sohl2015deep}, and variational autoencoders (VAEs) \citep{kingma2013auto}. 
However, unlike deep generative models, stochastic simulation models are often considered \emph{inexact}, with an inherent model discrepancy where the simulation does not match the target system under any parameter. No existing frequentist methods are well-equipped for stochastic simulation calibration under model inexactness. 
We further discuss the key differences between deep generative models and stochastic simulation models in Section~\ref{sec:diff_deep_gen} of the Supplementary Material.

\paragraph{Contribution and Outline}
In this paper, we propose the first frequentist method to learn and quantify the uncertainties of a differentiable simulation parameter with output-level data under model inexactness. 
Our method, termed the \emph{kernel optimum score estimation}, leverages kernel score minimization to estimate the simulation parameter.
Kernel score is a strictly proper scoring rule that induces the maximum mean discrepancy (MMD) \citep{gretton2012kernel}. 
Notably, kernel score gradients can be unbiasedly estimated with U-statistic approximations using simulated samples, allowing stochastic gradient descent (SGD) optimization.
Our proposed learning and confidence set procedure hence applies to learning and quantifying uncertainties of a differentiable simulation parameter.
We validate the conditions outlined in \citep{binkowski2018demystifying} to ensure the unbiasedness of U-statistic gradient estimates in the stochastic simulation context, focusing on the G/G/1 queueing model.
We then present the first asymptotic normality result for the kernel optimum score estimator under model inexactness. 
Based on this result, we propose a novel confidence set procedure for the simulation parameter.
We evaluate the proposed method on both exact and inexact G/G/1 queueing models, demonstrating its effectiveness in learning and quantifying uncertainties of a simulation parameter.

The remainder of the paper is organized as follows. 
Section~\ref{sec:lit_review} concludes the introduction with a literature review.
Section \ref{sec:method} introduces the notions of scoring rules and kernel optimum score estimation, validates the unbiasedness of the gradient estimate, demonstrates the asymptotic normality of kernel optimum score estimators, and introduces a confidence set procedure for the simulation parameter. 
Section \ref{sec:simulation} studies the performance of the proposed method on exact and inexact G/G/1 queueing models with extensive simulation experiments.
Section \ref{sec:conclusion} concludes the paper.

\subsection{Literature Review}
\label{sec:lit_review}
Our proposed kernel optimum score estimator is closely related to several works in the frequentist and Bayesian literature that leverages MMD or scoring rules. 
In the frequentist literature, it is akin to the minimum MMD estimator for generative models \citep{briol2019statistical, cherief2022finite}, such as MMD-GANs \citep{dziugaite2015training}. 
\cite{dziugaite2015training} propose training generative neural networks with MMD as the discriminator and utilize SGD for network training. 
\cite{binkowski2018demystifying} provide the regularity conditions for the unbiasedness of the gradient estimate in the context of deep generative models. 
\cite{briol2019statistical} comprehensively investigate the statistical properties of minimum MMD estimators under model exactness. \cite{cherief2022finite} further investigate minimum MMD estimators under data dependence and outliers. 
\cite{oates2022minimum} offers the most general conditions for the strong consistency of minimum kernel discrepancy estimators, with the minimum MMD estimator as a special case.
In the Bayesian literature, several works leverage MMD \citep{cherief2020mmd} and scoring rules \citep{pacchiardi2021generalized, pacchiardi2024generalized} for generalized Bayesian inference. 
In particular, \cite{pacchiardi2024generalized} provides a posterior consistency result similar to our frequentist asymptotic normality result. 
However, our result guarantees asymptotically correct coverage for our confidence set procedure, which their posterior consistency result cannot provide. 

Another primary approach to likelihood-free inference for simulation models is the approximate Bayesian computation (ABC) \citep{beaumont2002approximate}. However, ABC is not robust to model inexactness. Several works address this issue by revising the original ABC \citep{frazier2020model, frazier2020robust, fujisawa2021gamma}, using synthetic likelihood \citep{frazier2021synthetic, frazier2021robust}, posterior bootstrap and minimum MMD estimators \citep{dellaporta2022robust}, or neural network approximations of the likelihood \citep{kelly2023misspecification} or posterior \citep{ward2022robust, wehenkel2024addressing}.
These works focus on contamination models where the data is noisy but the model is well-specified, while our approach considers the inexact case where the data is not noisy but the model is misspecified. Moreover, these works do not assess whether their methods can produce credible sets with valid coverage. In fact, \cite{frazier2020model} demonstrates that under model inexactness, ABC posteriors can produce credible sets with arbitrary coverage levels.
Another key difference is that existing works do not specifically consider inexact queueing models, which are a focus of our study. By addressing model inexactness in the context of queueing systems, our work fills an important gap in the literature and provides a frequentist approach to quantify the uncertainty of the simulation parameter.

Our work also relates to existing research in stochastic operations research. 
The discrepancy between simulation model outputs and the target system is traditionally handled via iterative model \emph{validation} and \emph{calibration} \citep{kleijnen1995verification,sargent2010verification}. Validation confirms the simulation's accuracy by comparing output data from the target system and the simulation, often using statistical tests such as the Schruben-Turing test \citep{schruben1980establishing} and the two-sample mean-difference test \citep{balci1982some}. 
Calibration aligns the simulation with the target system by adjusting parameters using data. 
Existing calibration works investigate computing bounds on input simulation parameters \citep{bai2020calibrating}, bounds on input-dependent quantities \citep{goeva2019optimization}, or learning model discrepancy \citep{plumlee2017uncertainty}. 
Similar to our approach, \cite{bai2020calibrating} construct confidence sets for the simulation parameter via empirical matching between the simulation and target system output using the Kolmogorov-Smirnov statistic. 
However, their method assumes model exactness, potentially leading to empty confidence sets in inexact model cases.

Lastly, extensive literature addresses calibration of deterministic computer models from both Bayesian \citep{kennedy2001bayesian, storlie2015calibration, plumlee2017bayesian} and frequentist \citep{tuo2015efficient, tuo2016theoretical, tuo2019adjustments, plumlee2019computer} perspectives. However, these deterministic calibration approaches are not directly applicable to stochastic simulation calibration, where both target system and simulation model outputs are represented by probabilistic measures.

\section{METHODOLOGY}
\label{sec:method}
\subsection{Setting and Formal Model}
\label{sec:setting}
\paragraph{Scoring Rules and Kernel Score} A scoring rule is a function $S(P, x): \mathcal{P} \times \mathbb{R}^d \to \mathbb{R}$ that evaluates a model based on its distribution $P \in \mathcal{P}$ and a realization $x \in \mathbb{R}^d$ of the $d$-dimensional random variable $X$ \citep{gneiting2007strictly}. 
Set $\mathcal{P}$ is the domain of the score function such that $S(P, x) < \infty$ for all $P \in \mathcal{P}$ and $x \in \mathbb{R}^d$.
Denoting the distribution of $X$ by $Q \in \mathcal{P}$, the \emph{expected score} of $P$ under $Q$ is denoted $\mathbb{E}_{X \sim Q} S(P, X)$.
It is assumed throughout this paper that $\mathbb{E}_{X \sim Q} S(P, X) < \infty$ for all $P, Q \in \mathcal{P}$. 
A scoring rule is strictly proper relative to $\mathcal{P}$ if and only if $\mathbb{E}_{X \sim Q} S(Q, X) = \mathbb{E}_{X \sim Q} S(P, X)$
implies $P = Q$.
The statistical distance that the strictly proper scoring rule induces is the difference between expected scores, $d_S(P, Q) = \mathbb{E}_{X \sim Q} \left[S(P, X) - S(Q, X)\right]$, 
termed the \emph{divergence function} associated with $S$. 
When a scoring rule is strictly proper, $d_S(P, Q) > 0$ for all $P \neq Q$.

Examples of strictly proper scoring rules include the logarithmic score, the \emph{energy score} (ES) \citep{szekely2013energy}, and the \emph{kernel score} (KS). 
The energy score is defined as:
$\operatorname{ES}(P, x)=\mathbb{E}_{Y}\|Y-x\|^\beta - \frac{1}{2} \mathbb{E}_{Y, Y^{\prime}}\left\|Y-Y^{\prime}\right\|^\beta, $
where $\|\cdot \|$ is the Euclidean norm, $\beta \in (0, 2)$, and $Y, Y^{\prime}$ are identically distributed, independent random variables with distribution $P$ i.e. $Y, Y^{\prime} \stackrel{\mathrm{i.i.d.}}{\sim} P$. 
The energy score is a special case of the kernel score $\operatorname{KS}(P, x) = \mathbb{E}_{Y, Y^{\prime}} [k(Y, Y^{\prime})]- 2 \mathbb{E}_{Y} \left[k(x, Y)\right]$, 
where $k(\cdot, \cdot): \mathbb{R}^d \times \mathbb{R}^d \to \mathbb{R}$ is a measurable kernel on $\mathbb{R}^d$, and $Y, Y^{\prime} \stackrel{\mathrm{i.i.d.}}{\sim} P$. 
The energy score is the kernel score with the Riesz kernel $k(x, y) = -\frac{1}{2}\|x-y\|^\beta$.
The kernel score is strictly proper when the kernel $k$ is \emph{characteristic}.
Examples of characteristic kernels include the Gaussian kernel $k(x, y) = \exp \left(-\frac{\left\|x-y\right\|_2^2}{2 \sigma}\right)$, where $\sigma > 0$ is the bandwidth parameter, and $\|\cdot\|_2$ is the Euclidean ($L_2$) distance on $\mathbb{R}^d$, and the Laplacian kernel $k(x, y) = \exp \left(-\frac{\left\|x-y\right\|_1}{\sigma}\right)$, where $\sigma > 0$ and $\|\cdot\|_1$ is the $L_1$ distance on $\mathbb{R}^d$. 
While the Riesz kernel is not characteristic, the fractional Brownian motion kernel \citep{sejdinovic2013equivalence} is a characteristic kernel whose corresponding squared MMD is the energy distance (see Section~\ref{sec:MMD} of the Supplementary Material):
$k(x, y) = \frac{1}{2}\left(\|x\|^\beta + \|y\|^\beta - \|x-y\|^\beta\right)$.
\paragraph{Kernel Optimum Score Estimation}
\label{sec:opt_score_est}
Given the output distribution of the stochastic simulation model $P_{\theta}$ and the output data $X_{1}, \ldots, X_{m} \stackrel{\mathrm{i.i.d.}}{\sim} P_\star$, recall that the optimum score estimator is defined as:
$\widehat{\theta}_{m} \in \arg \min_{\theta \in \Theta} L_m(\theta) =  \frac{1}{m} \sum_{i=1}^{m} S\left(P_{\theta}, X_{i}\right), 
$
where $S(\cdot, \cdot)$ is a strictly proper scoring rule, and $L_m(\theta)$ is the optimum score.
Optimum score estimation is challenging since the output distribution of the stochastic simulation model $P_{\theta}$ is not directly observable.
Instead, the simulation model output $P_{\theta}$ can only be assessed by means of the simulated sample $Y_1(\theta), \ldots, Y_n(\theta) \stackrel{\mathrm{i.i.d.}}{\sim} P_\theta$. 
The idea is to pick a good strictly proper scoring rule that allows for an empirical unbiased approximation of its optimum score with the simulated sample.
To this end, we propose to use kernel scores:
\begin{equation}
\begin{aligned}
\widehat{\theta}_{m}^{\operatorname{KS}} \in \arg \min _{\theta \in \Theta} L_{m}^{\operatorname{KS}}(\theta) = \mathbb{E}_{Y(\theta), Y^{\prime}(\theta)} [k(Y(\theta), Y^{\prime}(\theta))] \\ - \frac{2}{m} \sum_{i=1}^{m} \mathbb{E}_{Y(\theta)} \left[k(Y(\theta), X_i)\right], \nonumber 
\end{aligned}
\end{equation}
where $Y(\theta), Y^\prime(\theta) \stackrel{\mathrm{i.i.d.}}{\sim} P_\theta$.
We call $\widehat{\theta}_{m}^{\operatorname{KS}}$ \emph{kernel optimum score estimator}.
We term the U-statistic approximation of the kernel optimum score with the simulated sample the \emph{kernel simulated score}: 
\begin{equation}
\begin{aligned}
\widehat{L}_{m, n}^{\operatorname{KS}}(\theta) = \frac{1}{n(n-1)}  \sum_{1 \leq i \neq j \leq n} k(Y_i(\theta), Y_j(\theta)) \\ - \frac{2}{mn}  \sum_{i=1}^{m} \sum_{j=1}^{n} k(Y_j(\theta), X_i),  \nonumber 
\end{aligned}
\end{equation}
and we call the minimizer $\widehat{\theta}_{m, n}^{\operatorname{KS}}$ the \emph{kernel simulated score estimator}.
The gradient of the kernel simulated score \citep{dziugaite2015training, briol2019statistical} is:
\begin{equation}
\begin{aligned}
\nabla_\theta \widehat{L}_{m, n}^{\operatorname{KS}}(\theta) = \frac{1}{n(n-1)} \sum_{1 \leq i \neq j \leq n} \nabla_\theta k(Y_i(\theta), Y_j(\theta)) \\ - \frac{2}{mn}  \sum_{i=1}^{m} \sum_{j=1}^{n} \nabla_\theta k(Y_j(\theta), X_i),\nonumber       
\end{aligned}
\end{equation}
where $\nabla_\theta(\cdot)$ denotes the gradient with respect to $\theta$.
Section~\ref{sec:cons_asymp_norm} delves into the asymptotic properties of the kernel optimum score estimator. 
\cite{briol2019statistical} presents a generalization bound for the kernel optimum score estimator when the kernel $k$ is bounded (Theorem~\ref{thm:generalize} in Section~\ref{sec:lemma} of the Supplementary Material).
The generalization bound for unbounded kernels such as the Riesz kernel is still an open problem.

\paragraph{Unbiased Gradient Estimation}
The gradient of the kernel simulated score is an unbiased estimate of the gradient of the kernel optimum score under certain regularity conditions.
\cite{binkowski2018demystifying} first establishes these conditions for deep generative models, and we adapt them in the rest of Section~\ref{sec:setting} to stochastic simulation models. 
We further validate these conditions for G/G/1 queueing models in Section~\ref{sec:diff_GG1} of the Supplementary Material.

We require an alternative representation of the stochastic simulation model output $P_\theta$, referred to as the ‘pushforward' representation \citep{binkowski2018demystifying, briol2019statistical}, to facilitate our discussion.
Let $P$ be a Borel probability measure on a measurable space $\mathcal{Z}$, typically a subspace of $\mathbb{R}^d$ or $\mathbb{R}^d$ itself.
The simulation model output can be expressed as $P_\theta = G^{\#}_\theta P$, where $G_\theta: \mathcal{Z} \to \mathbb{R}^d$ is a measurable parametric map, and $G^{\#}_\theta P$ is the \emph{pushforward} of $P$ through $G_\theta$. 

In stochastic simulation models, $G_\theta$ can be interpreted as the input model or the deterministic part of the input-output map, with $\theta$ as its (input) simulation parameter. 
The pushforward entails simulating $n$ i.i.d. samples $Y_1(\theta), \ldots, Y_n(\theta) \stackrel{\mathrm{i.i.d.}}{\sim} P_\theta$ by first simulating $n$ i.i.d. samples $Z_1, \ldots, Z_n \stackrel{\mathrm{i.i.d.}}{\sim} P$ and then setting $Y_j(\theta) = G_\theta(Z_j)$ for $j = 1, \ldots, n$.
Function $G_\theta$ is usually a composition of multiple nonparametric component functions $\phi_1, \ldots, \phi_i, \ldots$ and parametric component functions $\psi_{\theta, 1}, \ldots, \psi_{\theta, j}, \ldots$.
For instance, consider simulating a random sample $Y$ from an exponential distribution with rate $\theta$ denoted $\operatorname{Exp}(\theta)$ using inverse transform sampling. 
We first simulate pseudo-random number $Z \sim P = U(0, 1)$, then set $Y = G_\theta(Z) = \psi_{\theta, 1}(\phi_1(Z)) = \frac{\log(1-Z)}{\theta}$, where $\phi_1(z) = \log(1-z)$ and $\psi_{\theta, 1}(z) = \frac{z}{\theta}$.
The reference measure $P$, as in inverse transform sampling, is typically the uniform measure on $(0, 1)^d$, though it can be specified differently based on requirements. 
Different choice of $P$ yields different $G_\theta$. 
The selection of $P$ hence plays a key role in ensuring unbiasedness.
The unbiased gradient estimation relies on the following regularity conditions.
\begin{assumption}
\label{ass:ubg_reg_conditions}
1. $\mathbb{E}_{X} \|X\|^\alpha, \mathbb{E}_{Z} \|Z\|^\alpha < \infty$ for some $\alpha \geq 1$.        
2. \textbf{Bounded kernel:} There exist constants $C_0, C_1$ such that $|k(x, y)| \leq C_0\left(\left(\|x\|^2+\|y\|^2\right)^{\alpha / 2}+1\right)$, 
$\left\|\nabla_{x, y} k(x, y)\right\| \leq C_1\left(\left(\|x\|^2+\|y\|^2\right)^{(\alpha-1) / 2}+1\right)$, 
where $\alpha$ is the same $\alpha$ in Assumption 1.      
3. \textbf{Lipschitzness:} Each non-parametric component of $G_\theta$, $\phi_i$, is M-Lipschitz.
4. \textbf{Piecewise analytic:} For each $\phi_i$, there are $K_i$ functions $\phi_i^k$, $k = 1, \ldots, K_i$, each real analytic on its input space, and these functions agree with $\phi_i$ on the closure of a set $\mathcal{D}^k$, and the sets $\mathcal{D}_i^k$ cover the whole input space.
5. \textbf{Differentiability:} Each parametric component of $G_\theta$, $\psi_{\theta, j}$, is almost everywhere differentiable on the parameter space $\Theta$.
\end{assumption}
Assumption 1 in Assumption~\ref{ass:ubg_reg_conditions} naturally hold.
Assumption 2 (Assumption~E in \cite{binkowski2018demystifying}) holds for a wide range of kernels, including the Gaussian kernel and the Riesz kernel for $1 \leq \beta < 2$. 
Assumptions 3 and 4 (Assumptions~C and D in \cite{binkowski2018demystifying}) hold for the vast majority of the activation functions used in deep neural networks, such as softmax and ReLU.
However, Assumptions 3, 4, and 5 deal with components of $G_\theta$, they are hence context-dependent. 
We leverage the Lindley equation \citep{lindley1952theory} to discuss these conditions for G/G/1 queueing models in Section~\ref{sec:diff_GG1} of the Supplementary Material.

\begin{proposition}[Unbiased Gradient Estimate]
\label{thm:unbiased_gradient}
    If Assumption~\ref{ass:ubg_reg_conditions} holds, then $\mathbb{E}_{Y_1(\theta), \cdots, Y_n(\theta)} \nabla_\theta \widehat{L}_{m, n}^{\operatorname{KS}}(\theta) = \nabla_\theta L_{m}^{\operatorname{KS}}(\theta)$.
\end{proposition}
The proof follows from Theorem~5 in \cite{binkowski2018demystifying} and the chain rule. 
Proposition~\ref{thm:unbiased_gradient} enables the design of an SGD algorithm in the model inexactness setting as outlined in Algorithm~\ref{alg:SGD}. 
The loop is the proposed SGD algorithm and the remaining steps are about the confidence set procedure; see Section~\ref{sec:conf_proc} for details.
Note that \cite{briol2019statistical} offers a similar SGD algorithm under model exactness.

\paragraph{Time Complexity of Kernel Optimum Score Estimation} Applying SGD in Algorithm~\ref{alg:SGD} can be computationally costly. Following from \cite{briol2019statistical}, the time complexity for SGD in Algorithm~\ref{alg:SGD} in each iteration is $O((n^2 + mn)dp)$, where $d$ is the dimension of data and $p$ is the dimension of the parameter space $\Theta$.
The quadratic time complexity in the simulated sample size $n$ renders the use of large $n$ impractical within limited computational resources.
To determine the most computationally efficient $n$, the optimal ratio $\lambda = \frac{n}{m+n}$ corresponds to the highest statistical efficiency,  minimizing the asymptotic variance-covariance for the kernel simulated score estimator.
The simulated score estimator is not used in practice as it needs to resample from $P_\theta$ to obtain $Y_1(\theta), \ldots, Y_n(\theta)$ in each SGD iteration.
Section~\ref{sec:cons_asymp_norm} provides the asymptotic normality of both the kernel simulated score estimator and the kernel optimum score estimator, followed by a brief discussion of the optimal ratio $\lambda$.

\paragraph{Optimal Simulation Parameter}
\label{sec:opt_score_est}
Minimizing the optimum score asymptotically amounts to minimizing the statistical distance $d_S(P_\theta, P_\star)$ between $P_\theta$ and $P_\star$.
The parameter that minimizes $d_S(P_\theta, P_\star)$ is termed the \emph{optimal simulation parameter}: $\theta_\star \in \arg \min _{\theta \in \Theta} d_S(P_\theta, P_\star) = \mathbb{E}_{X \sim P_\star} \left[S(P_\theta, X) - S(P_\star, X)\right].$
The choice of the kernel influences both the optimal simulation parameter and the statistical properties of the kernel optimum score estimator, discussed in Section~\ref{sec:conf_proc} and Section~\ref{sec:lambda_kernel} of the Supplementary Material.
The optimal simulation parameter may not be unique due to the complex nature of the input-output probabilistic map. 
This issue is known as model non-identifiability, a common challenge in inverse problems where the input is calibrated with the output \citep{tarantola2005inverse}.
This challenge is exacerbated when the scoring rule used for estimation lacks the strict propriety. An example is the Dawid-Sebastiani score \citep{dawid1999coherent}, which is proper but not strictly proper, relying solely on the first two moments. Multiple $\theta \in \Theta$ may exist such that $P_\theta$ aligns with the first two moments of $P_\star$.
Using strictly proper scoring rules for estimation helps alleviate model non-identifiability problems.
In the case of an exact ($P_{\star} \in \{P_{\theta}: \theta \in \Theta \}$) and an identifiable stochastic simulation model, the optimal simulation parameter $\theta_\star$ is the true simulation parameter $\theta_0$ (Proposition~\ref{prop:opt_unique}).

\subsection{New Results on Asymptotic Normality under Model Inexactness}
\label{sec:cons_asymp_norm}
We denote the optimal simulation parameter with the kernel score by $\theta_\star^{\operatorname{KS}}$.
The strong consistency property of the kernel optimum score estimator, i.e., $\widehat{\theta}_{m}^{\operatorname{KS}}  \stackrel{\mathrm{a.s.}}{\to} \theta_\star^{\operatorname{KS}}$, is given in the Supplementary Material (Proposition~\ref{thm:consistency}).
Here, we provide the speed of convergence in terms of asymptotic variance-covariance with asymptotic normality of the kernel optimum score estimator. We also exhibit asymptotic normality of the kernel simulated score estimator to facilitate the discussion of the choice of the ratio $\lambda = \frac{n}{m+n}$.
The following assumption gives the regularity conditions for the asymptotic normality of the kernel optimum score estimator and kernel simulated score estimator.
\begin{assumption}
\label{ass:combined_assumption}
1. There exists an open and convex neighborhood of $\theta_{\star}^{\operatorname{KS}}$ denoted $\mathcal{N}$ such that 
$\mathbb{E}_{Z, X} \sup_{\theta \in \mathcal{N}} \left\|\nabla_{\theta} k(G_\theta(Z), X)\right\| < \infty$ and $\mathbb{E}_{Z, Z^\prime} \sup_{\theta \in \mathcal{N}} \left\|\nabla_{\theta} k(G_\theta(Z), k(G_\theta(Z^\prime))\right\| < \infty$ for all $\theta \in \mathcal{N}$.
2. $\mathbb{E}_{Z, X} \left\|\nabla_{\theta} k(G_\theta(Z), X)\right\|^2 \left.\right|_{\theta=\theta_{\star}^{\operatorname{KS}}} < \infty$ and $\mathbb{E}_{Z, Z^\prime} \left\|\nabla_{\theta} k(G_\theta(Z), k(G_\theta(Z^\prime))\right\|^2 \left.\right|_{\theta=\theta_{\star}^{\operatorname{KS}}} < \infty$.
3. For all $r, s \in \{1, \ldots, p\}$ and $\theta = (\theta_1, \ldots, \theta_p) \in \mathbb{R}^p$: 
\begin{equation}
\begin{aligned}
\mathbb{E}_{Z, X} &\left[ \left|\nabla_{\theta_r \theta_s} k(G_\theta(Z), X) \left.\right|_{\theta=\theta_{\star}^{\operatorname{KS}}}\right| \right. \\ & \left. \log^{+}\left(\left|\nabla_{\theta_r \theta_s} k(G_\theta(Z), X) \left.\right|_{\theta=\theta_{\star}^{\operatorname{KS}}}\right|\right) \right], 
\\  \mathbb{E}_{Z, Z^\prime} &\left[ \left|\nabla_{\theta_r \theta_s} k(G_\theta(Z), G_\theta(Z^\prime)) \left.\right|_{\theta=\theta_{\star}^{\operatorname{KS}}}\right|  \right] < \infty. \nonumber   
\end{aligned}
\end{equation}
4. $\nabla_\theta k(G_\theta(z), x)$, $\nabla_\theta k(G_\theta(z), G_\theta(z^\prime))$ are differentiable on $\mathcal{N}$ for $P$-almost all $z, z^\prime \in Z$ and $P_\star$-almost all $x \in \mathbb{R}^d$.
5. $\nabla_{\theta \theta} k(G_\theta(z), x)$, $\nabla_{\theta \theta} k(G_\theta(z), G_\theta(z^\prime))$ are uniformly continuous at $\theta_{\star}^{\operatorname{KS}}$ for $P$-almost all $z, z^\prime \in Z$ and $P_\star$-almost all $x \in \mathbb{R}^d$, where $\nabla_{\theta \theta}(\cdot)$ denotes the Hessian with respect to $\theta$.
6. The Hessian $H = \mathbb{E}_{Z, Z^\prime} \left[\nabla_{\theta \theta} k(G_\theta(Z), G_\theta(Z^\prime)) \left.\right|_{\theta=\theta_{\star}^{\operatorname{KS}}} \right] - 2\mathbb{E}_{Z, X} \left[\nabla_{\theta \theta} k(G_\theta(Z), X) \left.\right|_{\theta=\theta_{\star}^{\operatorname{KS}}} \right]$ is positive definite.
\end{assumption}
\begin{theorem}[Asymptotic Normality]
\label{thm:asymp_norm}
Let us suppose $\widehat{\theta}_{m}^{\operatorname{KS}} \stackrel{\mathrm{a.s.}}{\to} \theta_{\star}^{\operatorname{KS}}$, $\widehat{\theta}_{m, n}^{\operatorname{KS}} \stackrel{\mathrm{a.s.}}{\to} \theta_{\star}^{\operatorname{KS}}$ as $m, n \to \infty$, $\lim_{m, n \to \infty} \frac{n}{m+n} = \lambda \in (0, 1)$, and Assumption~\ref{ass:combined_assumption} holds. 
Then as $m \to \infty$,
\begin{equation}
    \sqrt{m}(\widehat{\theta}_{m}^{\operatorname{KS}} - \theta_{\star}^{\operatorname{KS}}) \stackrel{\mathrm{d}}{\to} N(0, C), \nonumber
\end{equation}
where $\stackrel{\mathrm{d}}{\to}$ denotes convergence in distribution.
The variance-covariance matrix is the Godambe matrix $C = H^{-1} \Sigma H^{-1}$, where $\Sigma = 4 \mathbb{C}_{X}\left[\mathbb{E}_{Z} \left[\nabla_\theta k(G_{\theta}(Z), X)\left.\right|_{\theta=\theta_{\star}^{\operatorname{KS}}}\right]\right]$, $\mathbb{C}[\cdot]$ denotes the variance-covariance matrix.
Furthermore, 
\begin{equation}
    \sqrt{m+n}(\widehat{\theta}_{m, n}^{\operatorname{KS}} - \theta_{\star}^{\operatorname{KS}}) \stackrel{\mathrm{d}}{\to} N(0, C_\lambda), \nonumber
\end{equation}
where $C_\lambda = H^{-1} \Sigma_\lambda H^{-1}$, $\Sigma_\lambda = \frac{\mathbb{C}_{Z} \left[2h(Z) - g_{0,1}(Z)\right]}{\lambda} + \frac{\mathbb{C}_{X} \left[g_{1, 0}(X)\right]}{1-\lambda}$, $h(z) = \mathbb{E}_{Z^\prime} \nabla_\theta k(G_\theta(z), G_\theta(Z^\prime)) \left.\right|_{\theta=\theta_{\star}^{\operatorname{KS}}}$, $g_{0,1}(z) = 2\mathbb{E}_{X} \nabla_\theta k(G_\theta(z), X) \left.\right|_{\theta=\theta_{\star}^{\operatorname{KS}}}$, and $g_{1,0}(x) = 2\mathbb{E}_{Z} \nabla_\theta k(G_\theta(Z), x) \left.\right|_{\theta=\theta_{\star}^{\operatorname{KS}}}$.
\end{theorem}
See Section~\ref{proof:asymp_norm} in the Supplementary Material for the proof.
Note that $\Sigma = \mathbb{C}_{X} \left[g_{1, 0}(X)\right]$.
When the model is exact and identifiable, 
Theorem~\ref{thm:asymp_norm} reduces to the asymptotic normality under model exactness as in Theorem 2 of \cite{briol2019statistical}, where the asymptotic variance-covariance of $\widehat{\theta}_{m, n}^{\operatorname{KS}}$ is $\frac{C}{\lambda(1-\lambda)}$.
The asymptotic variance-covariance is hence minimized at $\lambda = \frac{1}{2}$ i.e. the simulated sample size $n$ equals the output data size from the target system $m$, meaning that using $n$ much larger than $m$ is not computationally efficient.

However, under model inexactness, the determination of an optimal $\lambda$ becomes complex due to the instance-dependent nature of the asymptotic variance-covariance $C_\lambda$, influenced by various factors, such as the kernel $k$, the reference measure of $Z$, $P$, and the input model $G_\theta$, rendering it computationally intractable. 
A nuanced trade-off between the variance-covariance terms from the simulated sample $Z$ and the data $X$ further complicates the selection of $\lambda$. 
It is also possible to use asymptotic variance-covariance matrices $C$ and $C_\lambda$ to choose the kernel that minimizes the asymptotic variance-covariance \citep{briol2019statistical}.
While both $C_\lambda$ and $C$ are computationally intractable as the unknown parameter $\theta_{\star}^{\operatorname{KS}}$, consistent estimations can be obtained (See Section~\ref{sec:conf_proc}). 
Section~\ref{sec:lambda_kernel} in the Supplementary further discusses $\lambda$ and kernel selection.

\subsection{New Confidence Set Procedure and Analyses}
\label{sec:conf_proc}
Theorem~\ref{thm:asymp_norm} allows the construction of an asymptotically valid confidence set for the optimal simulation parameter $\theta_\star^{\operatorname{KS}}$. 
Such a confidence procedure requires a consistent estimator for the asymptotic variance-covariance matrix $C = H^{-1} \Sigma H^{-1}$. 
Both $H$ and $\Sigma$ have expectation and covariance terms concerning $X$ and $Z$, and depend on $\theta_\star^{\operatorname{KS}}$.
To obtain consistent estimators for $H$ and $\Sigma$, we can estimate expectation and covariance terms with empirical estimates, and replace $\theta_\star^{\operatorname{KS}}$ with the obtained kernel optimum score estimator $\widehat{\theta}_{m}^{\operatorname{KS}}$.
We can use a larger simulation sample size $n_c$ to estimate expectation and covariance terms.
Let $\widehat{H}_{m, n_c}$, $\widehat{\Sigma}_{m, n_c}$ denote the consistent estimators to $H$ and $\Sigma$, respectively. Then 
\begin{equation}
\label{eq:cons_hessian}
\begin{aligned}
    \widehat{H}_{m, n_c} & = \frac{1}{n_c(n_c-1)} \sum_{1 \leq i \neq j \leq n_c} \nabla_{\theta \theta} k(G_\theta(Z_i), G_\theta(Z_j)) \\ &\left.\right|_{\theta=\widehat{\theta}_{m}^{\operatorname{KS}}}  - \frac{2}{mn_c}  \sum_{i=1}^{m}   \sum_{j=1}^{n_c} \nabla_{\theta \theta} k(G_\theta(Z_j), X_i)\left.\right|_{\theta=\widehat{\theta}_{m}^{\operatorname{KS}}}.    
\end{aligned}
\end{equation}
Let $\widehat{\mu}_{m, n_c} = \frac{1}{mn_c} \sum_{i=1}^{m} \sum_{j=1}^{n_c} \nabla_\theta k\left(G_\theta(Z_j), X_i\right) \left.\right|_{\theta=\widehat{\theta}_{m}^{\operatorname{KS}}}$, $\widehat{\mu}_{i} = \frac{1}{n_c} \sum_{j=1}^{n_c}\nabla_\theta k\left(G_\theta(Z_j), X_i\right)\left.\right|_{\theta=\widehat{\theta}_{m}^{\operatorname{KS}}}.$
Then we have
\begin{equation}
\label{eq:cons_jac}
\widehat{\Sigma}_{m, n_c} = \frac{4}{m-1} \sum_{i=1}^{m} (\widehat{\mu}_i - \widehat{\mu}_{m, n_c})^{T}(\widehat{\mu}_i - \widehat{\mu}_{m, n_c}).    
\end{equation}
The following assumption is regarding the consistency of $\widehat{\Sigma}_{m, n_c}$.
\begin{assumption}
\label{ass:unif_cont_bdd_jacobian}
    $\nabla_{\theta} k(G_\theta(z), x)$ is uniformly continuous and uniformly bounded at $\theta_{\star}^{\operatorname{KS}}$ for $P$-almost all $z, z^\prime \in Z$ and $P_\star$-almost all $x \in \mathbb{R}^d$.
\end{assumption}
We denote the consistent estimator for $C$ by $\widehat{C}_{m, n_c} = \widehat{H}_{m, n_c}^{-1} \widehat{\Sigma}_{m, n_c} \widehat{H}_{m, n_c}^{-1}$.
\begin{theorem}
\label{thm:cons_var_est}
If Assumptions 2 and 3 in Assumption~\ref{ass:combined_assumption}, and Assumption~\ref{ass:unif_cont_bdd_jacobian} hold, then $\widehat{H}_{m, n_c} \stackrel{\mathrm{p}}{\to} H$, $\widehat{\Sigma}_{m, n_c} \stackrel{\mathrm{p}}{\to} \Sigma$, as $m, n_c \to \infty$, and $\widehat{C}_{m, n_c} = \widehat{H}_{m, n_c}^{-1} \widehat{\Sigma}_{m, n_c} \widehat{H}_{m, n_c}^{-1} \stackrel{\mathrm{p}}{\to} H^{-1} \Sigma H^{-1}$ as $m, n_c \to \infty$.
\end{theorem}
See Section~\ref{proof:cons_var_est} of the Supplementary Material for the proof. Theorem~\ref{thm:cons_var_est} guarantees that the ellipsoid
\begin{equation}
\label{eq:conf_set}
 \left\{\theta \in \Theta:\left\|\sqrt{m} \widehat{\Sigma}_{m, n_c}^{-\frac{1}{2}} \widehat{H}_{m, n_c} \left(\theta-\widehat{\theta}_{m}^{\operatorname{KS}}\right)\right\|^2 \leq \chi^{2}_{1-\alpha}(p)\right\}
\end{equation}
denoted $\operatorname{CS}_{m, n_c}^{\alpha}$
is an asymptotically valid $100(1-\alpha)\%$ confidence set for $\theta_{\star}^{\operatorname{KS}}$, where $\chi^{2}_{1-\alpha}(p)$ is the $(1-\alpha)$-quantile of the chi-squared distribution with $p$ degrees of freedom where $p$ is the dimension of the parameter space $\Theta$.
The procedure to compute $\operatorname{CS}_{m, n_c}^{\alpha}$ is summarized in Algorithm~\ref{alg:SGD}.
The time complexity for computing $\widehat{H}_{m, n_c}$ and $\widehat{\Sigma}_{m, n_c}$ in total is $O((n_c^2 + mn_c)pd)$ if using automatic differentiation for Hessian estimation, and $O((n_c^2 + mn_c)p^2d)$ if using finite difference for Hessian estimation.
The time complexity for checking whether an arbitrary point $\theta$ is in $\widehat{C}_{m, n_c}^{\alpha}$ is $O(p^3)$.

\begin{algorithm}[tb]
   \caption{Kernel Optimum Score Estimation and Confidence Set Estimation}
   \label{alg:SGD}
   \begin{algorithmic}
      \STATE {\bfseries Input:} Initial $\widehat{\theta}_{m, n}^{(0)} \in \Theta$, output data $X_1, \ldots, X_m \stackrel{\mathrm{i.i.d.}}{\sim} P_\star$, step sizes $\{\eta_t\}_{t \geq 0}$, simulation sample size $n$ per iteration, simulation sample size $n_c$ for confidence set estimation, significance level $\alpha$. 
      \STATE Initialize $t = 0$.
      \STATE {\bfseries repeat} Sample $Z_1, \ldots, Z_n \stackrel{\mathrm{i.i.d.}}{\sim} P$ and set $Y_j(\widehat{\theta}_{m, n}^{(t)}) = G_{\widehat{\theta}_{m, n}^{(t)}}(Z_j)$ for $j = 1, \ldots, n$. Compute $\widehat{\theta}_{m, n}^{(t+1)} = \Pi_{\Theta}\left[\widehat{\theta}_{m, n}^{(t)} - \eta_t \nabla_\theta \widehat{L}_{m, n}^{\operatorname{KS}}(\widehat{\theta}_{m, n}^{(t)})\right].$ Set $t = t + 1$.
      \STATE {\bfseries until} convergence criterion is met.
      \STATE Let $\widehat{\theta}_{m}^{\operatorname{KS}} = \widehat{\theta}_{m, n}^{(t)}$. Sample $Z_1, \ldots, Z_{n_c} \stackrel{\mathrm{i.i.d.}}{\sim} P$. Compute $\widehat{\Sigma}_{m, n_c}$ with \eqref{eq:cons_jac} and $\widehat{H}_{m, n_c}$ with \eqref{eq:cons_hessian}. Compute $\operatorname{CS}_{m, n_c}^{\alpha}$ with \eqref{eq:conf_set}.
      \STATE {\bfseries Output:} $\widehat{\theta}_{m}^{\operatorname{KS}}$ and $\operatorname{CS}_{m, n_c}^{\alpha}$.
   \end{algorithmic}
\end{algorithm}

\section{NUMERICAL EXPERIMENTS}
\label{sec:simulation}
This section studies the performance of Algorithm~\ref{alg:SGD} on exact and inexact stochastic simulation models of G/G/1 queueing models. Throughout the experiments, unless otherwise specified, we use the Riesz kernel with $\beta=1$ (KOSE-Riesz) and the Gaussian kernel with the bandwidth determined by the median heuristic \citep{gretton2012kernel} (KOSE-Gaussian). 
We consider the case where a sample of average waiting times $X_1, \ldots, X_m$ from the target G/G/1 system output $P_\star$ is available. 
The corresponding G/G/1 queueing model output is $P_\theta = G_\theta(Z)$, where $\theta$ is the input simulation parameter. Table~\ref{tab: exp} gives the details of the settings for Experiments 1 to 4.
We use Experiments 2 and 4 to explore how the performance changes with increasing model inexactness. 
Specifically, we set the shape parameter of the service time distribution to be $a=1, 0.8, 0.6, 0.4, 0.2$, where $a=1$ is the model exact case, and smaller values of $a$ indicate higher model inexactness.
Further experiment details are reported in Section~\ref{sec:exp_details} in the Supplementary Material.
Our SGD-based approach requires the differentiability of $\theta$ and the unbiasedness of the gradient estimate. 
We discuss and validate these requirements for G/G/1 queueing models and for other models in Sections~\ref{sec:diff_GG1} and~\ref{sec:app_other_model} in the Supplementary Material, respectively.

\begin{table}[htb]
\centering
\caption{Experiment settings. Arr., Serv. give the distributions of the inter-arrival time and service time in the target G/G/1 queueing system and the G/G/1 simulation model. $\operatorname{Gam.}$ means Gamma.} \label{tab: exp}
\vspace{1ex}
\begin{tabular}{|c|c|c|c|c|}
\hline
No. & Dist. & Target & Model & $\theta$ \\ \hline
1 & Arr. & $\operatorname{Exp}(1)$ & $\operatorname{Exp}(1)$ & $\mu$\\ 
 & Serv. & $\operatorname{Exp}(1.2)$ & $\operatorname{Exp}(\mu)$ &\\ \hline
2 & Arr. & $\operatorname{Exp}(1)$ & $\operatorname{Exp}(1)$ & $\mu$\\
 & Serv. & $\operatorname{Gam.}(a, 1.2)$ & $\operatorname{Exp}(\mu)$ &\\ \hline
3 & Arr. & $\operatorname{Gam.}(0.5, 1)$ & $\operatorname{Gam.}(0.5, \lambda)$ & $(\mu, \lambda)$\\
 & Serv. & $\operatorname{Exp}(1)$ & $\operatorname{Exp}(\mu)$ &\\ \hline
4 & Arr. & $\operatorname{Gam.}(0.5, 1)$ & $\operatorname{Gam.}(0.5, \lambda)$ & $(\mu, \lambda)$\\
 & Serv. & $\operatorname{Gam.}(a, 2.5)$ & $\operatorname{Exp}(\mu)$ &\\ \hline
\end{tabular}
\end{table}

We compare our method with three baselines: accept-reject ABC (ABC-AR) \citep{beaumont2002approximate} with MMD, MMD posterior bootstrap (NPL-MMD) \citep{dellaporta2022robust}, and eligibility set (ESet) \citep{bai2020calibrating}. For each baseline method and our proposed method, we record performance metrics such as mean-squared error (MSE), Monte Carlo coverage of credible or confidence sets, width of credible or confidence sets, and average run time. These results are presented in Tables~\ref{tab: 1}, \ref{tab: 2}, \ref{tab: 3}, \ref{tab: 4} for Experiments 1, 2, 3, 4, respectively.
We also provide the plots for the confidence set produced in Experiment 4 for KOSE-Riesz (Figures~\ref{fig:ci-1-2d-inexact} and \ref{fig:ci-06-2d-inexact}) and KOSE-Gaussian (Figures~\ref{fig:ci-1-2d-inexact-G} and \ref{fig:ci-06-2d-inexact-G}) in $a=1, 0.6$.

We conclude from Tables~\ref{tab: 2} and \ref{tab: 4} that KOSE-Riesz consistently produces valid confidence sets (despite a slight undercoverage issue for $a=1$ in Table~\ref{tab: 2}), except for $a=0.4, 0.2$ in Experiment 4, both more extreme model inexactness scenarios. 
We use $R=1,000$ in most cases except when the run time is larger we use $R=100$.
In all model inexactness scenarios, KOSE-Riesz produces stable optimum score estimates with constantly small MSE.
KOSE-Gaussian suffers from small undercoverage issues for $a=0.8, 0.2$ in Experiment 2, and the produced optimum score estimates appear less stable than those of KOSE-Riesz, while all the benchmarks do not produce valid credible or confidence intervals under model inexactness. 
We also observe from Tables~\ref{tab: 1} and \ref{tab: 3} in the Supplementary Material that our method consistently produces valid confidence sets, for both KOSE-Riesz and KOSE-Gaussian under model exactness.
And both KOSE-Riesz and KOSE-Gaussian yield better MSE for the optimum score estimates and smaller confidence set width as $m$ increases, as expected. 
ABC-AR and ESet methods also produce valid credible or confidence sets for larger $m$ under model exactness, while NPL-MMD does not produce valid credible sets even under model exactness. 
\begin{table}[h!]
\centering
\caption{Experiment 2 results; Value $a$ is the shape parameter of the service time distribution in the target system; $R$ is the number of independent runs; $m=n=500$. MSE is the average mean-squared error, Cov. is the coverage of the $95\%$ confidence or credible set, Width is the average length of the confidence or credible set, and T is the average time per run in seconds. Empty is the number of empty confidence sets. Parameter $\theta_\star^{\operatorname{KS}}$ is the optimal simulation parameter obtained using KOSE-Gaussian.}\label{tab: 2}
\vspace{1ex}
\begin{tabular}{|c|cccc|c|}
\hline
& \multicolumn{4}{c|}{KOSE-Riesz, $R=1,000$} & \\
$a$ & MSE & Cov. & Width & T & $\theta_\star^{\operatorname{KS}}$\\
\hline
$1$ & $4.0\cdot 10^{-4}$ & $0.920$ & $0.07$ & $3$ & $1.2$\\
$0.8$ & $1.6\cdot 10^{-4}$ & $\mathbf{0.996}$ & $0.07$ & $3$ & $1.5$\\
$0.6$ & $2.6\cdot 10^{-4}$ & $\mathbf{0.999}$ & $0.10$ & $3$ & $1.9$\\
$0.4$ & $6.7\cdot 10^{-4}$ & $\mathbf{0.999}$ & $0.15$ & $3$ & $2.5$\\
$0.2$ & $3.0\cdot 10^{-3}$ & $\mathbf{1.000}$ & $0.34$ & $3$ & $4.1$\\
\hline
\end{tabular}

\vspace{0.5em}

\begin{tabular}{|c|cccc|c|}
\hline
& \multicolumn{4}{c|}{KOSE-Gaussian, $R=1,000$} & \\
$a$ & MSE & Cov. & Width & T & $\theta_\star^{\operatorname{KS}}$\\
\hline
$1$ & $1.8\cdot 10^{-4}$ & $\mathbf{0.990}$ & $0.06$ & $4$ & $1.2$\\
$0.8$ & $7.4\cdot 10^{-4}$ & $0.842$ & $0.08$ & $4$ & $1.5$\\
$0.6$ & $8.0\cdot 10^{-7}$ & $\mathbf{0.942}$ & $0.10$ & $3$ & $1.9$\\
$0.4$ & $1.0\cdot 10^{-3}$ & $\mathbf{0.997}$ & $0.17$ & $4$ & $2.5$\\
$0.2$ & $1.9\cdot 10^{-2}$ & $0.833$ & $0.37$ & $4$ & $4.1$\\
\hline
\end{tabular}
\vspace{0.5em}

\begin{tabular}{|c|ccc|ccc|}
\hline
 & \multicolumn{3}{c|}{ABC-AR, $R=1,000$} & \multicolumn{3}{c|}{NPL-MMD, $R=100$} \\
$a$ & Cov. & Width & T & Cov. & Width & T \\
\hline
$1$ & $\mathbf{0.994}$ & $0.04$ & $3$ &  $0.63$ & $0.09$ & $81$ \\
$0.8$ & $0.000$ & $0.06$ & $3$ &  $0.00$ & $0.10$ & $77$ \\
$0.6$ & $0.000$ & $0.13$ & $3$ &  $0.00$ & $0.12$ & $79$ \\
$0.4$ & $0.000$ & $0.08$ & $3$ &  $0.00$ & $0.15$ & $76$ \\
$0.2$ & $0.000$ & $0.24$ & $3$ &  $0.00$ & $0.22$ & $78$ \\
\hline
\end{tabular}

\vspace{0.5em}

\begin{tabular}{|c|ccc|}
\hline
 & \multicolumn{3}{c|}{ESet, $R=1,000$} \\
$a$ & Cov. & Empty & T \\
\hline
$1$ & $\mathbf{0.989}$ & $0$ & $3$\\
$0.8$ & $0.000$ & $0$ & $3$\\
$0.6$ & $0.000$ & $0$ & $3$\\
$0.4$ & NA & $1,000$ & $3$\\
$0.2$ & NA & $1,000$ & $3$\\
\hline
\end{tabular}
\end{table}
\begin{table}[htb]
\centering
\caption{Experiment 4 results; Here, $m=n=1,000$. MSE, $\mu$ and MSE, $\lambda$ are the average mean-squared errors for the service and arrival rates.}\label{tab: 4}
\vspace{1ex}
\begin{tabular}{|c|cccc|c|}
\hline
& \multicolumn{4}{c|}{KOSE-Riesz, $R=1,000$} & \\
$a$ & MSE, $\mu$ & MSE, $\lambda$ & Cov. & T & $\theta_\star^{\operatorname{KS}}$\\
\hline
$1$ & $1.3\cdot 10^{-2}$ & $4.7\cdot 10^{-2}$ & $\mathbf{1}$ & $10$ & $(2.5, 1.0)$\\
$0.8$ & $1.2\cdot 10^{-2}$ & $3.1\cdot 10^{-4}$ & $\mathbf{1}$ & $9$ & $(6.0, 2.9)$\\
$0.6$ & $4.2\cdot 10^{-3}$ & $4.7\cdot 10^{-2}$ & $\mathbf{1}$ & $9$ & $(7.9, 3.1)$\\
$0.4$ & $2.3\cdot 10^{-4}$ & $7.6\cdot 10^{-4}$ & $0$ & $9$ & $(10, 2.9)$\\
$0.2$ & $1.2\cdot 10^{-5}$ & $4.3\cdot 10^{-3}$ & $0$ & $9$ & $(9.5, 1.0)$\\
\hline
\end{tabular}

\vspace{0.5em}

\begin{tabular}{|c|cccc|c|}
\hline
& \multicolumn{4}{c|}{KOSE-Gaussian, $R=1,000$} & \\
$a$ & MSE, $\mu$ & MSE, $\lambda$ & Cov. & T & $\theta_\star^{\operatorname{KS}}$\\
\hline
$1$ & $1.1\cdot 10^{-2}$ & $4.0\cdot 10^{-3}$ & $\mathbf{1}$ & $9$ & $(2.5, 1.0)$\\
$0.8$ & $4.6\cdot 10^{-2}$ & $2.0 \cdot 10^{-2}$ & $\mathbf{1}$ & $9$ & $(6.0, 2.9)$\\
$0.6$ & $2.0\cdot 10^{-2}$ & $1.1\cdot 10^{-2}$ & $\mathbf{1}$ & $9$ & $(7.9, 3.1)$\\
$0.4$ & $4.6\cdot 10^{-2}$ & $8.4\cdot 10^{-3}$ & $0$ & $9$ & $(10, 2.9)$\\
$0.2$ & $6.7\cdot 10^{-5}$ & $3.9\cdot 10^{-1}$ & $0$ & $9$ & $(9.5, 1.0)$\\
\hline
\end{tabular}

\vspace{0.5em}

\begin{tabular}{|c|ccc|}
\hline
 & \multicolumn{3}{c|}{ABC-AR, NPL-MMD, $R=100$} \\
$a$ & Cov. & Width, $\mu, \lambda$ & T \\
\hline
$1$ & $0.41; 0.59$ & $0.54, 0.69; 0.74, 0.62$ & $88; 66$ \\
$0.8$ & $0.00; 0.59$ & $0.24, 0.25; 1.99, 1.44$ & $88; 64$ \\
$0.6$ & $0.00; 0.32$ & $0.26, 0.24; 3.11, 1.65$ & $88; 63$ \\
$0.4$ & $0.00; 0.66$ & $0.32, 0.24; 8.00, 4.67$ & $88; 58$ \\
$0.2$ & $0.00; 0.27$ & $0.23, 0.35; 0.06, 4.91$ & $87; 55$ \\
\hline
\end{tabular}

\vspace{0.5em}

\begin{tabular}{|c|ccc|}
\hline
 & \multicolumn{3}{c|}{ESet, $R=1,000$} \\
$a$ & Cov. & Empty & T \\
\hline
$1$ & $0$ & $3$ & $2$\\
$0.8$ & NA & $1,000$ & $3$\\
$0.6$ & NA & $1,000$ & $3$\\
$0.4$ & NA & $1,000$ & $3$\\
$0.2$ & NA & $1,000$ & $3$\\
\hline
\end{tabular}
\end{table}

Additional experiments on different choices of $\beta$ in KOSE-Riesz, sensitivity of the simulation sample size $n$, bias in uncertainty evaluation, and contamination models are reported in Sections~\ref{sec:sens_beta}, ~\ref{sec:sens_n}, ~\ref{sec:unc_ev}, and ~\ref{sec:extra_numericals} in the Supplementary Material, respectively.

\section{CONCLUSION}
\label{sec:conclusion}
In this paper, we introduce the first frequentist method to learn and quantify the uncertainties of differentiable simulation parameters with output-level data under model inexactness.
Our approach is based on kernel score minimization and a new asymptotic normality result.
We demonstrate the superior numerical performance of our approach for both exact and inexact G/G/1 queueing models.
Our approach produces accurate estimators with low mean-squared error (MSE) and more valid confidence sets, outperforming both Bayesian \citep{beaumont2002approximate, dellaporta2022robust} and frequentist \citep{bai2020calibrating} methods.

In future work, we plan to explore methods to reduce the cost of optimization, such as the weighted MMD approach \citep{bharti2023optimally} for MMD estimation or using natural gradient descent \citep{briol2019statistical}.
We also aim to explore avenues for simplifying our assumptions, particularly in the i.i.d. assumption and Assumption~\ref{ass:combined_assumption}.
Extending our results to non-i.i.d. scenarios presents a significant challenge, primarily due to the absence of a law of large numbers for generalized U-statistics in such settings.
Moreover, many of the stringent conditions in Assumption~\ref{ass:combined_assumption} stem from Theorem~\ref{thm:SLLN_Ustat}.
One potential approach to simplify these assumptions is establishing asymptotic normality through the Hilbert space version of the central limit theorem  \citep{oates2022minimum}.

\bibliographystyle{apalike}
\bibliography{sample_paper}

\begin{thebibliography}{}

\bibitem[Bai and Lam, 2020]{bai2020calibrating}
Bai, Y. and Lam, H. (2020).
\newblock Calibrating input parameters via eligibility sets.
\newblock In {\em 2020 Winter Simulation Conference (WSC)}, pages 2114--2125. IEEE.

\bibitem[Balci and Sargent, 1982]{balci1982some}
Balci, O. and Sargent, R.~G. (1982).
\newblock Some examples of simulation model validation using hypothesis testing.
\newblock Technical report, Institute of Electrical and Electronics Engineers (IEEE).

\bibitem[Baringhaus and Franz, 2004]{baringhaus2004new}
Baringhaus, L. and Franz, C. (2004).
\newblock On a new multivariate two-sample test.
\newblock {\em Journal of multivariate analysis}, 88(1):190--206.

\bibitem[Barton et~al., 2022]{barton_input_2022}
Barton, R.~R., Lam, H., and Song, E. (2022).
\newblock Input {Uncertainty} in {Stochastic} {Simulation}.
\newblock In Salhi, S. and Boylan, J., editors, {\em The {Palgrave} {Handbook} of {Operations} {Research}}, pages 573--620. Springer International Publishing, Cham.

\bibitem[Basu et~al., 2011]{basu2011statistical}
Basu, A., Shioya, H., and Park, C. (2011).
\newblock {\em Statistical inference: the minimum distance approach}.
\newblock CRC press.

\bibitem[Beaumont et~al., 2002]{beaumont2002approximate}
Beaumont, M.~A., Zhang, W., and Balding, D.~J. (2002).
\newblock Approximate bayesian computation in population genetics.
\newblock {\em Genetics}, 162(4):2025--2035.

\bibitem[Bharti et~al., 2023]{bharti2023optimally}
Bharti, A., Naslidnyk, M., Key, O., Kaski, S., and Briol, F.-X. (2023).
\newblock Optimally-weighted estimators of the maximum mean discrepancy for likelihood-free inference.
\newblock In {\em International Conference on Machine Learning}, pages 2289--2312. PMLR.

\bibitem[Biller et~al., 2022]{biller2022simulation}
Biller, B., Jiang, X., Yi, J., Venditti, P., and Biller, S. (2022).
\newblock Simulation: The critical technology in digital twin development.
\newblock In {\em 2022 Winter Simulation Conference (WSC)}, pages 1340--1355. IEEE.

\bibitem[Bińkowski et~al., 2018]{binkowski2018demystifying}
Bińkowski, M., Sutherland, D.~J., Arbel, M., and Gretton, A. (2018).
\newblock Demystifying {MMD} {GAN}s.
\newblock In {\em International Conference on Learning Representations}.

\bibitem[Briol et~al., 2019]{briol2019statistical}
Briol, F.-X., Barp, A., Duncan, A.~B., and Girolami, M. (2019).
\newblock Statistical inference for generative models with maximum mean discrepancy.
\newblock {\em arXiv preprint arXiv:1906.05944}.

\bibitem[Chambers and Mount-Campbell, 2002]{chambers2002process}
Chambers, M. and Mount-Campbell, C. (2002).
\newblock Process optimization via neural network metamodeling.
\newblock {\em International Journal of Production Economics}, 79(2):93--100.

\bibitem[Ch{\'e}rief-Abdellatif and Alquier, 2020]{cherief2020mmd}
Ch{\'e}rief-Abdellatif, B.-E. and Alquier, P. (2020).
\newblock Mmd-bayes: Robust bayesian estimation via maximum mean discrepancy.
\newblock In {\em Symposium on Advances in Approximate Bayesian Inference}, pages 1--21. PMLR.

\bibitem[Ch{\'e}rief-Abdellatif and Alquier, 2022]{cherief2022finite}
Ch{\'e}rief-Abdellatif, B.-E. and Alquier, P. (2022).
\newblock Finite sample properties of parametric mmd estimation: robustness to misspecification and dependence.
\newblock {\em Bernoulli}, 28(1):181--213.

\bibitem[Cranmer et~al., 2020]{cranmer2020frontier}
Cranmer, K., Brehmer, J., and Louppe, G. (2020).
\newblock The frontier of simulation-based inference.
\newblock {\em Proceedings of the National Academy of Sciences}, 117(48):30055--30062.

\bibitem[Dawid, 2007]{dawid2007geometry}
Dawid, A.~P. (2007).
\newblock The geometry of proper scoring rules.
\newblock {\em Annals of the Institute of Statistical Mathematics}, 59:77--93.

\bibitem[Dawid and Sebastiani, 1999]{dawid1999coherent}
Dawid, A.~P. and Sebastiani, P. (1999).
\newblock Coherent dispersion criteria for optimal experimental design.
\newblock {\em Annals of Statistics}, pages 65--81.

\bibitem[Dellaporta et~al., 2022]{dellaporta2022robust}
Dellaporta, C., Knoblauch, J., Damoulas, T., and Briol, F.-X. (2022).
\newblock Robust bayesian inference for simulator-based models via the mmd posterior bootstrap.
\newblock In {\em International Conference on Artificial Intelligence and Statistics}, pages 943--970. PMLR.

\bibitem[Dellaporta et~al., 2024]{npl_mmd_project}
Dellaporta, C., Knoblauch, J., Damoulas, T., and Briol, F.-X. (Accessed 04-17-2024).
\newblock npl\_mmd\_project.
\newblock \url{https://github.com/haritadell/npl_mmd_project}.

\bibitem[Dziugaite et~al., 2015]{dziugaite2015training}
Dziugaite, G.~K., Roy, D.~M., and Ghahramani, Z. (2015).
\newblock Training generative neural networks via maximum mean discrepancy optimization.
\newblock In {\em Proceedings of the Thirty-First Conference on Uncertainty in Artificial Intelligence}, pages 258--267.

\bibitem[Frazier and Drovandi, 2021]{frazier2021robust}
Frazier, D.~T. and Drovandi, C. (2021).
\newblock Robust approximate bayesian inference with synthetic likelihood.
\newblock {\em Journal of Computational and Graphical Statistics}, 30(4):958--976.

\bibitem[Frazier et~al., 2020a]{frazier2020robust}
Frazier, D.~T., Drovandi, C., and Loaiza-Maya, R. (2020a).
\newblock Robust approximate bayesian computation: An adjustment approach.
\newblock {\em arXiv preprint arXiv:2008.04099}.

\bibitem[Frazier et~al., 2021]{frazier2021synthetic}
Frazier, D.~T., Drovandi, C., and Nott, D.~J. (2021).
\newblock Synthetic likelihood in misspecified models: Consequences and corrections.
\newblock {\em arXiv preprint arXiv:2104.03436}.

\bibitem[Frazier et~al., 2020b]{frazier2020model}
Frazier, D.~T., Robert, C.~P., and Rousseau, J. (2020b).
\newblock Model misspecification in approximate bayesian computation: consequences and diagnostics.
\newblock {\em Journal of the Royal Statistical Society Series B: Statistical Methodology}, 82(2):421--444.

\bibitem[Frazier et~al., 2022]{frazier2022modeling}
Frazier, P.~I., Cashore, J.~M., Duan, N., Henderson, S.~G., Janmohamed, A., Liu, B., Shmoys, D.~B., Wan, J., and Zhang, Y. (2022).
\newblock Modeling for covid-19 college reopening decisions: Cornell, a case study.
\newblock {\em Proceedings of the National Academy of Sciences}, 119(2):e2112532119.

\bibitem[Fujisawa et~al., 2021]{fujisawa2021gamma}
Fujisawa, M., Teshima, T., Sato, I., and Sugiyama, M. (2021).
\newblock $\gamma$-abc: Outlier-robust approximate bayesian computation based on a robust divergence estimator.
\newblock In {\em International Conference on Artificial Intelligence and Statistics}, pages 1783--1791. PMLR.

\bibitem[Fukumizu et~al., 2007]{fukumizu2007kernel}
Fukumizu, K., Gretton, A., Sun, X., and Sch{\"o}lkopf, B. (2007).
\newblock Kernel measures of conditional dependence.
\newblock {\em Advances in Neural Information Processing Systems}, 20.

\bibitem[Gneiting and Raftery, 2007]{gneiting2007strictly}
Gneiting, T. and Raftery, A.~E. (2007).
\newblock Strictly proper scoring rules, prediction, and estimation.
\newblock {\em Journal of the American statistical Association}, 102(477):359--378.

\bibitem[Goeva et~al., 2019]{goeva2019optimization}
Goeva, A., Lam, H., Qian, H., and Zhang, B. (2019).
\newblock Optimization-based calibration of simulation input models.
\newblock {\em Operations Research}, 67(5):1362--1382.

\bibitem[Good, 1952]{good1952rational}
Good, I. (1952).
\newblock Rational decisions.
\newblock {\em Journal of the Royal Statistical Society. Series B (Methodological)}, pages 107--114.

\bibitem[Goodfellow et~al., 2014]{goodfellow2014generative}
Goodfellow, I., Pouget-Abadie, J., Mirza, M., Xu, B., Warde-Farley, D., Ozair, S., Courville, A., and Bengio, Y. (2014).
\newblock Generative adversarial nets.
\newblock {\em Advances in Neural Information Processing Systems}, 27.

\bibitem[Gretton et~al., 2012]{gretton2012kernel}
Gretton, A., Borgwardt, K.~M., Rasch, M.~J., Sch{\"o}lkopf, B., and Smola, A. (2012).
\newblock A kernel two-sample test.
\newblock {\em The Journal of Machine Learning Research}, 13(1):723--773.

\bibitem[Kelly et~al., 2023]{kelly2023misspecification}
Kelly, R.~P., Nott, D.~J., Frazier, D.~T., Warne, D.~J., and Drovandi, C. (2023).
\newblock Misspecification-robust sequential neural likelihood.
\newblock {\em arXiv preprint arXiv:2301.13368}.

\bibitem[Kennedy and O'Hagan, 2001]{kennedy2001bayesian}
Kennedy, M.~C. and O'Hagan, A. (2001).
\newblock Bayesian calibration of computer models.
\newblock {\em Journal of the Royal Statistical Society: Series B (Statistical Methodology)}, 63(3):425--464.

\bibitem[Kingma and Ba, 2015]{KingmaB14}
Kingma, D.~P. and Ba, J. (2015).
\newblock Adam: {A} method for stochastic optimization.
\newblock In {\em International Conference on Learning Representations}.

\bibitem[Kingma and Welling, 2014]{kingma2013auto}
Kingma, D.~P. and Welling, M. (2014).
\newblock {Auto-Encoding Variational Bayes}.
\newblock In {\em International Conference on Learning Representations}.

\bibitem[Kleijnen, 1995]{kleijnen1995verification}
Kleijnen, J.~P. (1995).
\newblock Verification and validation of simulation models.
\newblock {\em European Journal of Operational Research}, 82(1):145--162.

\bibitem[Li et~al., 2015]{li2015generative}
Li, Y., Swersky, K., and Zemel, R. (2015).
\newblock Generative moment matching networks.
\newblock In {\em International Conference on Machine Learning}, pages 1718--1727. PMLR.

\bibitem[Lindley, 1952]{lindley1952theory}
Lindley, D.~V. (1952).
\newblock The theory of queues with a single server.
\newblock In {\em Mathematical proceedings of the Cambridge philosophical society}, volume~48, pages 277--289. Cambridge University Press.

\bibitem[Liu et~al., 2020]{liu2020learning}
Liu, F., Xu, W., Lu, J., Zhang, G., Gretton, A., and Sutherland, D.~J. (2020).
\newblock Learning deep kernels for non-parametric two-sample tests.
\newblock In {\em International Conference on Machine Learning}, pages 6316--6326. PMLR.

\bibitem[Oates et~al., 2022]{oates2022minimum}
Oates, C. et~al. (2022).
\newblock Minimum kernel discrepancy estimators.
\newblock {\em arXiv preprint arXiv:2210.16357}.

\bibitem[Ojeda et~al., 2021]{ojeda2021learning}
Ojeda, C., Cvejoski, K., Georgiev, B., Bauckhage, C., Schuecker, J., and S{\'a}nchez, R.~J. (2021).
\newblock Learning deep generative models for queuing systems.
\newblock In {\em Proceedings of the AAAI Conference on Artificial Intelligence}, volume~35, pages 9214--9222.

\bibitem[Pacchiardi and Dutta, 2021]{pacchiardi2021generalized}
Pacchiardi, L. and Dutta, R. (2021).
\newblock Generalized bayesian likelihood-free inference using scoring rules estimators.
\newblock {\em arXiv preprint arXiv:2104.03889}, 2(8).

\bibitem[Pacchiardi et~al., 2024]{pacchiardi2024generalized}
Pacchiardi, L., Khoo, S., and Dutta, R. (2024).
\newblock Generalized bayesian likelihood-free inference.
\newblock {\em Electronic Journal of Statistics}, 18(2):3628--3686.

\bibitem[Pan et~al., 2021]{pan2021simulation}
Pan, Y., Xu, Z., Guang, J., Chen, X., Dai, J., Wang, C., Zhang, X., Sun, J., Shi, P., Ding, Y., Wu, S., Yang, K., and Pan, H. (2021).
\newblock A high-fidelity, machine-learning enhanced queueing network simulation model for hospital ultrasound operations.
\newblock In {\em 2021 Winter Simulation Conference (WSC)}, pages 1--12.

\bibitem[Paszke et~al., 2019]{paszke2019pytorch}
Paszke, A., Gross, S., Massa, F., Lerer, A., Bradbury, J., Chanan, G., Killeen, T., Lin, Z., Gimelshein, N., Antiga, L., et~al. (2019).
\newblock Pytorch: An imperative style, high-performance deep learning library.
\newblock {\em Advances in Neural Information Processing Systems}, 32.

\bibitem[Plumlee, 2017]{plumlee2017bayesian}
Plumlee, M. (2017).
\newblock Bayesian calibration of inexact computer models.
\newblock {\em Journal of the American Statistical Association}, 112(519):1274--1285.

\bibitem[Plumlee, 2019]{plumlee2019computer}
Plumlee, M. (2019).
\newblock Computer model calibration with confidence and consistency.
\newblock {\em Journal of the Royal Statistical Society: Series B (Statistical Methodology)}, 81(3):519--545.

\bibitem[Plumlee and Lam, 2017]{plumlee2017uncertainty}
Plumlee, M. and Lam, H. (2017).
\newblock An uncertainty quantification method for inexact simulation models.
\newblock {\em arXiv preprint arXiv:1707.06544}.

\bibitem[Polyak and Juditsky, 1992]{polyak1992acceleration}
Polyak, B.~T. and Juditsky, A.~B. (1992).
\newblock Acceleration of stochastic approximation by averaging.
\newblock {\em SIAM Journal on Control and Optimization}, 30(4):838--855.

\bibitem[Sargent, 2010]{sargent2010verification}
Sargent, R.~G. (2010).
\newblock Verification and validation of simulation models.
\newblock In {\em Proceedings of the 2010 Winter Simulation Conference}, pages 166--183. IEEE.

\bibitem[Schruben, 1980]{schruben1980establishing}
Schruben, L.~W. (1980).
\newblock Establishing the credibility of simulations.
\newblock {\em Simulation}, 34(3):101--105.

\bibitem[Sejdinovic et~al., 2013]{sejdinovic2013equivalence}
Sejdinovic, D., Sriperumbudur, B., Gretton, A., and Fukumizu, K. (2013).
\newblock Equivalence of distance-based and rkhs-based statistics in hypothesis testing.
\newblock {\em The annals of statistics}, pages 2263--2291.

\bibitem[Sen, 1977]{sen1977almost}
Sen, P.~K. (1977).
\newblock Almost sure convergence of generalized u-statistics.
\newblock {\em The Annals of Probability}, pages 287--290.

\bibitem[Smola et~al., 2007]{smola2007hilbert}
Smola, A., Gretton, A., Song, L., and Sch{\"o}lkopf, B. (2007).
\newblock A hilbert space embedding for distributions.
\newblock In {\em International conference on algorithmic learning theory}, pages 13--31. Springer.

\bibitem[Sohl-Dickstein et~al., 2015]{sohl2015deep}
Sohl-Dickstein, J., Weiss, E., Maheswaranathan, N., and Ganguli, S. (2015).
\newblock Deep unsupervised learning using nonequilibrium thermodynamics.
\newblock In {\em International conference on machine learning}, pages 2256--2265. PMLR.

\bibitem[Steinwart and Christmann, 2008]{steinwart2008support}
Steinwart, I. and Christmann, A. (2008).
\newblock {\em Support vector machines}.
\newblock Springer Science \& Business Media.

\bibitem[Storlie et~al., 2015]{storlie2015calibration}
Storlie, C.~B., Lane, W.~A., Ryan, E.~M., Gattiker, J.~R., and Higdon, D.~M. (2015).
\newblock Calibration of computational models with categorical parameters and correlated outputs via bayesian smoothing spline anova.
\newblock {\em Journal of the American Statistical Association}, 110(509):68--82.

\bibitem[Sutherland et~al., 2017]{sutherland2017generative}
Sutherland, D.~J., Tung, H.-Y., Strathmann, H., De, S., Ramdas, A., Smola, A., and Gretton, A. (2017).
\newblock Generative models and model criticism via optimized maximum mean discrepancy.
\newblock In {\em International Conference on Learning Representations}.

\bibitem[Sz{\'e}kely and Rizzo, 2013]{szekely2013energy}
Sz{\'e}kely, G.~J. and Rizzo, M.~L. (2013).
\newblock Energy statistics: A class of statistics based on distances.
\newblock {\em Journal of statistical planning and inference}, 143(8):1249--1272.

\bibitem[Tarantola, 2005]{tarantola2005inverse}
Tarantola, A. (2005).
\newblock {\em Inverse problem theory and methods for model parameter estimation}.
\newblock SIAM.

\bibitem[Tuo, 2019]{tuo2019adjustments}
Tuo, R. (2019).
\newblock Adjustments to computer models via projected kernel calibration.
\newblock {\em SIAM/ASA Journal on Uncertainty Quantification}, 7(2):553--578.

\bibitem[Tuo and Wu, 2015]{tuo2015efficient}
Tuo, R. and Wu, C.~J. (2015).
\newblock Efficient calibration for imperfect computer models.
\newblock {\em The Annals of Statistics}, 43(6):2331--2352.

\bibitem[Tuo and Wu, 2016]{tuo2016theoretical}
Tuo, R. and Wu, C.~J. (2016).
\newblock A theoretical framework for calibration in computer models: Parametrization, estimation and convergence properties.
\newblock {\em SIAM/ASA Journal on Uncertainty Quantification}, 4(1):767--795.

\bibitem[Van~der Vaart, 2000]{van2000asymptotic}
Van~der Vaart, A.~W. (2000).
\newblock {\em Asymptotic statistics}, volume~3.
\newblock Cambridge university press.

\bibitem[Ward et~al., 2022]{ward2022robust}
Ward, D., Cannon, P., Beaumont, M., Fasiolo, M., and Schmon, S. (2022).
\newblock Robust neural posterior estimation and statistical model criticism.
\newblock {\em Advances in Neural Information Processing Systems}, 35:33845--33859.

\bibitem[Wehenkel et~al., 2024]{wehenkel2024addressing}
Wehenkel, A., Gamella, J.~L., Sener, O., Behrmann, J., Sapiro, G., Cuturi, M., and Jacobsen, J.-H. (2024).
\newblock Addressing misspecification in simulation-based inference through data-driven calibration.
\newblock {\em arXiv preprint arXiv:2405.08719}.

\bibitem[Zhang et~al., 2021]{zhang2021analysis}
Zhang, Y., Lu, H., Zhou, Z., Yang, Z., and Xu, S. (2021).
\newblock Analysis and optimisation of perishable inventory with stocks-sensitive stochastic demand and two-stage pricing: A discrete-event simulation study.
\newblock {\em Journal of Simulation}, 15(4):326--337.

\end{thebibliography}


\newpage
\appendix
\onecolumn

\section{SUPPLEMENTARY MATERIAL}
\label{sec:appendix}
\subsection{G/G/1 Queueing System}
\label{sec:queue}
We provide a brief introduction to G/G/1 queues for interested readers. 
In probability theory, the G/G/1 queue is a single-server queue with the first-in-first-out (FIFO) discipline, meaning that the customer who first enters the server is also the first to exit the server. 
In G/G/1 queues, the interarrival times of customers follow a general distribution, and service times follow a different general distribution.
The dynamics of a G/G/1 queue are described by the Lindley equation (see Section~\ref{sec:diff_GG1} for details).

The M/M/1 queue and M/G/1 queue are two special cases of the G/G/1 queue. In M/M/1 queues, both the interarrival times and the service times are exponentially distributed. In M/G/1 queues, the interarrival times are exponentially distributed, while the service times follow a general distribution.

\subsection{Stochastic Simulation Models versus Deep Generative Models}
\label{sec:diff_deep_gen}
The key difference between deep generative models and stochastic simulation models lies in the deterministic structure of the latter, determined by target system mechanics. For instance, if the target system is an email server queue governed by a Poisson process for arrivals and unknown service time distributions, the stochastic simulation model would be an M/G/1 queueing model. Unlike deep generative models, altering the structure of stochastic simulation models, such as increasing network depth, is not applicable to enhance their approximation capabilities.

Consider the previous instance of modeling an email server queue. Given the unknown service time distribution of the target system, the model service time distribution likely does not align with that of the target system for any simulation parameter $\theta$. As a result, the stochastic simulation model is inherently inexact.

\subsection{Energy Distance and Maximum Mean Discrepancy (MMD)}
\label{sec:MMD}
The associated divergence function of the energy score is the energy distance 
\citep{baringhaus2004new, szekely2013energy}
\begin{equation}
\label{eq:Energy_dist}
\mathcal{E}^{(\beta)}(P, Q) = \mathbb{E}_{X, Y} \|X-Y\|^\beta - \frac{1}{2} \mathbb{E}_{X, X^{\prime}}\left\|X-X^{\prime}\right\|^\beta - \frac{1}{2} \mathbb{E}_{Y, Y^{\prime}}\left\|Y-Y^{\prime}\right\|^\beta,   \nonumber
\end{equation}
where $X, X^{\prime} \stackrel{\mathrm{i.i.d.}}{\sim} Q$.

The associated divergence function of the kernel score is the squared maximum mean discrepancy \citep{gretton2012kernel}:
\begin{equation}
\label{eq:sq_pop_MMD}
\operatorname{MMD}^2_k(P, Q)=\mathbb{E}_{X, X^{\prime}}\left[k\left(X, X^{\prime}\right)\right]-2 \mathbb{E}_{X, Y}[k(X, Y)]+\mathbb{E}_{Y, Y^{\prime}}\left[k\left(Y, Y^{\prime}\right)\right]. \nonumber
\end{equation}

\cite{sejdinovic2013equivalence} establishes the equivalence of the energy distance and the MMD.
The kernel score is strictly proper, and the MMD is a metric on the space of Borel probability measures such that $\mathbb{E}_{Y}\sqrt{k(Y, Y)} < \infty$, denoted $\mathcal{P}_k$, if the kernel $k$ is \emph{characteristic} \citep{fukumizu2007kernel}.

A positive-definite, measurable kernel $k$ is said to be characteristic if the \emph{kernel mean embedding} $\mu_k: \mathcal{P}_k \to \mathcal{H}_k$ \citep{smola2007hilbert} is injective, where $\mathcal{H}_k$ is the reproducing kernel Hilbert space (RKHS) associated with $k$ and equipped with inner product $\langle\cdot, \cdot \rangle_{\mathcal{H}_k}$ and norm $\|\cdot\|_{\mathcal{H}_k}$.

The kernel mean embedding $\mu_k$ is a continuous map that embeds a probability measure $P$ into an element $\mu_k^P = \mathbb{E}_{Y} \left[k(\cdot, Y)\right]$  of the RKHS $\mathcal{H}_k$.
The MMD can be expressed as the distance between kernel mean embeddings in $\mathcal{H}_k$: $\operatorname{MMD}_k(P, Q)=\|\mu_k^P - \mu_k^Q\|_{\mathcal{H}_k}$ \citep{gretton2012kernel}.
As a result, when $k$ is characteristic (i.e. $\mu_k$ is injective), $\|\mu_k^P - \mu_k^Q\|_{\mathcal{H}_k} = 0$ implies $P = Q$, the MMD is a metric on $\mathcal{P}_k$ and the kernel score is strictly proper.

\subsection{Useful Assumptions}
The following assumptions are those summarized in Assumption~\ref{ass:ubg_reg_conditions}.
\begin{assumption}[Finite moments]
\label{ass:finite_moment}
    $\mathbb{E}_{X} \|X\|^\alpha, \mathbb{E}_{Z} \|Z\|^\alpha < \infty$ for some $\alpha \geq 1$.
\end{assumption}
Assumption~\ref{ass:finite_moment} corresponds to Assumption~A in \cite{binkowski2018demystifying}.
\begin{assumption}
\label{ass:bounded_kernel}
    There exists constants $C_0, C_1$ such that 
\begin{equation}
\begin{gathered}
|k(x, y)| \leq C_0\left(\left(\|x\|^2+\|y\|^2\right)^{\alpha / 2}+1\right) \\
\left\|\nabla_{x, y} k(x, y)\right\| \leq C_1\left(\left(\|x\|^2+\|y\|^2\right)^{(\alpha-1) / 2}+1\right), \nonumber
\end{gathered}
\end{equation}
where $\alpha$ is the same $\alpha$ in Assumption~\ref{ass:finite_moment}.
\end{assumption}

The following two conditions (Assumption C and D in \cite{binkowski2018demystifying}) are imposed on the non-parametric components of $G_\theta$.
\begin{assumption}[Lipschitzness]
\label{ass:lipschitz_non_lin_layers}
Each non-parametric component of $G_\theta$, $\phi_i$, is M-Lipschitz.
\end{assumption}
\begin{assumption}[Piecewise analytic]
\label{ass:ana_pieces}
For $\phi_i$, there are $K_i$ functions $\phi_i^k$, $k = 1, \ldots, K_i$, each real analytic on the input space $\mathbb{R}^{d_{\phi_i}}$, where $d_{\phi_i}$ denotes the dimension of the input space, and agrees with $\phi_i$ on the closure of a set $\mathcal{D}^k$: $\phi_i(x)=\phi_i^k(x) \quad \forall x \in \overline{\mathcal{D}}_i^k.$
The sets $\mathcal{D}_i^k$ are disjoint, and cover the whole input space: $\bigcup_{k=1}^{K_i} \overline{\mathcal{D}}_i^k=\mathbb{R}^{d_{\pi}}$. 
Each $\mathcal{D}_i^k$ is defined by $S_{i, k}$ real analytic functions $G_{i, k, s}: \mathbb{R}^{d_{\phi_i}} \rightarrow \mathbb{R}$ as
$$
\mathcal{D}_i^k=\left\{x \in \mathbb{R}^{d_{\phi_i}} \mid  G_{i, k, s}(x)>0 \quad \forall s = 1, \ldots, S_{i, k}\right\} .
$$
\end{assumption}
\begin{assumption}[Differentiability]
\label{ass:para_diff}
Each parametric component of $G_\theta$, $\psi_{\theta, j}$, is almost everywhere differentiable with respect to $\theta$.  
\end{assumption}

The following regularity conditions are summarized in Assumption~\ref{ass:separable_kernel_bound}.
\begin{assumption}[Separability]
\label{ass:separable}
    The reproducing kernel Hilbert space $\mathcal{H}_k$ is separable.
\end{assumption}
\begin{assumption}
\label{ass:kernel_bound}
$\mathbb{E}_{X} \sqrt{k(X, X)} < \infty$, where $X \sim P_\star$.
\end{assumption}

The following regularity conditions are summarized in Assumption~\ref{ass:combined_assumption}.
\begin{assumption}
\label{ass:unif_finite_jacobian}
    There exists a open and convex neighbourhood of $\theta_{\star}^{\operatorname{KS}}$ denoted $\mathcal{N}$ such that $\mathbb{E}_{Z, X} \sup_{\theta \in \mathcal{N}} \left\|\nabla_{\theta} k(G_\theta(Z), X)\right\|  < \infty$ and $\mathbb{E}_{Z, Z^\prime} \sup_{\theta \in \mathcal{N}} \left\|\nabla_{\theta} k(G_\theta(Z), k(G_\theta(Z^\prime)\right\| < \infty$ for all $\theta \in \mathcal{N}$.
\end{assumption}
\begin{assumption}
\label{ass:finite_jacobian_l2}
$\mathbb{E}_{Z, X} \left\|\nabla_{\theta} k(G_\theta(Z), X)\right\|^2 \left.\right|_{\theta=\theta_{\star}^{\operatorname{KS}}}, \mathbb{E}_{Z, Z^\prime} \left\|\nabla_{\theta} k(G_\theta(Z), k(G_\theta(Z^\prime)\right\|^2 \left.\right|_{\theta=\theta_{\star}^{\operatorname{KS}}} < \infty$.
\end{assumption}
\begin{assumption}
\label{ass:finite_hessian_l2}
$\mathbb{E}_{Z, X} \left[ \left|\nabla_{\theta_r \theta_s} k(G_\theta(Z), X) \left.\right|_{\theta=\theta_{\star}^{\operatorname{KS}}}\right|  \log^{+}\left(\left|\nabla_{\theta_r \theta_s} k(G_\theta(Z), X) \left.\right|_{\theta=\theta_{\star}^{\operatorname{KS}}}\right|\right) \right], \\ \mathbb{E}_{Z, Z^\prime} \left[ \left|\nabla_{\theta_r \theta_s} k(G_\theta(Z), G_\theta(Z^\prime)) \left.\right|_{\theta=\theta_{\star}^{\operatorname{KS}}}\right|  \right] < \infty$ for all $r, s \in \{1, \ldots, p\}$, where $\theta = (\theta_1, \ldots, \theta_p) \in \mathbb{R}^p$.
\end{assumption}
\begin{assumption}
\label{ass:jacobian_diff}
    $\nabla_\theta k(G_\theta(z), x)$, $\nabla_\theta k(G_\theta(z), G_\theta(z^\prime))$ are differentiable on $\mathcal{N}$ for all $P$-almost all $z, z^\prime \in Z$ and $P_\star$-almost all $x \in \mathbb{R}^d$.
\end{assumption}
\begin{assumption}
\label{ass:unif_cont_hessian}
    $\nabla_{\theta \theta} k(G_\theta(z), x)$, $\nabla_{\theta \theta} k(G_\theta(z), G_\theta(z^\prime))$ are uniformly continuous at $\theta_{\star}^{\operatorname{KS}}$ for all $P$-almost all $z, z^\prime \in Z$ and $P_\star$-almost all $x \in \mathbb{R}^d$, where $\nabla_{\theta \theta}(\cdot)$ denotes the Hessian with respect to $\theta$
\end{assumption}
\begin{assumption}
\label{ass:pos_def}
The Hessian $H = \mathbb{E}_{Z, Z^\prime} \left[\nabla_{\theta \theta} k(G_\theta(Z), G_\theta(Z^\prime)) \left.\right|_{\theta=\theta_{\star}^{\operatorname{KS}}} \right] - 2\mathbb{E}_{Z, X} \left[\nabla_{\theta \theta} k(G_\theta(Z), X) \left.\right|_{\theta=\theta_{\star}^{\operatorname{KS}}} \right]$
is positive definite.
\end{assumption}

\subsection{Known Results}
\label{sec:lemma}
When a stochastic simulation model is identifiable, the (kernel) optimum score estimator converges to the optimal simulation parameter almost surely.
\cite{oates2022minimum} provides the following regularity conditions.
\begin{assumption}[Separability and Kernel Bound]
\label{ass:separable_kernel_bound}
    The reproducing kernel Hilbert space $\mathcal{H}_k$ is separable, and $\mathbb{E}_{X} \sqrt{k(X, X)} < \infty$, where $X \sim P_\star$.
\end{assumption}

Note that $\mathcal{H}_k$ is separable if the kernel $k$ corresponding to the RKHS $\mathcal{H}_k$ is continuous and its input space is separable \citep{steinwart2008support}. 
Since the input space $\mathbb{R}^d$ considered here is separable, $\mathcal{H}_k$ is separable for all continuous kernels on $\mathbb{R}^d$, which include the Gaussian, Laplacian, and Riesz kernel.
Condition $\mathbb{E}_{X} \sqrt{k(X, X)} < \infty$ ensures that $P_\star$ is in the space of Borel probability measures on which the corresponding MMD of kernel $k$ is a metric.
\begin{proposition}[Strong Consistency \citep{oates2022minimum}]
\label{thm:consistency}
If Assumption~\ref{ass:separable_kernel_bound} hold, and the optimal simulation parameter $\theta_\star^{\operatorname{KS}}$ is unique, then $\widehat{\theta}_{m}^{\operatorname{KS}}  \stackrel{\mathrm{a.s.}}{\to} \theta_\star^{\operatorname{KS}}$, where $\stackrel{\mathrm{a.s.}}{\to}$ denotes almost sure convergence.
\end{proposition}
Proposition~\ref{thm:consistency} provides the convergence guarantee for the kernel optimum score estimator without the speed of convergence. 

\begin{proposition}[Strong consistency under model non-identifiability \cite{oates2022minimum}]
If Assumptions~\ref{ass:separable} and \ref{ass:kernel_bound} hold, and the divergence function $d_{\operatorname{KS}}(P_\theta, P_\star)$ is continuous on $\Theta$, then any accumulation point of $\{\widehat{\theta}_m^{\operatorname{KS}}\}_{m \in \mathbb{N}}$ is almost surely an element of $\arg \min_{\theta \in \Theta} d_{\operatorname{KS}}(P_\theta, P_\star)$.
\end{proposition}
Note that Theorem~10 in \cite{oates2022minimum} gives the sufficient conditions for the existence of accumulation points of the sequence $\{\widehat{\theta}_m^{\operatorname{KS}}\}_{m \in \mathbb{N}}$.

The generalization bounds for MMD under bounded kernel require the following assumption \citep{briol2019statistical}.
\begin{assumption}
\label{ass:generalize}
\begin{enumerate}
    \item For every $Q \in \mathcal{P}_k$, there exists $c > 0$ such that the set $\{\theta \in \Theta: \operatorname{MMD}_k \left(P_\theta, Q\right) \leq \inf_{\theta^{\prime} \in \Theta} \operatorname{MMD}_k \left(P_{\theta^{\prime}}, Q\right)+c\}$ is bounded.
    \item For every $n \in \mathbb{N}$ and $Q \in \mathcal{P}_k$, there exists $c_n > 0$ such that the set $\{\theta \in \Theta: \operatorname{MMD}_k \left(\widehat{P}^{n}_\theta, Q\right) \leq \inf_{\theta^{\prime} \in \Theta} \operatorname{MMD}_k \left(\widehat{P}^{n}_{\theta^{\prime}}, Q\right)+c_n\}$ is bounded, where $\widehat{P}^{n}_\theta$ and $\widehat{P}^{n}_{\theta^{\prime}}$ are empirical measures.
\end{enumerate}
\end{assumption}
\begin{theorem}[Generalization bounds \citep{briol2019statistical}]
\label{thm:generalize}
    Suppose that the kernel $k$ is bounded, and Assumption~\ref{ass:generalize} holds, then with probability at least $1-\delta$,
    $$
    \operatorname{MMD}\left(P_{\widehat{\theta}_m^{\operatorname{KS}}}, Q\right) \leq \inf _{\theta \in \Theta} \operatorname{MMD}\left(P_\theta, Q\right)+2 \sqrt{\frac{2}{m} \sup _{x \in \mathbb{R}^d} k(x, x)}\left(2+\sqrt{\log \left(\frac{1}{\delta}\right)}\right),
    $$
    and
    $$
    \operatorname{MMD}\left(P_{\widehat{\theta}_{m, n}^{\operatorname{KS}}}, Q\right) \leq \inf _{\theta \in \Theta} \operatorname{MMD}\left(P_\theta, Q\right)+2\left(\sqrt{\frac{2}{n}}+\sqrt{\frac{2}{m}}\right) \sqrt{\sup _{x \in \mathbb{R}^d} k(x, x)}\left(2+\sqrt{\log \left(\frac{2}{\delta}\right)}\right).
    $$
\end{theorem}

The following results are used in proving Theorem~\ref{thm:asymp_norm}.
\begin{proposition}[Differentiation lemma \citep{binkowski2018demystifying}]
\label{prop:diff_prop}
    Let $\Theta \in \mathbb{R}^p$ be a non-trivial open set, $Q$ be a probability measure on $\mathbb{R}^m$ and $V$ is an $m$-dimensional random variable such that $V \sim Q$.
    Define a map $h: \mathbb{R}^m \times \Theta \mapsto \mathbb{R}^n$ with the following properties:
    \begin{enumerate}
        \item For any $\theta \in \Theta, \mathbb{E}_{V}\left[\left\|h_\theta(V)\right\|\right]<\infty$.
        \item For $Q$-almost all $v \in \mathbb{R}^m$, the map $\Theta \rightarrow \mathbb{R}^n, \theta \mapsto h_\theta(v)$ is differentiable.
        \item There exists a $Q$-integrable function $g: \mathbb{R}^m \mapsto \mathbb{R}$ such that $\left\|\nabla_\theta h_\theta(v)\right\| \leq g(v)$ for all $\theta \in \Theta$.
    \end{enumerate}
    Then, for any $\theta \in \Theta, \mathbb{E}_{V}\left\|\nabla_\theta h_\theta(V)\right\|<\infty$ and the function $\theta \mapsto \mathbb{E}_{V}\left[h_\theta(V)\right]$ is differentiable with differential
    $$
    \nabla_\theta \mathbb{E}_{V}\left[h_\theta(V)\right]=\mathbb{E}_{V}\left[\nabla_\theta h_\theta(V)\right].
    $$
\end{proposition}

\begin{theorem}[Strong law of large numbers for generalized $k$-sample U-statistics \citep{sen1977almost}]
\label{thm:SLLN_Ustat}
Let $\{X_{i,j}, i = 1, \ldots, k, j = 1, \ldots, n_i\}$ be a sequence of independent and identically distributed $d$-dimensional random variables with each $X_{i, j}$ having measure $P_i$.
For $m = (m_1, \ldots, m_k)$, consider the generalized U-statistic
$$
U(n) = \prod_{i=1}^k\left(\begin{array}{c}
n_i \\
m_i
\end{array}\right)^{-1} \sum \phi\left(X_{i, j_1}, \ldots, X_{i, j_{m_i}}, i = 1, \ldots, k\right),
$$
where $m_i \leq n_i$ is the degree vector and the summation $\sum$ extends over all possible $1\leq j_1 < \ldots < j_{m_i} \leq n_i, i = 1, \ldots, k$.
Suppose that $\mathbb{E}_{X_{i, j_1}, \ldots, X_{i, j_{m_i}}}\left[ \left|\phi\left(X_{i, j_1}, \ldots, X_{i, j_{m_i}}\right)\right|(\log^{+}\left|\phi\left(X_{i, j_1}, \ldots, X_{i, j_{m_i}}\right)\right|)^{k-1} \right] < \infty$, where $\log^{+}(x) = \max(0, \log(x))$. Then, as $n_1, \ldots, n_k \to \infty$, 
$$
U(n) \stackrel{\mathrm{a. s.}}{\to} \mathbb{E}_{X_{i, j_1}, \ldots, X_{i, j_{m_i}}}\left[\phi\left(X_{i, j_1}, \ldots, X_{i, j_{m_i}}\right)\right].
$$
\end{theorem}

\subsection{Optimal and True Simulation Parameter}
The following proposition gives the equivalence of the optimal simulation parameter and the true simulation parameter under model exactness and identifiability.
\begin{proposition}
\label{prop:opt_unique}
Suppose that the stochastic simulation model is exact and identifiable. 
If the scoring rule $S$ is strictly proper relative to $\mathcal{P}$, and $\{P_{\theta}: \theta \in \Theta \} \subset \mathcal{P}$, then $\theta_\star = \theta_0$.
\end{proposition} 
\begin{proof}
We prove by contradiction assuming $\theta_\star \neq \theta_0$. By uniqueness of $\theta_0$,  $P_{\star} \neq P_{\theta_\star}$. Since $S$ is strictly proper, $\mathbb{E}_{X \sim P_{\star}} S(P_{\star}, X) < \mathbb{E}_{X \sim P_{\star}} S(P_{\theta^\star}, X)$.  
Thus if $\theta_\star \neq \theta_0$, we have 
\begin{equation}
     \mathbb{E}_{X \sim P_{\star}} S(P_{\star}, X) < \mathbb{E}_{X \sim P_{\star}}S(P_{\theta_\star}, X) = L_{\star}(\theta_\star) \leq L_{\star}(\theta_0) = \mathbb{E}_{X \sim P_{\star}} S(P_{\theta_0}, X) = \mathbb{E}_{X \sim P_{\star}} S(P_{\star}, X). \nonumber
\end{equation}
\end{proof}

\subsection{Model Inexactness versus Model Contamination}
There are two types of model discrepancy: model contamination (correct model, contaminated data) and model inexactness (uncontaminated data, misspecified model). 
Our experiments varying the shape parameter $a$ of the service time distribution, such as Experiment 2 and 4, address the latter case, with $a$ providing a simple measure of model inexactness (smaller $a$, greater inexactness or discrepancy).
Section~\ref{sec:extra_numericals} addresses the former case.

Note that queueing and contamination models are distinct concepts. 
Queueing models are a specific class of stochastic simulation models, while any class of models can be contamination models if the data is contaminated. 

\subsection{Asymptotic Normality versus Posterior Consistency}
We want to emphasize the key distinctions between our asymptotic normality result (Theorem~\ref{thm:asymp_norm}) and Theorem~2 in \cite{pacchiardi2024generalized}. 
Our result is grounded in frequentist statistics, whereas \cite{pacchiardi2024generalized} provide a Bayesian posterior consistency result. 
Our theorem demonstrates that the distribution of the kernel optimum score estimator (KOSE) converges to a normal distribution centered at the optimal simulation parameter. 
In contrast, \cite{pacchiardi2024generalized} show that their posterior distribution concentrates around the optimal simulation parameter. 
Crucially, our result guarantees asymptotically correct coverage for our proposed confidence set procedure. This property is not achievable with the posterior consistency result of \cite{pacchiardi2024generalized}. As they note in Section 3.1 of their paper, their credible sets do not provide correct frequentist coverage, even under ideal conditions of strictly proper scoring rules and well-specified models. 

\subsection{Proof of Theorem~\ref{thm:asymp_norm}}
\label{proof:asymp_norm}
\begin{proof}
    We shall only prove the asymptotic normality of the kernel simulated score estimator $\widehat{\theta}_{m, n}^{\operatorname{KS}}$, as the proof for the asymptotic normality of the kernel optimum score estimator $\widehat{\theta}_{m}^{\operatorname{KS}}$ follows an entirely analogous way.

    The first order optimality condition implies that $\nabla_\theta \widehat{L}_{m, n}^{\operatorname{KS}}(\theta) \left.\right|_{\theta=\widehat{\theta}_{m, n}^{\operatorname{KS}}} = 0$.
    As $\widehat{\theta}_{m, n}^{\operatorname{KS}} \stackrel{\mathrm{a.s.}}{\to} \theta_{\star}^{\operatorname{KS}}$ as $m, n \to \infty$, $\widehat{\theta}_{m, n}^{\operatorname{KS}} \in \mathcal{N}$ a.s. for sufficiently large $m, n$. Then the mean value theorem implies that there a.s. exists a $\tilde{\theta}_{m, n}^{\operatorname{KS}} \in \mathcal{N}$ between $\theta_{\star}^{\operatorname{KS}}$ and $\widehat{\theta}_{m, n}^{\operatorname{KS}}$ such that
    \begin{equation}
        0 = \nabla_\theta \widehat{L}_{m, n}^{\operatorname{KS}}(\theta) \left.\right|_{\theta=\widehat{\theta}_{m, n}^{\operatorname{KS}}} =  \nabla_\theta \widehat{L}_{m, n}^{\operatorname{KS}}(\theta) \left.\right|_{\theta=\theta_{\star}^{\operatorname{KS}}} + \nabla_{\theta \theta} \widehat{L}_{m, n}^{\operatorname{KS}}(\theta) \left.\right|_{\theta=\tilde{\theta}_{m, n}^{\operatorname{KS}}} (\widehat{\theta}_{m, n}^{\operatorname{KS}} - \theta_{\star}^{\operatorname{KS}}). \nonumber
    \end{equation}
    If $\nabla_{\theta \theta} \widehat{L}_{m, n}^{\operatorname{KS}}(\theta) \left.\right|_{\theta=\tilde{\theta}_{m, n}^{\operatorname{KS}}}$ is non-singular,  $\sqrt{m+n}(\widehat{\theta}_{m, n}^{\operatorname{KS}} - \theta_{\star}^{\operatorname{KS}}) = -\nabla_{\theta \theta} \widehat{L}_{m, n}^{\operatorname{KS}}(\theta)^{-1} \left.\right|_{\theta=\tilde{\theta}_{m, n}^{\operatorname{KS}}} \cdot \sqrt{m+n} \nabla_\theta \widehat{L}_{m, n}^{\operatorname{KS}}(\theta) \left.\right|_{\theta=\theta_{\star}^{\operatorname{KS}}}$.

    We first show that $\nabla_{\theta \theta} \widehat{L}_{m, n}^{\operatorname{KS}}(\theta) \left.\right|_{\theta=\tilde{\theta}_{m, n}^{\operatorname{KS}}} \stackrel{\mathrm{p}}{\to} H$.
    We prove the stronger almost sure convergence.
    Observe that 
    \begin{equation}
    \label{eq:decomp_g}
        \nabla_{\theta \theta} \widehat{L}_{m, n}^{\operatorname{KS}}(\theta) \left.\right|_{\theta=\tilde{\theta}_{m, n}^{\operatorname{KS}}} = \nabla_{\theta \theta} \widehat{L}_{m, n}^{\operatorname{KS}}(\theta) \left.\right|_{\theta=\theta_{\star}^{\operatorname{KS}}} + \nabla_{\theta \theta} \widehat{L}_{m, n}^{\operatorname{KS}}(\theta) \left.\right|_{\theta=\tilde{\theta}_{m, n}^{\operatorname{KS}}} - \nabla_{\theta \theta} \widehat{L}_{m, n}^{\operatorname{KS}}(\theta) \left.\right|_{\theta=\theta_{\star}^{\operatorname{KS}}},
    \end{equation}
    where
    \begin{equation}
    \begin{aligned}
        \nabla_{\theta \theta} \widehat{L}_{m, n}^{\operatorname{KS}}(\theta) \left.\right|_{\theta=\theta_{\star}^{\operatorname{KS}}} = \frac{1}{n(n-1)} \sum_{1 \leq i \neq j \leq n} \nabla_{\theta \theta} k(&G_\theta(Z_i), G_\theta(Z_j)) \left.\right|_{\theta=\theta_{\star}^{\operatorname{KS}}} \\
        &- \frac{2}{mn}  \sum_{i=1}^{m} \sum_{j=1}^{n} \nabla_{\theta \theta} k(G_\theta(Z_j), X_i) \left.\right|_{\theta=\theta_{\star}^{\operatorname{KS}}}. \nonumber
    \end{aligned}
    \end{equation}

    Recall $\theta = (\theta_1, \ldots, \theta_p) \in \mathbb{R}^p$. 
    As Assumption~\ref{ass:finite_hessian_l2} holds, the strong law of large numbers for U-statistics (Theorem~\ref{thm:SLLN_Ustat}) implies that 
    \begin{equation}
    \frac{1}{n(n-1)} \sum_{1 \leq i \neq j \leq n} \nabla_{\theta_r \theta_s} k(G_\theta(Z_i), G_\theta(Z_j)) \left.\right|_{\theta=\theta_{\star}^{\operatorname{KS}}} \stackrel{\mathrm{a.s.}}{\to} \mathbb{E}_{Z, Z^\prime} \nabla_{\theta_r \theta_s} k(G_\theta(Z), G_\theta(Z^\prime)) \left.\right|_{\theta=\theta_{\star}^{\operatorname{KS}}}, \nonumber      
    \end{equation}
    and
    \begin{equation}
    \frac{2}{mn}  \sum_{i=1}^{m} \sum_{j=1}^{n} \nabla_{\theta_r \theta_s} k(G_\theta(Z_j), X_i) \left.\right|_{\theta=\theta_{\star}^{\operatorname{KS}}} \stackrel{\mathrm{a.s.}}{\to} 2 \mathbb{E}_{Z, X} \nabla_{\theta_r \theta_s} k(G_\theta(Z), X) \left.\right|_{\theta=\theta_{\star}^{\operatorname{KS}}}.  \nonumber      
    \end{equation}
    Therefore, 
    \begin{equation}
    \nabla_{\theta_r \theta_s} \widehat{L}_{m, n}^{\operatorname{KS}}(\theta) \left.\right|_{\theta=\theta_{\star}^{\operatorname{KS}}} \stackrel{\mathrm{a.s.}}{\to} \mathbb{E}_{Z, Z^\prime} \nabla_{\theta_r \theta_s} k(G_\theta(Z), G_\theta(Z^\prime)) \left.\right|_{\theta=\theta_{\star}^{\operatorname{KS}}} -  2 \mathbb{E}_{Z, X} \nabla_{\theta_r \theta_s} k(G_\theta(Z), X) \left.\right|_{\theta=\theta_{\star}^{\operatorname{KS}}} \nonumber 
    \end{equation}
    for all $r, s \in \{1, \ldots, p\}$. Then $\nabla_{\theta \theta} \widehat{L}_{m, n}^{\operatorname{KS}}(\theta) \left.\right|_{\theta=\theta_{\star}^{\operatorname{KS}}} \stackrel{\mathrm{a.s.}}{\to} H$.

    Now we just need to show that $\left\|\nabla_{\theta \theta} \widehat{L}_{m, n}^{\operatorname{KS}}(\theta) \left.\right|_{\theta=\tilde{\theta}_{m, n}^{\operatorname{KS}}} - \nabla_{\theta \theta} \widehat{L}_{m, n}^{\operatorname{KS}}(\theta) \left.\right|_{\theta=\theta_{\star}^{\operatorname{KS}}}\right\| \stackrel{\mathrm{p}}{\to} 0$.
    
    Assumption~\ref{ass:unif_cont_hessian} implies that for any $\epsilon > 0$ there exists an open neighborhood $\mathcal{N}_\epsilon \subset \mathcal{N}$ such that 
    \begin{equation}
    \left\|\nabla_{\theta \theta} k(G_\theta(z), x) - \nabla_{\theta \theta} k(G_\theta(z), x) \left.\right |_{\theta=\theta_{\star}^{\operatorname{KS}}} \right\|, \left\|\nabla_{\theta \theta} k(G_\theta(z), G_\theta(z^\prime)) - \nabla_{\theta \theta} k(G_\theta(z), G_\theta(z^\prime))\left.\right|_{\theta=\theta_{\star}^{\operatorname{KS}}} \right\| < \epsilon \nonumber
    \end{equation}
    for all $\theta \in \mathcal{N}_\epsilon$, $z, z^\prime \in \mathcal{Z}, x \in \mathbb{R}^d$.
    Since $\widehat{\theta}_{m, n}^{\operatorname{KS}} \stackrel{\mathrm{a.s.}}{\to} \theta_{\star}^{\operatorname{KS}}$ as $m, n \to \infty$, $\tilde{\theta}_{m, n}^{\operatorname{KS}} \stackrel{\mathrm{a.s.}}{\to} \theta_{\star}^{\operatorname{KS}}$ as $m, n \to \infty$, then $\tilde{\theta}_{m, n}^{\operatorname{KS}} \in \mathcal{N}_\epsilon$ a.s. for sufficiently large $m, n$.
    Therefore, 
    \begin{equation}
    \begin{aligned}
    &\left\|\nabla_{\theta \theta} k(G_\theta(z), x) \left.\right|_{\theta=\tilde{\theta}_{m, n}^{\operatorname{KS}}} - \nabla_{\theta \theta} k(G_\theta(z), x)\left.\right|_{\theta=\theta_{\star}^{\operatorname{KS}}} \right\|, \\ &\left\|\nabla_{\theta \theta} k(G_\theta(z), G_\theta(z^\prime)) \left.\right|_{\theta=\tilde{\theta}_{m, n}^{\operatorname{KS}}} - \nabla_{\theta \theta} k(G_\theta(z), G_\theta(z^\prime))\left.\right|_{\theta=\theta_{\star}^{\operatorname{KS}}} \right\| < \epsilon  \nonumber 
    \end{aligned}
    \end{equation}
    a.s. for all $z, z^\prime \in \mathcal{Z}, x \in \mathbb{R}^d$, which in turn implies that
    \begin{equation}
    \begin{aligned}
        &\left\|\nabla_{\theta \theta} \widehat{L}_{m, n}^{\operatorname{KS}}(\theta) \left.\right|_{\theta=\tilde{\theta}_{m, n}^{\operatorname{KS}}} - \nabla_{\theta \theta} \widehat{L}_{m, n}^{\operatorname{KS}}(\theta) \left.\right|_{\theta=\theta_{\star}^{\operatorname{KS}}}\right\| \\ =& \left\|\frac{1}{n(n-1)} \sum_{1 \leq i \neq j \leq n} \left(\nabla_{\theta \theta} k(G_\theta(Z_i), G_\theta(Z_j)) \left.\right|_{\theta=\tilde{\theta}_{m, n}^{\operatorname{KS}}} - k(G_\theta(Z_i), G_\theta(Z_j)) \left.\right|_{\theta=\theta_{\star}^{\operatorname{KS}}} \right) \right.\\ & \hspace{0.5in} \left.- \frac{2}{mn} \left(\sum_{i=1}^{m} \sum_{j=1}^{n} \nabla_{\theta \theta} k(G_\theta(Z_j), X_i)\left.\right|_{\theta=\tilde{\theta}_{m, n}^{\operatorname{KS}}} - k(G_\theta(Z_j), X) \left.\right|_{\theta=\theta_{\star}^{\operatorname{KS}}}\right)\right\| \\
        & \leq \frac{1}{n(n-1)} \sum_{1 \leq i \neq j \leq n} \left\|\nabla_{\theta \theta} k(G_\theta(Z_i), G_\theta(Z_j)) \left.\right|_{\theta=\tilde{\theta}_{m, n}^{\operatorname{KS}}} - k(G_\theta(Z_i), G_\theta(Z_j)) \left.\right|_{\theta=\theta_{\star}^{\operatorname{KS}}}\right\| \\ & \hspace{0.5in} + \frac{2}{mn} \sum_{i=1}^{m} \sum_{j=1}^{n} \left\|\nabla_{\theta \theta} k(G_\theta(Z_j), X_i)\left.\right|_{\theta=\tilde{\theta}_{m, n}^{\operatorname{KS}}} - k(G_\theta(Z_j), X) \left.\right|_{\theta=\theta_{\star}^{\operatorname{KS}}} \right\| \\
        & < 3 \epsilon. \nonumber
    \end{aligned}
    \end{equation}
Therefore, $\left\|\nabla_{\theta \theta} \widehat{L}_{m, n}^{\operatorname{KS}}(\theta) \left.\right|_{\theta=\tilde{\theta}_{m, n}^{\operatorname{KS}}} - \nabla_{\theta \theta} \widehat{L}_{m, n}^{\operatorname{KS}}(\theta) \left.\right|_{\theta=\theta_{\star}^{\operatorname{KS}}}\right\| \stackrel{\mathrm{a.s.}}{\to} 0$. Given $\nabla_{\theta \theta} \widehat{L}_{m, n}^{\operatorname{KS}}(\theta) \left.\right|_{\theta=\theta_{\star}^{\operatorname{KS}}} \stackrel{\mathrm{a.s.}}{\to} H$, \eqref{eq:decomp_g} implies that $\nabla_{\theta \theta} \widehat{L}_{m, n}^{\operatorname{KS}}(\theta) \left.\right|_{\theta=\tilde{\theta}_{m, n}^{\operatorname{KS}}} \stackrel{\mathrm{a.s.}}{\to} H$ as $m, n \to \infty$.
Since $H$ is positive-definite from Assumption~\ref{ass:pos_def}, $\nabla_{\theta \theta} \widehat{L}_{m, n}^{\operatorname{KS}}(\theta) \left.\right|_{\theta=\tilde{\theta}_{m, n}^{\operatorname{KS}}}$ is non-singular for sufficiently large $m, n$.

We now show that $\sqrt{m+n} \nabla_\theta \widehat{L}_{m, n}^{\operatorname{KS}}(\theta) \left.\right|_{\theta=\theta_{\star}^{\operatorname{KS}}} \stackrel{\mathrm{d}}{\to} N(0, C_\lambda)$. 
Observe that
\begin{equation}
    \begin{aligned}
     \nabla_\theta \widehat{L}_{m, n}^{\operatorname{KS}}(\theta) \left.\right|_{\theta=\theta_{\star}^{\operatorname{KS}}} = \frac{1}{n(n-1)} \sum_{1 \leq i \neq j \leq n} \nabla_\theta k(G_\theta(Z_i),& G_\theta(Z_j)) \left.\right|_{\theta=\theta_{\star}^{\operatorname{KS}}} \\&- \frac{2}{mn}  \sum_{i=1}^{m} \sum_{j=1}^{n} \nabla_\theta k(G_\theta(Z_j), X_i) \left.\right|_{\theta=\theta_{\star}^{\operatorname{KS}}}. \nonumber   
    \end{aligned}
\end{equation}
Denote 
\begin{equation}
U_1 = \frac{1}{n(n-1)} \sum_{1 \leq i \neq j \leq n} \nabla_\theta k(G_\theta(Z_i), G_\theta(Z_j)) \left.\right|_{\theta=\theta_{\star}^{\operatorname{KS}}} \nonumber    
\end{equation}
and 
\begin{equation}
U_2 = \frac{2}{mn}  \sum_{i=1}^{m} \sum_{j=1}^{n} \nabla_\theta k(G_\theta(Z_j), X_i) \left.\right|_{\theta=\theta_{\star}^{\operatorname{KS}}}.   \nonumber 
\end{equation}
Both $U_1$ and $U_2$ are U-statistics.
Assumption~\ref{ass:unif_finite_jacobian} implies that there exists integrable functions $b_1: Z \times \mathbb{R}^d \to \mathbb{R}$ and $b_2: Z \times Z \to \mathbb{R}$ such that $\left\|\nabla_\theta k(G_\theta(Z), X)\right\| \leq b_1(Z, X)$ and $\left\|\nabla_\theta k(G_\theta(Z), G_\theta(Z^\prime))\right\| \leq b_2(Z, Z^\prime)$, $\mathbb{E}_{Z, X} \left|b_1(Z, X)\right|, \mathbb{E}_{Z, Z^\prime} \left|b_2(Z, Z^\prime)\right| < \infty$.

As Assumptions~\ref{ass:finite_jacobian_l2} and \ref{ass:jacobian_diff} hold, Proposition~\ref{prop:diff_prop} implies that 
\begin{equation}
\mathbb{E}_{Z_1, \ldots, Z_n, X}  \nabla_\theta \widehat{L}_{m, n}^{\operatorname{KS}}(\theta) \left.\right|_{\theta=\theta_{\star}^{\operatorname{KS}}} = \nabla_\theta \mathbb{E}_{Z_1, \ldots, Z_n, X} \widehat{L}_{m, n}^{\operatorname{KS}}(\theta) \left.\right|_{\theta=\theta_{\star}^{\operatorname{KS}}} = \nabla_\theta \operatorname{MMD}_k(P_\theta, P_\star)\left.\right|_{\theta=\theta_{\star}^{\operatorname{KS}}}.   
\end{equation}
By the first order optimality condition, $\nabla_\theta \operatorname{MMD}_k(P_\theta, P_\star)\left.\right|_{\theta=\theta_{\star}^{\operatorname{KS}}} = 0$.

Therefore, $\mathbb{E}_{Z_1, \ldots, Z_n, X}  \nabla_\theta \widehat{L}_{m, n}^{\operatorname{KS}}(\theta) \left.\right|_{\theta=\theta_{\star}^{\operatorname{KS}}} = 0$, which implies that 
\begin{equation}
\begin{aligned}
\mathbb{E} U_1 &= \mathbb{E}_{Z_1, \ldots, Z_n} \frac{1}{n(n-1)} \sum_{1 \leq i \neq j \leq n} \nabla_\theta k(G_\theta(Z_i), G_\theta(Z_j)) \left.\right|_{\theta=\theta_{\star}^{\operatorname{KS}}} \\
&= \frac{2}{mn}  \sum_{i=1}^{m} \sum_{j=1}^{n} \mathbb{E}_{Z_1, \ldots, Z_n, X} \nabla_\theta k(G_\theta(Z_j), X_i) \left.\right|_{\theta=\theta_{\star}^{\operatorname{KS}}} \\
&= \mathbb{E} U_2. \nonumber
\end{aligned}
\end{equation}
Now denote $\mathbb{E} U_1 = \mathbb{E} U_2$ by $M$. 
We need to show that
$\sqrt{n}(U_1 - \widehat{U}_1) \stackrel{\mathrm{p}}{\to} 0$ and $\sqrt{m+n}(U_2 - \widehat{U}_2) \stackrel{\mathrm{p}}{\to} 0$, where $\widehat{U}_1 = \frac{2}{n} \sum_{j=1}^{n} h(Z_j), \widehat{U}_2 = \frac{1}{m} \sum_{i=1}^{m} g_{1, 0}(X_i) + \frac{1}{n} \sum_{j=1}^{n} g_{0, 1}(Z_j)$ are the Hájek projections of $U_1$ and $U_2$, respectively.
As $\sqrt{m+n}(U_1 - U_2) = \sqrt{m+n}\left(\widehat{U}_1 - \widehat{U}_2 + (U_1 - \widehat{U}_1) + (U_2 - \widehat{U}_2)\right)$,
the asymptotic distribution of $\sqrt{m+n}(U_1 - U_2)$ is the same as the asymptotic distribution of $\sqrt{m+n}(\widehat{U}_1 - \widehat{U}_2)$ if $\sqrt{n}(U_1 - \widehat{U}_1) \stackrel{\mathrm{p}}{\to} 0$ and $\sqrt{m+n}(U_2 - \widehat{U}_2) \stackrel{\mathrm{p}}{\to} 0$.

We only prove $\sqrt{m+n}(U_2 - \widehat{U}_2) \stackrel{\mathrm{p}}{\to} 0$ as the proof of $\sqrt{n}(U_1 - \widehat{U}_1) \stackrel{\mathrm{p}}{\to} 0$ follows an entirely analogous way.
We prove the stronger $L_2$ convergence result $\mathbb{E}\left[\left(\sqrt{m+n}(U_2 - \widehat{U}_2)\right)^{T} \left(\sqrt{m+n}(U_2 - \widehat{U}_2)\right)\right] \to 0$ as $m, n \to \infty$.
Trivial algebra gives that $\mathbb{E} \widehat{U}_1 = \mathbb{E} \widehat{U}_2 = M$.
Then $\mathbb{E}(U_2 - \widehat{U}_2) = 0$, and hence
\begin{equation}
\begin{aligned}
    &\mathbb{E}\left[\left(\sqrt{m+n}(U_2 - \widehat{U}_2)\right)^{T} \left(\sqrt{m+n}(U_2 - \widehat{U}_2)\right)\right] \\ =& (m+n) \mathbb{E} \left[\left(U_2 - \widehat{U}_2\right)^{T} \left(U_2 - \widehat{U}_2\right)\right] \\
    =& (m+n) \operatorname{tr}\left(\mathbb{C}\left[U_2 - \widehat{U}_2\right]\right) \\
    =& (m+n) \operatorname{tr}\left(\mathbb{C}\left[U_2\right] + \mathbb{C}\left[\widehat{U}_2\right] - 2  \mathbb{C}\left[U_2, \widehat{U}_2\right]\right). \nonumber
\end{aligned}
\end{equation}
As Assumption~\ref{ass:finite_hessian_l2} holds, Theorem 12.6 in \cite{ van2000asymptotic} implies that both $(m+n) \operatorname{tr}\left(\mathbb{C}\left[U_2\right]\right)$ and $(m+n) \operatorname{tr}\left(\mathbb{C}\left[\widehat{U}_2\right]\right)$ converges to $\frac{\operatorname{tr}\left( \mathbb{C}_{Z} \left[g_{0, 1}(Z)\right]\right)}{\lambda} + \frac{\operatorname{tr}\left( \mathbb{C}_{X} \left[g_{1, 0}(X)\right]\right)}{1-\lambda}$.
We just need to show that $(m+n) \operatorname{tr}\left(\mathbb{C}\left[U_2, \widehat{U}_2\right]\right)$ converges to $\frac{\operatorname{tr}\left( \mathbb{C}_{Z} \left[g_{0, 1}(Z)\right]\right)}{\lambda} + \frac{\operatorname{tr}\left( \mathbb{C}_{X} \left[g_{1, 0}(X)\right]\right)}{1-\lambda}$ as well.

By definition, 
\begin{equation}
\begin{aligned}
    &(m+n) \operatorname{tr}\left(\mathbb{C}\left[U_2, \widehat{U}_2\right]\right) \\=& (m+n) \operatorname{tr}\left(\mathbb{E}\left[\left(\frac{2}{mn}  \sum_{i=1}^{m} \sum_{j=1}^{n} \nabla_\theta k(G_\theta(Z_j), X_i) \left.\right|_{\theta=\theta_{\star}^{\operatorname{KS}}} - M\right)^{T} \right.\right.\\
    &\hspace{1.5in}\left.\left.\left(\frac{1}{m} \sum_{i=1}^{m} g_{1,0}(X_i) + \frac{1}{n} \sum_{j=1}^{n} g_{0,1}(Z_j) - M\right)\right]\right)  \\
    =& \operatorname{tr}\left(\frac{m+n}{m} \mathbb{E}_{X_1, \ldots, X_m}\left[\sqrt{m}\left(\frac{1}{m} \sum_{i=1}^{m} g_{1,0}(X_i) - \frac{1}{2}M\right)^{T} \sqrt{m} \left(\frac{1}{m} \sum_{i=1}^{m} g_{1,0}(X_i) - \frac{1}{2}M\right)\right] \right.\\
    &\hspace{0.5in}\left. + \frac{m+n}{n} \mathbb{E}_{Y_1, \ldots, Y_n}\left[\sqrt{n}\left(\frac{1}{n} \sum_{j=1}^{n} g_{0,1}(Z_j) - \frac{1}{2}M\right)^{T} \sqrt{n}\left(\frac{1}{n} \sum_{j=1}^{n} g_{0,1}(Z_j) - \frac{1}{2}M\right)\right]\right) \\
     \to& \frac{\operatorname{tr}\left( \mathbb{C}_{Z} \left[g_{0, 1}(Z)\right]\right)}{\lambda} + \frac{\operatorname{tr}\left( \mathbb{C}_{X} \left[g_{1, 0}(X)\right]\right)}{1-\lambda}, \nonumber
\end{aligned}
\end{equation}
leveraging the classical central limit theorem.
Hence, we have $\sqrt{n}(U_1 - \widehat{U}_1) \stackrel{\mathrm{p}}{\to} 0$ and $\sqrt{m+n}(U_2 - \widehat{U}_2) \stackrel{\mathrm{p}}{\to} 0$ as $m, n \to \infty$.

Plug in the expressions of $\widehat{U}_1$ and $\widehat{U}_2$ to derive
\begin{equation}
\begin{aligned}
    \sqrt{m+n}(\widehat{U}_1 - \widehat{U}_2) &= \sqrt{m+n}\left(\frac{2}{n} \sum_{j=1}^{n} h_1(Z_j) - \frac{1}{m} \sum_{i=1}^{m} g_{1, 0}(X_i) - \frac{1}{n} \sum_{j=1}^{n} g_{0, 1}(Z_j)\right) \\
    &= \sqrt{m+n}\left(\frac{1}{n} \sum_{j=1}^{n} \left(2 h_1(Z_j) - g_{0, 1}(Z_j)\right) - \frac{1}{m} \sum_{i=1}^{m} g_{1, 0}(X_i) \right) \\
    &= \sqrt{\frac{m+n}{n}} \sqrt{n} \left(\frac{1}{n} \sum_{j=1}^{n} \left(2 h_1(Z_j) - g_{0, 1}(Z_j)\right) - \frac{1}{2} M \right)\\
    &  \hspace{0.5in} - \sqrt{\frac{m+n}{m}} \sqrt{m}  \left(\frac{1}{m} \sum_{i=1}^{m} g_{1, 0}(X_i) - \frac{1}{2} M \right). \nonumber
\end{aligned}
\end{equation}
As Assumption~\ref{ass:finite_hessian_l2} holds, the classical central limit theorem implies that 
\begin{equation}
\begin{gathered}
    \sqrt{n} \left(\frac{1}{n} \sum_{j=1}^{n} \left(2 h_1(Z_j) - g_{0, 1}(Z_j)\right) - \frac{1}{2} M \right) \stackrel{\mathrm{d}}{\to} N\left(0, \mathbb{C}_{Z} \left[2 h_1(Z) - g_{0, 1}(Z)\right]\right) \\
    \sqrt{m}  \left(\frac{1}{m} \sum_{i=1}^{m} g_{1, 0}(X_i) - \frac{1}{2} M \right) \stackrel{\mathrm{d}}{\to} N\left(0, \mathbb{C}_{X} \left[g_{1, 0}(X)\right]\right). \nonumber
\end{gathered}
\end{equation}
Furthermore, as $\lim_{m, n \to \infty} \frac{n}{m+n} = \lambda$,  $X_i$ and $Z_j$ are independent, we get
$$\sqrt{m+n}(\widehat{U}_1 - \widehat{U}_2) \stackrel{\mathrm{d}}{\to}  N\left(0, \frac{\mathbb{C}_{Z} \left[2 h_1(Z) - g_{0, 1}(Z)\right]}{\lambda} + \frac{\mathbb{C}_{X} \left[g_{1, 0}(X)\right]}{1-\lambda}\right) = N(0, \Sigma_\lambda).$$
Hence $\sqrt{m+n} \nabla_\theta \widehat{L}_{m, n}^{\operatorname{KS}}(\theta) \left.\right|_{\theta=\theta_{\star}^{\operatorname{KS}}} = \sqrt{m+n}(U_1 - U_2) \stackrel{\mathrm{d}}{\to}  N(0, \Sigma_\lambda)$.
As $\nabla_{\theta \theta} \widehat{L}_{m, n}^{\operatorname{KS}}(\theta) \left.\right|_{\theta=\tilde{\theta}_{m, n}^{\operatorname{KS}}} \stackrel{\mathrm{p}}{\to} H$, the Slutsky's theorem implies that $\sqrt{m+n}(\widehat{\theta}_{m, n}^{\operatorname{KS}} - \theta_{\star}^{\operatorname{KS}}) = -\nabla_{\theta \theta} \widehat{L}_{m, n}^{\operatorname{KS}}(\theta)^{-1} \left.\right|_{\theta=\tilde{\theta}_{m, n}^{\operatorname{KS}}} \cdot \sqrt{m+n} \nabla_\theta \widehat{L}_{m, n}^{\operatorname{KS}}(\theta) \left.\right|_{\theta=\theta_{\star}^{\operatorname{KS}}} \stackrel{\mathrm{d}}{\to} N(0, H^{-1} \Sigma_\lambda H^{-1})$.
\end{proof}

\subsection{Proof of Theorem~\ref{thm:cons_var_est}}
\label{proof:cons_var_est}
\begin{proof}
To simplify notation, $n$ in this proof is the simulation sample size for confidence set estimation that equals to $n_c$ in Theorem~\ref{thm:cons_var_est}.
We first show that $\widehat{H}_{m, n} \stackrel{\mathrm{p}}{\to} H$ as $m, n \to \infty$.
Recall that $\widehat{H}_{m, n} = \nabla_{\theta \theta} \widehat{L}_{m, n}^{\operatorname{KS}}(\theta) \left.\right|_{\theta=\widehat{\theta}_{m}^{\operatorname{KS}}} $.
The proof follows an entirely analogous way as the proof of $\nabla_{\theta \theta} \widehat{L}_{m, n}^{\operatorname{KS}}(\theta) \left.\right|_{\theta=\tilde{\theta}_{m, n}^{\operatorname{KS}}} \stackrel{\mathrm{p}}{\to} H$ in \ref{proof:asymp_norm}.

As Assumption~\ref{ass:finite_hessian_l2} holds, leveraging the strong law of large numbers for generalized $k$-sample U-statistics (Theorem~\ref{thm:SLLN_Ustat}) gives $\nabla_{\theta \theta} \widehat{L}_{m, n}^{\operatorname{KS}}(\theta) \left.\right|_{\theta=\theta_{\star}^{\operatorname{KS}}} \stackrel{\mathrm{a.s.}}{\to} H$.
We just need to show that $\left\|\nabla_{\theta \theta} \widehat{L}_{m, n}^{\operatorname{KS}}(\theta) \left.\right|_{\theta=\widehat{\theta}_{m, n}^{\operatorname{KS}}} - \nabla_{\theta \theta} \widehat{L}_{m, n}^{\operatorname{KS}}(\theta) \left.\right|_{\theta=\theta_{\star}^{\operatorname{KS}}}\right\| \stackrel{\mathrm{a.s.}}{\to} 0$.

Assumption~\ref{ass:unif_cont_hessian} implies that for any $\epsilon > 0$ there exists an open neighborhood $\mathcal{N}_\epsilon \subset \mathcal{N}$ such that
\begin{equation}
\begin{aligned}
&\left\|\nabla_{\theta \theta} k(G_\theta(z), x) -  \nabla_{\theta \theta} k(G_\theta(z), x) \left.\right|_{\theta=\theta_{\star}^{\operatorname{KS}}} \right\|, \\ &\left\|\nabla_{\theta \theta} k(G_\theta(z), G_\theta(z^\prime)) - \nabla_{\theta \theta} k(G_\theta(z), G_\theta(z^\prime))\left.\right|_{\theta=\theta_{\star}^{\operatorname{KS}}} \right\| < \epsilon \nonumber   
\end{aligned}
\end{equation}
for all $\theta \in \mathcal{N}_\epsilon$, $z, z^\prime \in \mathcal{Z}, x \in \mathbb{R}^d$. 
Since $\widehat{\theta}_{m, n}^{\operatorname{KS}} \stackrel{\mathrm{a.s.}}{\to} \theta_{\star}^{\operatorname{KS}}$ as $m, n \to \infty$, then $\widehat{\theta}_{m, n}^{\operatorname{KS}} \in \mathcal{N}_\epsilon$ a.s. for sufficiently large $m, n$. 

Therefore, 
\begin{equation}
\begin{aligned}
&\left\|\nabla_{\theta \theta} k(G_\theta(z), x) \left.\right|_{\theta=\widehat{\theta}_{m, n}^{\operatorname{KS}}} - \nabla_{\theta \theta} k(G_\theta(z), x) \left.\right|_{\theta=\theta_{\star}^{\operatorname{KS}}} \right\|, \\ &\left\|\nabla_{\theta \theta} k(G_\theta(z), G_\theta(z^\prime)) \left.\right|_{\theta=\widehat{\theta}_{m, n}^{\operatorname{KS}}} - \nabla_{\theta \theta} k(G_\theta(z), G_\theta(z^\prime))\left.\right|_{\theta=\theta_{\star}^{\operatorname{KS}}} \right\| < \epsilon \nonumber 
\end{aligned}
\end{equation}
a.s. for all $z, z^\prime \in \mathcal{Z}, x \in \mathbb{R}^d$, which in turn implies that 
\begin{equation}
\left\|\nabla_{\theta \theta} \widehat{L}_{m, n}^{\operatorname{KS}}(\theta) \left.\right|_{\theta=\widehat{\theta}_{m, n}^{\operatorname{KS}}} - \nabla_{\theta \theta} \widehat{L}_{m, n}^{\operatorname{KS}}(\theta) \left.\right|_{\theta=\theta_{\star}^{\operatorname{KS}}}\right\| < 3 \epsilon.  \nonumber
\end{equation}
Therefore, $\left\|\nabla_{\theta \theta} \widehat{L}_{m, n}^{\operatorname{KS}}(\theta) \left.\right|_{\theta=\widehat{\theta}_{m, n}^{\operatorname{KS}}} - \nabla_{\theta \theta} \widehat{L}_{m, n}^{\operatorname{KS}}(\theta) \left.\right|_{\theta=\theta_{\star}^{\operatorname{KS}}}\right\| \stackrel{\mathrm{a.s.}}{\to} 0$.

Now we show that $\widehat{\Sigma}_{m, n} \stackrel{\mathrm{p}}{\to} \Sigma$ as $m, n \to \infty$.
We slightly abuse notation to define $\widehat{\Sigma}_{m, n}(\theta)$ as a function of $\theta$:
\begin{equation}
    \widehat{\Sigma}_{m, n}(\theta) = \frac{4}{m-1} \sum_{i=1}^{m} \left(\widehat{\mu}_i(\theta) - \widehat{\mu}_{m, n}(\theta)\right)^{T}\left(\widehat{\mu}_i(\theta) - \widehat{\mu}_{m, n}(\theta)\right), \nonumber
\end{equation}
where
\begin{equation}
\begin{aligned}
\widehat{\mu}_{m, n}(\theta) = \frac{1}{mn} &\sum_{i=1}^{m} \sum_{j=1}^{n} \nabla_\theta k\left(G_\theta(Z_j), X_i\right) \\
\widehat{\mu}_{i}(\theta) = \frac{1}{n} &\sum_{j=1}^{n}\nabla_\theta k\left(G_\theta(Z_j), X_i\right). \nonumber
\end{aligned}
\end{equation}
Then $\widehat{\Sigma}_{m, n}(\widehat{\theta}_{m, n}^{\operatorname{KS}}) = \widehat{\Sigma}_{m, n}$.
We first prove that $\widehat{\Sigma}_{m, n}(\theta_{\star}^{\operatorname{KS}}) \stackrel{\mathrm{p}}{\to} \Sigma$.
By definition,
\begin{equation}
\begin{aligned}
    &\widehat{\Sigma}_{m, n}(\theta_{\star}^{\operatorname{KS}}) \\ =& \frac{4}{m-1} \sum_{i=1}^{m} \left(\widehat{\mu}_i(\theta_{\star}^{\operatorname{KS}}) - \widehat{\mu}_{m, n}(\theta_{\star}^{\operatorname{KS}})\right)^{T}\left(\widehat{\mu}_i(\theta_{\star}^{\operatorname{KS}}) - \widehat{\mu}_{m, n}(\theta_{\star}^{\operatorname{KS}})\right) \\
    =& \frac{4}{m-1} \sum_{i=1}^{m} \left(\widehat{\mu}_i(\theta_{\star}^{\operatorname{KS}})^{T} \widehat{\mu}_i(\theta_{\star}^{\operatorname{KS}}) - \widehat{\mu}_{m, n}(\theta_{\star}^{\operatorname{KS}})^{T} \widehat{\mu}_i(\theta_{\star}^{\operatorname{KS}}) - \widehat{\mu}_i(\theta_{\star}^{\operatorname{KS}})^{T} \widehat{\mu}_{m, n}(\theta_{\star}^{\operatorname{KS}}) +  \widehat{\mu}_{m, n}(\theta_{\star}^{\operatorname{KS}})^{T} \widehat{\mu}_{m, n}(\theta_{\star}^{\operatorname{KS}}) \right) \\
    =& \frac{4}{(m-1)n^2} \sum_{i=1}^{m} \sum_{j=1}^{n} \sum_{l=1}^{n} \nabla_\theta k\left(G_\theta(Z_j), X_i\right)^{T} \left.\right|_{\theta=\theta_{\star}^{\operatorname{KS}}} \nabla_\theta k\left(G_\theta(Z_l), X_i\right) \left.\right|_{\theta=\theta_{\star}^{\operatorname{KS}}}\\
    & \hspace{0.5in}- \frac{4}{m(m-1)n^2} \sum_{i=1}^{m} \sum_{p=1}^{m} \sum_{j=1}^{n} \sum_{l=1}^{n} \nabla_\theta k\left(G_\theta(Z_j), X_i\right)^{T} \left.\right|_{\theta=\theta_{\star}^{\operatorname{KS}}} \nabla_\theta k\left(G_\theta(Z_l), X_p\right) \left.\right|_{\theta=\theta_{\star}^{\operatorname{KS}}}. \nonumber
\end{aligned}
\end{equation}
As Assumption~\ref{ass:finite_jacobian_l2} holds, the strong law of large numbers for generalized $k$-sample U-statistics (Theorem~\ref{thm:SLLN_Ustat}) implies that as $m, n \to \infty$,
\begin{equation}
\begin{aligned}
\frac{4}{(m-1)n^2} \sum_{i=1}^{m} \sum_{j=1}^{n} \sum_{l=1}^{n} &\nabla_\theta k\left(G_\theta(Z_j), X_i\right)^{T} \left.\right|_{\theta=\theta_{\star}^{\operatorname{KS}}}  \nabla_\theta k\left(G_\theta(Z_l), X_i\right) \left.\right|_{\theta=\theta_{\star}^{\operatorname{KS}}} \stackrel{\mathrm{a.s.}}{\to} \\
& 4 \mathbb{E}_{X, Z, Z^\prime} \left[\nabla_\theta k\left(G_\theta(Z), X\right)^{T} \left.\right|_{\theta=\theta_{\star}^{\operatorname{KS}}} \nabla_\theta k\left(G_\theta(Z^\prime), X\right) \left.\right|_{\theta=\theta_{\star}^{\operatorname{KS}}}\right] \\
\frac{4}{m(m-1)n^2} \sum_{i=1}^{m} \sum_{p=1}^{m} \sum_{j=1}^{n} \sum_{l=1}^{n} &\nabla_\theta k\left(G_\theta(Z_j), X_i\right)^{T} \left.\right|_{\theta=\theta_{\star}^{\operatorname{KS}}}  \nabla_\theta k\left(G_\theta(Z_l), X_p\right) \left.\right|_{\theta=\theta_{\star}^{\operatorname{KS}}} \stackrel{\mathrm{a.s.}}{\to} \\
& 4 \mathbb{E}_{X, X^\prime, Z, Z^\prime} \left[\nabla_\theta k\left(G_\theta(Z), X\right)^{T} \left.\right|_{\theta=\theta_{\star}^{\operatorname{KS}}} \nabla_\theta k\left(G_\theta(Z^\prime), X^\prime\right) \left.\right|_{\theta=\theta_{\star}^{\operatorname{KS}}}\right], \nonumber
\end{aligned}
\end{equation}
where $X, X^\prime \stackrel{\mathrm{i.i.d.}}{\sim} P_\star$, $Z, Z^\prime \stackrel{\mathrm{i.i.d.}}{\sim} P$. 
Therefore, 
\begin{equation}
\begin{aligned}
&\widehat{\Sigma}_{m, n}(\theta_{\star}^{\operatorname{KS}}) \\
\stackrel{\mathrm{a.s.}}{\to}& 4 \mathbb{E}_{X, Z, Z^\prime} \left[\nabla_\theta k\left(G_\theta(Z), X\right)^{T} \left.\right|_{\theta=\theta_{\star}^{\operatorname{KS}}} \nabla_\theta k\left(G_\theta(Z^\prime), X\right) \left.\right|_{\theta=\theta_{\star}^{\operatorname{KS}}}\right] \\ 
& \hspace{0.5in} - 4 \mathbb{E}_{X, X^\prime, Z, Z^\prime} \left[\nabla_\theta k\left(G_\theta(Z), X\right)^{T} \left.\right|_{\theta=\theta_{\star}^{\operatorname{KS}}} \nabla_\theta k\left(G_\theta(Z^\prime), X^\prime\right) \left.\right|_{\theta=\theta_{\star}^{\operatorname{KS}}}\right] \\
=& 4 \mathbb{E}_{X, X^\prime} \left[\mathbb{E}_{Z} \left[\nabla_\theta k\left(G_\theta(Z), X\right)^{T} \left.\right|_{\theta=\theta_{\star}^{\operatorname{KS}}} \right] \left(\mathbb{E}_{Z} \left[\nabla_\theta k\left(G_\theta(Z), X\right)\left.\right|_{\theta=\theta_{\star}^{\operatorname{KS}}}\right] - \mathbb{E}_{Z} \left[\nabla_\theta k\left(G_\theta(Z), X^\prime\right)\left.\right|_{\theta=\theta_{\star}^{\operatorname{KS}}}\right] \right)\right] \\
=& 4 \mathbb{E}_{X} \left[\mathbb{E}_{Z} \left[\nabla_\theta k\left(G_\theta(Z), X\right)^{T}\left.\right|_{\theta=\theta_{\star}^{\operatorname{KS}}}\right]\left(\mathbb{E}_{Z} \left[\nabla_\theta k\left(G_\theta(Z), X\right)\left.\right|_{\theta=\theta_{\star}^{\operatorname{KS}}}\right] - \mathbb{E}_{X, Z} \left[\nabla_\theta k\left(G_\theta(Z), X\right)\left.\right|_{\theta=\theta_{\star}^{\operatorname{KS}}}\right] \right)\right] \\
=& \mathbb{C}_{X}\left[\mathbb{E}_{Z}\nabla_\theta k\left(G_\theta(Z), X\right) \left.\right|_{\theta=\theta_{\star}^{\operatorname{KS}}} \right] \\
=& \Sigma. \nonumber
\end{aligned}
\end{equation}
Now we just need to show that $\left\|\widehat{\Sigma}_{m, n}(\theta_{\star}^{\operatorname{KS}}) - \widehat{\Sigma}_{m, n} \right\| \stackrel{\mathrm{p}}{\to} 0$.
We only prove that 
\begin{equation}
\left\|\frac{4}{m-1} \sum_{i=1}^{m} \left[\widehat{\mu}_i(\theta_{\star}^{\operatorname{KS}})^{T} \widehat{\mu}_i(\theta_{\star}^{\operatorname{KS}}) - \widehat{\mu}_i(\widehat{\theta}_{m, n}^{\operatorname{KS}})^{T} \widehat{\mu}_i(\widehat{\theta}_{m, n}^{\operatorname{KS}}) \right]\right\| \stackrel{\mathrm{a.s.}}{\to} 0, \nonumber
\end{equation}
as the almost sure convergence of all the other components of $\widehat{\Sigma}_{m, n}(\theta_{\star}^{\operatorname{KS}}) - \widehat{\Sigma}_{m, n}$ follows an entirely analogous way.

By definition,
\begin{equation}
\label{eq:cons_est}
\begin{aligned}
&\left\|\frac{4}{m-1} \sum_{i=1}^{m} \left[\widehat{\mu}_i(\theta_{\star}^{\operatorname{KS}})^{T} \widehat{\mu}_i(\theta_{\star}^{\operatorname{KS}}) - \widehat{\mu}_i(\widehat{\theta}_{m, n}^{\operatorname{KS}})^{T} \widehat{\mu}_i(\widehat{\theta}_{m, n}^{\operatorname{KS}}) \right]\right\|  \\
= &\frac{4}{m-1}  \left\|\sum_{i=1}^{m} \left[\frac{1}{n} \sum_{j=1}^{n} \nabla_\theta k \left(G_\theta(Z_j), X_i\right)^{T} \left.\right|_{\theta=\theta_{\star}^{\operatorname{KS}}} \cdot \frac{1}{n} \sum_{l=1}^{n} \nabla_\theta k \left(G_\theta(Z_l), X_i\right) \left.\right|_{\theta=\theta_{\star}^{\operatorname{KS}}} \right.\right. \\
& \hspace{1in} \left.\left. - \frac{1}{n} \sum_{j=1}^{n} \nabla_\theta k \left(G_\theta(Z_j), X_i\right)^{T} \left.\right|_{\theta=\widehat{\theta}_{m, n}^{\operatorname{KS}}} \cdot \frac{1}{n} \sum_{l=1}^{n} \nabla_\theta k \left(G_\theta(Z_l), X_i\right)\left.\right|_{\theta=\widehat{\theta}_{m, n}^{\operatorname{KS}}}\right]\right\| \\
=& \frac{4}{(m-1)n^2} \left\|\sum_{i=1}^{m} \sum_{j=1}^{n} \sum_{l=1}^{n} \left[ \nabla_\theta k \left(G_\theta(Z_j), X_i\right)^{T} \left.\right|_{\theta=\theta_{\star}^{\operatorname{KS}}} \nabla_\theta k \left(G_\theta(Z_l), X_i\right) \left.\right|_{\theta=\theta_{\star}^{\operatorname{KS}}} \right.\right. \\
& \hspace{1in} \left.\left. - \nabla_\theta k \left(G_\theta(Z_j), X_i\right)^{T} \left.\right|_{\theta=\widehat{\theta}_{m, n}^{\operatorname{KS}}}  \nabla_\theta k \left(G_\theta(Z_l), X_i\right)\left.\right|_{\theta=\widehat{\theta}_{m, n}^{\operatorname{KS}}}\right]\right\| \\
\leq &\frac{4}{(m-1)n^2} \sum_{i=1}^{m} \sum_{j=1}^{n} \sum_{l=1}^{n} \left\| \nabla_\theta k \left(G_\theta(Z_j), X_i\right)^{T} \left.\right|_{\theta=\theta_{\star}^{\operatorname{KS}}}\nabla_\theta k \left(G_\theta(Z_l), X_i\right) \left.\right|_{\theta=\theta_{\star}^{\operatorname{KS}}} \right.  \\
& \hspace{1in} \left. - \nabla_\theta k \left(G_\theta(Z_j), X_i\right)^{T} \left.\right|_{\theta=\widehat{\theta}_{m, n}^{\operatorname{KS}}}  \nabla_\theta k \left(G_\theta(Z_l), X_i\right)\left.\right|_{\theta=\widehat{\theta}_{m, n}^{\operatorname{KS}}}\right\| \\
\leq & \frac{4}{(m-1)n^2} \sum_{i=1}^{m} \sum_{j=1}^{n} \sum_{l=1}^{n} \left\{ \left\| \left(\nabla_\theta k \left(G_\theta(Z_j), X_i\right)^{T} \left.\right|_{\theta=\theta_{\star}^{\operatorname{KS}}}- \nabla_\theta k \left(G_\theta(Z_j), X_i\right)^{T} \left.\right|_{\theta=\widehat{\theta}_{m, n}^{\operatorname{KS}}} \right) \right.\right. \\
& \hspace{3in} \left.\left. \cdot \nabla_\theta k \left(G_\theta(Z_l), X_i\right) \left.\right|_{\theta=\theta_{\star}^{\operatorname{KS}}} \right\| \right.\\
&+\left.  \left\| \nabla_\theta k \left(G_\theta(Z_j), X_i\right)^{T} \left.\right|_{\theta=\widehat{\theta}_{m, n}^{\operatorname{KS}}}  \left(\nabla_\theta k \left(G_\theta(Z_l), X_i\right) \left.\right|_{\theta=\theta_{\star}^{\operatorname{KS}}} - \nabla_\theta k \left(G_\theta(Z_l), X_i\right)\left.\right|_{\theta=\widehat{\theta}_{m, n}^{\operatorname{KS}}}\right)\right\| \right\}  \\
= & \frac{4}{(m-1)n^2} \sum_{i=1}^{m} \sum_{j=1}^{n} \sum_{l=1}^{n} \left\{\left\|\nabla_\theta k \left(G_\theta(Z_l), X_i\right) \left.\right|_{\theta=\theta_{\star}^{\operatorname{KS}}} \right\| \left\|\nabla_\theta k \left(G_\theta(Z_j), X_i\right)\left.\right|_{\theta=\theta_{\star}^{\operatorname{KS}}} \right.\right.\\
&\hspace{3in} -\left.  \nabla_\theta k \left(G_\theta(Z_j), X_i\right)\left.\right|_{\theta=\widehat{\theta}_{m, n}^{\operatorname{KS}}}\right\| \\
& \left.+ \left\| \nabla_\theta k \left(G_\theta(Z_j), X_i\right)\left.\right|_{\theta=\widehat{\theta}_{m, n}^{\operatorname{KS}}}\right\| \left\|\nabla_\theta k \left(G_\theta(Z_l), X_i\right) \left.\right|_{\theta=\theta_{\star}^{\operatorname{KS}}} - \nabla_\theta k \left(G_\theta(Z_l), X_i\right)\left.\right|_{\theta=\widehat{\theta}_{m, n}^{\operatorname{KS}}}\right\| \right\}.
\end{aligned}
\end{equation}
As Assumption~\ref{ass:unif_cont_bdd_jacobian} holds, for any $\epsilon > 0$ there exists an open neighborhood $\mathcal{N}_\epsilon \subset \mathcal{N}$ such that 
$$
\left\|\nabla_\theta k \left(G_\theta(z), x\right) \left.\right|_{\theta=\theta_{\star}^{\operatorname{KS}}} - \nabla_\theta k \left(G_\theta(z), x\right) \right\| < \epsilon,
$$
and there exists a non-negative constant $M$ such that
$$
\left\|\nabla_\theta k \left(G_\theta(z), x\right) \left.\right|_{\theta=\theta_{\star}^{\operatorname{KS}}} \right\| < M
$$
for all $x \in \mathbb{R}^d$ and $z \in \mathcal{Z}$.
Since $\widehat{\theta}_{m, n}^{\operatorname{KS}} \stackrel{\mathrm{a.s.}}{\to} \theta_{\star}^{\operatorname{KS}}$ as $m, n \to \infty$, then $\widehat{\theta}_{m, n}^{\operatorname{KS}} \in \mathcal{N}_\epsilon$ a.s. for sufficiently large $m, n$.
Then for sufficiently large $m, n$, the right-hand side of \eqref{eq:cons_est} denoted RHS a.s.  has
$$
\operatorname{RHS} \leq \frac{4}{(m-1)n^2} \sum_{i=1}^{m} \sum_{j=1}^{n} \sum_{l=1}^{n} 2 M \epsilon \leq 8 M \epsilon.
$$
Therefore, we conclude that $\left\|\frac{4}{m-1} \sum_{i=1}^{m} \left[\widehat{\mu}_i(\theta_{\star}^{\operatorname{KS}})^{T} \widehat{\mu}_i(\theta_{\star}^{\operatorname{KS}}) - \widehat{\mu}_i(\widehat{\theta}_{m, n}^{\operatorname{KS}})^{T} \widehat{\mu}_i(\widehat{\theta}_{m, n}^{\operatorname{KS}}) \right]\right\| \stackrel{\mathrm{a.s.}}{\to} 0$.
Making the analogous arguments for all the other components of $\widehat{\Sigma}_{m, n}(\theta_{\star}^{\operatorname{KS}}) - \widehat{\Sigma}_{m, n}$ gives $\left\|\widehat{\Sigma}_{m, n}(\theta_{\star}^{\operatorname{KS}}) - \widehat{\Sigma}_{m, n} \right\| \stackrel{\mathrm{p}}{\to} 0$. Hence $\widehat{\Sigma}_{m, n} \stackrel{\mathrm{p}}{\to} \Sigma$ as $m, n \to \infty$.
Similarly, $\widehat{C}_{m, n} = \widehat{H}_{m, n}^{-1} \widehat{\Sigma}_{m, n} \widehat{H}_{m, n}^{-1} \stackrel{\mathrm{p}}{\to} H^{-1} \Sigma H^{-1}$ as $m, n \to \infty$ follows from the continuity of matrix inversion and multiplication.

\subsection{Optimal Sample Size and Kernel Selection}
\label{sec:lambda_kernel} 
\cite{briol2019statistical} discuss the choice of the simulation sample size $n$ under model exactness by finding the best ratio $\lambda = \frac{n}{m+n}$ in terms of computational efficiency.
Theorem~2 in \cite{briol2019statistical} establishes asymptotic variance-covariance of the kernel simulated score estimator $\widehat{\theta}_{m, n}^{\operatorname{KS}}$ to be $\frac{C}{\lambda(1-\lambda)}$.
The asymptotic variance-covariance is hence minimized at $\lambda = \frac{1}{2}$ (i.e. the size of the simulated sample $n$ equals that of the output data from the target system $m$).
This means that using $n$ much larger than $m$ is not computationally efficient.

The choice of optimal $\lambda$ in terms of computational efficiency under model inexactness is notably intricate.
Recall that the asymptotic variance-covariance of the kernel simulated score estimator is $C_\lambda = H^{-1} \Sigma_\lambda H^{-1}$, where  $\Sigma_\lambda = \frac{\mathbb{C}_{Z} \left[2h(Z) - g_{0,1}(Z)\right]}{\lambda} + \frac{\mathbb{C}_{X} \left[g_{1, 0}(X)\right]}{1-\lambda}$.
In addition to being dependent on $\lambda$ and the unknown $\theta_{\star}^{\operatorname{KS}}$, $\Sigma_\lambda$ is influenced by factors such as the chosen kernel, the reference measure of $Z$, the distribution of $X$ (i.e., the distribution of the target system output), and the input model $G_\theta$. 
Identifying the optimal $\lambda$ is thus highly instance-dependent and computationally intractable.

Furthermore, $\Sigma_\lambda$ represents a trade-off between the variance-covariance term arising from the simulated sample $Z$ and that arising from data $X$. 
Let $c_1 = \mathbb{C}_Z \left[2h(Z) - g_{0,1}(Z)\right]$ and $c_2 = \mathbb{C}_X \left[g_{1, 0}(X)\right]$.
In the case where $p=1$ (i.e., the parameter $\theta \in \Theta$ is one-dimensional), simple algebra yields that the asymptotic variance $C_\lambda$ is minimized at $\lambda = \frac{\sqrt{c_1}}{\sqrt{c_1} + \sqrt{c_2}}$.
When $c_1 = c_2$, including the model exactness case, $\lambda = \frac{1}{2}$. Heuristically, when the distributions of $X$ and $G_\theta(Z) \left.\right|_{\theta=\theta_{\star}^{\operatorname{KS}}} = Y(\theta_{\star}^{\operatorname{KS}})$ are relatively close (i.e., the model discrepancy is small), $\lambda$ is close to $\frac{1}{2}$. Therefore, using $n$ much larger than $m$ is not computationally efficient. However, when the variance of $Y(\theta_{\star}^{\operatorname{KS}})$ is larger than that of $X$ such that its corresponding variance-covariance $c_1$ dominates $c_2$, it becomes advantageous to use $n$ much larger than $m$ to reduce the variance of the estimator.
Such a phenomenon usually happens at the beginning of training.  

On the other hand, two main streams of approaches to kernel selection exist in the literature. 
The first stream focuses on investigating the asymptotic distribution of the estimator or the test statistics, such as in the kernel two-sample test \citep{gretton2012kernel}. 
The objective in this stream is to choose the kernel to maximize the power of the test \citep{sutherland2017generative, liu2020learning} or minimize the asymptotic variance-covariance \citep{briol2019statistical}.
In the latter case, while both $C_\lambda$ and $C$ are typically computationally intractable due to the unknown parameter $\theta_{\star}^{\operatorname{KS}}$, consistent estimations can be obtained (for example, $\widehat{C}_{m, n}$). 
The selection of the kernel for stochastic simulation models, guided by empirical estimations, is a subject for future research.

The second stream focuses on investigating the radial basis function (RBF) kernel, represented as $k(x, y) = r\left(\frac{\|x-y\|}{l}\right)$, where $r: \mathbb{R} \to \mathbb{R}^+$ is a positive-valued function, and $l$ is the bandwidth of the kernel. 
The Gaussian kernel is a notable instance of the RBF kernel. Several heuristics for RBF kernel selection, such as the median heuristic \citep{gretton2012kernel} and the kernel mixture heuristic \citep{li2015generative}, have demonstrated high power in kernel two-sample testing. 
However, the effectiveness of these heuristics in the context of stochastic simulation models requires further exploration in future research.
Additionally, the choice of the parameter $\beta$ for Riesz kernels presents another open problem to be investigated.
The Riesz kernel with $1 \leq \beta \leq 2$ complies with Assumption~\ref{ass:bounded_kernel}. However, there are currently no established guidelines for selecting an optimal value of $\beta$ within the range of $1$ to $2$.

\subsection{Differentiability and Unbiased Gradient for G/G/1 Queueing Models}
\label{sec:diff_GG1}
We validate the differentiability of $\theta$ and gradient unbiasedness for each experiment in this section.
We only discuss Experiment 4, as the case for all the other experiments follows an entirely analogous way.
Recall in Experiment 4, the inter-arrival time of the G/G/1 model follows the Gamma distribution with rate $\lambda$, denoted $\operatorname{Gamma}(0.5, \lambda)$, and the service time follows an exponential distribution with rate $\mu$, denoted $\operatorname{Exp}(\mu)$.
The simulation parameter is denoted $\theta = (\mu, \lambda)$, and the input model is denoted $G_\theta$.

For each $\theta$, the procedure for generating a single sample $Y(\theta)$ is as follows.
Sample $Y(\theta)$ is the average waiting times of a set of customers.
To simulate $Y(\theta)$, we need to simulate a set of customer waiting times.
We denote the waiting time of the $j$-th customer by $W_j(\theta)$.
The Lindley equation provides a representation of the waiting time in a G/G/1 queueing system:
\begin{equation}
    W_{j+1}(\theta) = \operatorname{ReLU}\left(W_j(\theta) + S_j(\mu) - T_j(\lambda)\right), \nonumber
\end{equation}
where the service time of the $j$-th customer is denoted by $S_j(\mu)$, and the interarrival time between the $j$-th and $(j+1)$-th customer is denoted by $T_j(\lambda)$.
We just need to simulate a set of customer service times $\{S_j(\mu)\}$ and waiting times $\{T_j(\lambda)\}$ and apply the Lindley equation to get $\{W_j(\theta)\}$.

The Lindley equation is akin to a ReLU activation function, hence simplifying validation of differentiability and unbiased gradient to Assumptions 3 and 4 in Assumption~\ref{ass:ubg_reg_conditions} on $S_j(\mu)$ and $T_j(\lambda)$. 
We focus solely on $T_j(\lambda)$, as the analysis for $S_j(\mu)$ follows an entirely analogous way.

Consider the ‘pushforward' representation $T_j(\lambda) = T_\lambda(Z_j)$, $Z_j \stackrel{\mathrm{i.i.d.}}{\sim} P$, where $P$ is the user-specified reference measure. 
If we use inverse-transform sampling, then $P = U(0, 1)$, $T_j(\lambda) = \frac{\log(1-Z_j)}{\lambda} = \psi_{\lambda, 1}(\phi_{1}(Z_j))$, where $\psi_{\lambda, 1}(z) = \frac{z}{\lambda}$ and $\phi_1(z) = \log(1-z)$.
Such usual pseudo-random number generation with $P = U(0, 1)$ violates Assumption 3 due to the non-M-Lipschitz inverse transform $\phi_1(z) = \log(1-z)$. 
Both assumptions hold if we select $P = \operatorname{Exp}(1)$ and apply the linear transform $S_\mu = \frac{Q_n}{\mu}$.
A similar decomposition holds for the rate parameter of a Gamma distribution: $P = \operatorname{Gamma}(0.5, 0.5)$, $Z_j \sim P$, $S_j(\mu) = \frac{Z_j}{\mu}$.
Hence, by picking a good reference measure $P$, $G_\theta$ is (almost everywhere) differentiable with respect to $\theta = (\mu, \lambda)$, and the gradient is unbiased.

Such decomposition of $G_\theta$ may not be universally applicable, as in the case of the shape parameter of a Gamma distribution.
There exists no reference measure $P$ such that the input model $G_\theta$ is differentiable when $\theta$ is the shape parameter of a Gamma distribution.
Hence, the shape parameter is not learnable with our approach.

Furthermore, $G_\theta$ can also be a meta-model that models the service and arrival time distribution with learnable parameter $\theta$.
Such meta-models are usually deep neural networks \citep{chambers2002process, ojeda2021learning} that satisfy Assumption 3 and 4 in Assumption~\ref{ass:ubg_reg_conditions}.

\subsection{Applicability to Other Models}
\label{sec:app_other_model}
Our approach can be applied to other stochastic simulation models, such as $(s, S)$ stochastic inventory models and time-series models with latent variables, such as the stochastic volatility model described in Section 5.2 of \citep{briol2019statistical}.
However, as in the case of G/G/1 queueing models, learning the simulation parameter $\theta$ indeed requires careful choice of reference measure $P$ to ensure differentiability of $\theta$ and unbiasedness of the gradient estimate.

Even when the simulation parameter is differentiable, there is no guarantee that the U-statistic gradient estimate is unbiased if Assumption~\ref{ass:ubg_reg_conditions} is violated. 
Consider the stochastic volatility model described in Section 5.2 of \citep{briol2019statistical}. 
While the simulation parameters in this model are differentiable, the transformation from unobserved latent variables $h_t$ to observed data $y_t$ involves a non-M-Lipschitz exponential function, violating Assumption 3 in Assumption~\ref{ass:ubg_reg_conditions}.
Nevertheless, exploring the applicability of our approach to such models is of great interest to us and we plan to address this in our future work.

Besides the differentiability and unbiased gradient issue, the applicability of our confidence set procedure relies on verifying Assumption~\ref{ass:combined_assumption}.
Despite the fact that the assumptions in Assumption~\ref{ass:combined_assumption} are standard assumptions for asymptotic normality \citep{briol2019statistical, oates2022minimum}, they are technical and unintuitive in nature and hard to verify.
Note that both \cite{briol2019statistical, oates2022minimum} do not verify these conditions in practice.

One case where Assumption~\ref{ass:combined_assumption} holds is when $X$ and $Z$ have bounded support (or are almost surely bounded), $G_\theta$ is an (almost everywhere, or $P$-almost surely) smooth and bounded function, and the kernel $k$ is bounded (e.g. the Gaussian kernel).
In queueing systems, this translates to bounded waiting and arrival times (e.g. from a $U(0, 1)$ distribution) and using the Gaussian kernel for KOSE.
While we do not provide an explicit example where Assumption~\ref{ass:combined_assumption} does not hold, potential violations might occur with unbounded $X$ and $Z$, when $G_\theta$ involves exponential transforms, or when using unbounded kernels like the Riesz or Polynomial kernels.

\subsection{Scalability in Parameter Dimension}
\label{sec:scal_para}
We acknowledge our experiments' limitation to small parameter dimensions. 
This stems from implementation constraints and the need to ensure differentiability and gradient unbiasedness for each parameter. 

Regarding scalability, while concentration and generalization bounds for MMD \citep{briol2019statistical} suggest convergence speed does not depend on the parameter dimension $p$, we expect the marginal asymptotic variance for each parameter to increase with $p$. 
However, the computationally intractable nature of the asymptotic variance-covariance $C$ limits further theoretical insights. 
Larger $p$ (and data dimension $d$) will also introduce challenges in Hessian and Jacobian estimation. 
Our plug-in estimators might produce confidence sets of incorrect coverage and the memory needed to compute the Hessian significantly increases.

\subsection{Experimental Details}
\label{sec:exp_details}
We provide details on the experimental setup required to produce Tables~\ref{tab: 2} - \ref{tab: 3}.

In Table 1, we carefully design the experiment settings to study both stationary and non-stationary queueing systems. 
For Experiments 1 and 2, we set the target service rate to $1.2$ to ensure system stationarity. Experiment 3 replicates the non-stationary queueing system setup from \cite{bai2020calibrating}, allowing us to compare our results with established benchmarks. 
For Experiment 4, we selected a target service rate of $2.5$ to maintain system stationarity while testing in a higher parameter dimension.

Unless otherwise specified, all the experiments are run individually on an NVIDIA GeForce RTX 2080 Ti GPU. For Experiments 1, 2, the waiting time from the target system $X$ and $Y(\theta)$ are the averages of $50$ Monte Carlo draws following a burn-in period of $10$. For Experiment 4, the waiting time from the target system $X$ and $Y(\theta)$ are the averages of $20$ Monte Carlo draws following a burn-in period of $10$. The target system in Experiment 3 is a non-stationary queueing system, and the waiting times are thus the averages of $10$ Monte Carlo draws with no burn-in period.

\paragraph{Kernel Optimum Score Estimation (KOSE) and Confidence Set Estimation} For all experiments with our method, we use PyTorch \citep{paszke2019pytorch} for kernel optimum score estimation and confidence set estimation, and the Adam optimizer \citep{KingmaB14} for minimizing the kernel optimum score. The learning rate at the $k$-th iteration is $\frac{\eta}{\sqrt{1+k}}$, where $\eta$ is the initial learning rate.
We set $\eta$ to be $1$ throughout our experiments.
We set the stopping criterion at $200$ iterations for Experiments 1 and 2, and $800$ iterations for Experiments $3$ and $4$.
We use $n_c = 5,000$ for confidence set estimation. 
The Hessian estimator $\widehat{H}_{m, n_c}$ is obtained with finite difference with interval $0.1$ in Experiments 1 and 2, where the parameter space is one-dimensional, and $\widehat{H}_{m, n_c}$ is obtained with the built-in torch.autograd.functional.hessian within Pytorch in Experiments 3 and 4.
For Experiments 2 and 4, where the model is inexact, the optimal simulation parameter $\theta_\star^{\operatorname{KS}}$ used to compute MSE and coverage for all methods except KOSE-Riesz is estimated with a sample of size $5,000$ from the target system using KOSE-Gaussian with learning rate $\eta = 1$ and stopping at iteration $800$. 
The obtained optimal simulation parameters are displayed in the last column of Tables~\ref{tab: 2} - \ref{tab: 3}.
For KOSE-Riesz, the optimal simulation parameter $\theta_\star^{\operatorname{KS}}$ used to compute MSE and coverage is estimated with a sample of size $5,000$ from the target system using KOSE-Riesz with learning rate $\eta = 1$ and stopping at iteration $1,000$.
We reinforce numerical stability of the estimated optimal simulation parameters with Polyak-Ruppert averaging \citep{polyak1992acceleration} for the last $100$ iteration.
However, in Experiment 4, for $a=0.2$, numerical instability causes $\theta_\star^{\operatorname{KS}}$ for KOSE-Gaussian to be $[0.00772042, 1.02951885]$, which is on the border of the parameter space. 
To address this issue, we report $\theta_\star^{\operatorname{KS}}$ for KOSE-Riesz instead.

\begin{figure}[htb]
    \centering
    \includegraphics[width=0.8\textwidth]{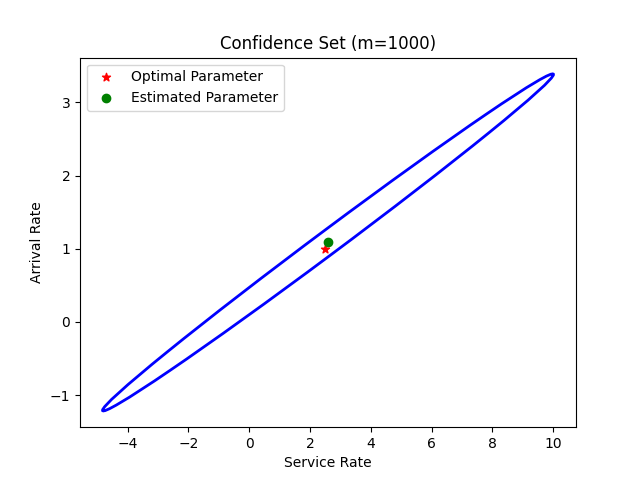}
    \caption{Confidence set for KOSE-Riesz, Experiment 4, $a=1$.}
    \label{fig:ci-1-2d-inexact}
\end{figure}

\begin{figure}[htb]
    \centering
    \includegraphics[width=0.8\textwidth]{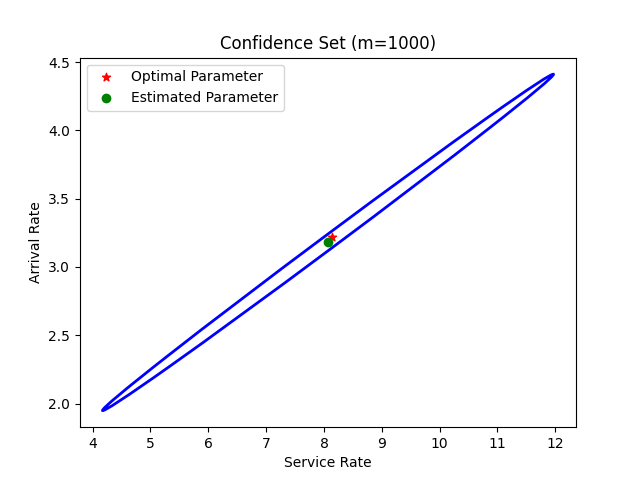}
    \caption{Confidence set for KOSE-Riesz, Experiment 4, $a=0.6$.}
    \label{fig:ci-06-2d-inexact}
\end{figure}

\begin{figure}[htb]
    \centering
    \includegraphics[width=0.8\textwidth]{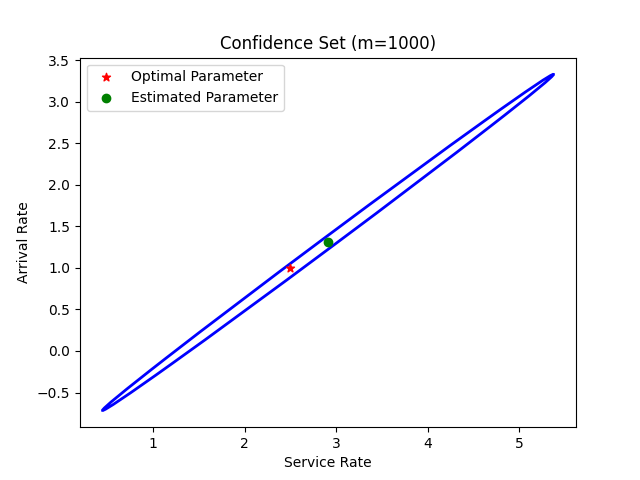}
    \caption{Confidence set for KOSE-Gaussian, Experiment 4, $a=1$.}
    \label{fig:ci-1-2d-inexact-G}
\end{figure}

\begin{figure}[htb]
    \centering
    \includegraphics[width=0.8\textwidth]{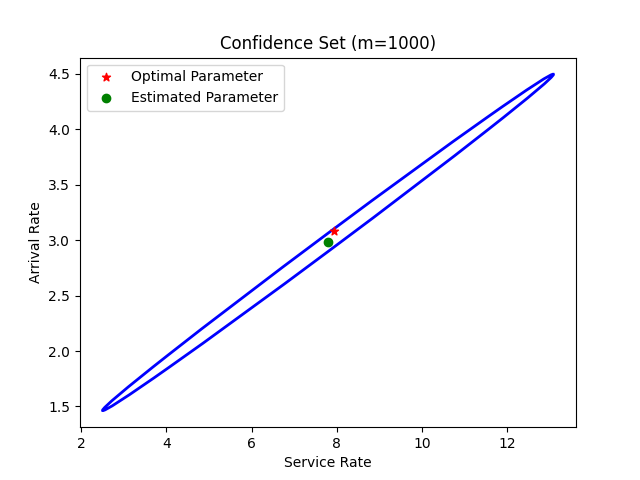}
    \caption{Confidence set for KOSE-Gaussian, Experiment 4, $a=0.6$.}
    \label{fig:ci-06-2d-inexact-G}
\end{figure}
\paragraph{MMD Posterior Bootstrap (NPL-MMD)} For the MMD posterior bootstrap method \citep{dellaporta2022robust}, we use the public GitHub repository by \cite{npl_mmd_project} that contains the experiment implementations. 
For all experiments with the MMD posterior bootstrap method, the number of points sampled from the model at each MMD optimization iteration is fixed at $30$ (except it is set at $20$ for $a=0.4, 0.2$ in Experiment 4 due to memory issues), and the number of bootstrap iterations is fixed at $1,000$. 
The bandwidth of the Gaussian kernel used in MMD is picked with the median heuristic \citep{gretton2012kernel}, and the prior is the non-informative Dirichlet Process (DP) prior. 
For Experiments 1 and 2, where the parameter space is one-dimensional, the $95\%$ credible set is the equal-tailed interval. The $95\%$ joint credible set is constructed as the union of two $97.5\%$ equal-tailed intervals on each dimension of the parameter space, for Experiments 3 and 4, from the joint posterior draw.

\paragraph{Accept-Reject ABC with MMD (ABC-AR)} For the accept-reject ABC with MMD, we set the tolerance to be the $1\%$ quantile of the simulated distances, as used in Experiment 1 of \cite{frazier2020model}.  
The bandwidth of the Gaussian kernel used in MMD is picked with the median heuristic \citep{gretton2012kernel}. 
The prior is the uniform distribution on the set of parameters that are all within $1$ in each dimension from the optimal simulation parameter, except in Experiment 1, where the optimal simulation parameter $\theta_\star^{\operatorname{KS}}=1.2$ and the prior is $U(0.5, 2.5)$. We construct the $95\%$ credible sets in the same way as the MMD posterior bootstrap method.
The number of draws from the prior is $1,000$ for Experiments 1 and 2, and the number of draws from the prior is $9,000$ for Experiments 3 and 4 to account for the extra dimension of the parameter.

\paragraph{Eligibility Set (ES)} For the eligibility set method \citep{bai2020calibrating}, we sample $1,000$ candidate parameters $\theta \sim U(0.5, 2.5)$ in Experiment 1, and from the uniform distribution on the set of parameters that are all within $1$ in each dimension from the optimal simulation parameter for Experiments 2 and 4.

\begin{table}[htb]
\centering
\caption{Experiment 1 results; $R$ is the number of independent runs; $m=n$. }\label{tab: 1}
\vspace{1ex}
\begin{tabular}{|c|cccc|cccc|}
\hline
& \multicolumn{4}{c|}{KOSE-Riesz, $R=1,000$} & \multicolumn{4}{c|}{KOSE-Gaussian, $R=1,000$} \\
$m, n$& MSE & Cov. & Width & T & MSE & Cov. & Width & T \\
\hline
$10$ & $6.0\cdot 10^{-3}$ & $\mathbf{0.973}$ & $0.45$ & $6$ & $2.0\cdot 10^{-1}$ & $\mathbf{0.959}$ & $0.51$ & $2$\\
$50$ & $1.2\cdot 10^{-3}$ & $\mathbf{0.990}$ & $0.20$ & $7$ & $1.5\cdot 10^{-3}$ & $\mathbf{0.986}$ & $0.21$ & $2$ \\
$100$ & $6.4 \cdot 10^{-4}$ & $\mathbf{0.995}$ & $0.14$ & $6$ & $6.0\cdot 10^{-4}$ & $\mathbf{1.000}$ & $0.15$ & $3$\\
$500$ & $1.4\cdot 10^{-4}$ & $\mathbf{0.993}$ & $0.06$ & $8$ & $1.4\cdot 10^{-4}$ & $\mathbf{0.991}$ & $0.07$ & $8$ \\
$1,000$ & $8.6\cdot 10^{-5}$ & $\mathbf{0.984}$ & $0.05$ & $10$ & $7.4\cdot 10^{-5}$ & $\mathbf{0.983}$ & $0.04$ & $15$ \\
\hline
\end{tabular}
\vspace{1em}
\begin{tabular}{|c|ccc|ccc|ccc|}
\hline
 & \multicolumn{3}{c|}{ABC-AR, $R=1,000$} & \multicolumn{3}{c|}{NPL-MMD, $R=100$} & \multicolumn{3}{c|}{ESet, $R=1,000$} \\
$m, n$ & Cov. & Width & T & Cov. & Width & T & Cov. & Width & T \\
\hline
$10$ & $0.885$ & $0.29$ & $3$ & $0.79$ & $0.77$ & $30$ & $\mathbf{1.000}$ & $0.42$ & $3$\\
$50$ & $0.854$ & $0.12$ & $3$ & $0.80$ & $0.38$ & $30$ & $\mathbf{1.000}$ & $0.19$ & $3$ \\
$100$ & $\mathbf{0.977}$ & $0.08$ & $3$ & $0.67$ & $0.27$ & $31$ & $\mathbf{1.000}$ & $0.13$ & $3$\\
$500$ & $\mathbf{0.990}$ & $0.05$ & $3$ & $0.44$ & $0.14$ & $34$ & $\mathbf{0.998}$ & $0.05$ & $3$\\
$1,000$ & $\mathbf{0.992}$ & $0.03$ & $3$ & $0.43$ & $0.09$ & $67$ & $\mathbf{0.998}$ & $0.04$ & $3$\\
\hline
\end{tabular}
\end{table}

\begin{table}[htb]
\centering
\caption{Experiment 3 results; $R$ is the number of independent runs; $m=n$; MSE, $\mu$ and MSE, $\lambda$ are the average mean-squared errors for the service and arrival rates. NR means that the experiment is not run due to excessive computation time.}\label{tab: 3}
\vspace{1ex}
\begin{tabular}{|c|cccc|cccc|}
\hline
& \multicolumn{4}{c|}{KOSE-Riesz, $R=1,000$} & \multicolumn{4}{c|}{KOSE-Gaussian, $R=1,000$} \\
$m, n$ & MSE, $\mu$ & MSE, $\lambda$ & Cov. & T & MSE, $\mu$ & MSE, $\lambda$ & Cov. & T \\
\hline
$10$ & $9.0\cdot 10^{-3}$ & $1.09$ & $\mathbf{1}$ & $3$ & $1.1\cdot 10^{-2}$ & $1.6\cdot 10^{-1}$ & $\mathbf{0.999}$ & $2$\\
$50$ & $2.5\cdot 10^{-3}$ & $2.5\cdot 10^{-3}$ & $\mathbf{1}$ & $3$ & $1.0\cdot 10^{-3}$ & $1.5 \cdot 10^{-2}$ & $\mathbf{1.000}$ & $3$\\
$100$ & $2.1\cdot 10^{-4}$ & $2.0\cdot 10^{-2}$ & $\mathbf{1}$ & $3$ & $1.9\cdot 10^{-3}$ & $4.9\cdot 10^{-3}$ & $\mathbf{1.000}$ & $4$\\
$500$ & $3.4\cdot 10^{-3}$ & $6.5\cdot 10^{-3}$ & $\mathbf{1}$ & $2$ & $7.1 \cdot 10^{-4}$ & $1.1 \cdot 10^{-2}$ & $\mathbf{1.000}$ & $15$\\
$1,000$ & $2.0\cdot 10^{-4}$ & $6.0\cdot 10^{-4}$ & $\mathbf{1}$ & $3$ & $4.1 \cdot 10^{-4}$ & $9.1 \cdot 10^{-5}$ & $\mathbf{1.000}$ & $36$\\
\hline
\end{tabular}
\vspace{1em}
\begin{tabular}{|c|ccc|ccc|}
\hline
 & \multicolumn{3}{c|}{ABC-AR, $R=100$} & \multicolumn{3}{c|}{NPL-MMD, $R=100$}  \\
$m, n$ & Cov. & Width, $\mu, \lambda$ & T & Cov. & Width, $\mu, \lambda$ & T \\
\hline
$10$ &  $0$ & $0.62, 0.59$ & $66$ &  $0.88$ & $1.06, 12.8$ & $30$ \\
$50$ & NR & NR & NR &  $0.86$ & $0.62, 7.07$ & $32$ \\
$100$ & $\mathbf{1}$ & $0.19, 0.65$ & $66$ &  $0.81$ & $0.50, 5.86$ & $38$ \\
$500$ & NR & NR & NR &  $0.73$ & $0.30, 3.29$ & $69$ \\
$1,000$ & NR & NR & NR &  $0.57$ & $0.22, 2.41$ & $161$ \\
\hline
\end{tabular}
\end{table}

\subsection{Extra Experiments on the Choice of $\beta$ in KOSE-Riesz}
\label{sec:sens_beta}
We conduct extra experiments on the choice of $\beta$ in the Riesz kernel for $\beta=1.25, 1.5, 1.75, 2$, where $\beta=2$ corresponds to a non-strictly proper scoring rule. 
All experiments, unless otherwise specified, are run with an NVIDIA TITAN Xp GPU.
The results under the setting of Table~\ref{tab: 2}, Experiment 2: $R = 1, 000$ are in Table~\ref{table:performance_combined}.
The results under the setting of Table~\ref{tab: 4}, Experiment 4 are in Table~\ref{table:performance_combined_exp4}.

Unlike the Gaussian kernel, there are no existing guidelines for selecting $\beta$ in the Riesz kernel to the best of our knowledge. 
For non-integer values of $\beta$, the energy score is not differentiable, necessitating a smoothing of the energy score computation.

\begin{table}[htb]
\centering
\caption{Under the setting of Table~\ref{tab: 2}, Experiment 2: $R = 1, 000$. Top: $\beta=1.25$ (left entry), $\beta=1.5$ (right entry). Bottom: $\beta=1.75$ (left entry), $\beta=2$ (right entry). Asymp.Var. is the estimated average asymptotic variance. NR means that the experiment is not run due to excessive computation time.}
\vspace{1ex}
\begin{tabular}{c|cccccc}
\hline
$m, n$ & $a$ & MSE & Cov. & Width & Asymp.Var. & T \\
\hline
$500$, $500$ & $1.0$ & $3.0 \times 10^{-4}$, $1.2 \times 10^{-4}$ & $0.683$, $0.729$ & $0.04$, $0.02$ & $1.1 \times 10^{-4}$, $4.1 \times 10^{-5}$ & $8$, $8$ \\
$500$, $500$ & $0.8$ & $1.5 \times 10^{-4}$, $4.2 \times 10^{-4}$ & $\mathbf{0.979}$, $0.638$ & $0.06$, $0.04$ & $2.2 \times 10^{-4}$, $1.3 \times 10^{-4}$ & $8$, $8$ \\
$500$, $500$ & $0.6$ & $2.3 \times 10^{-4}$, $3.4 \times 10^{-4}$ & $\mathbf{0.990}$, $\mathbf{0.961}$ & $0.09$, $0.08$ & $4.9 \times 10^{-4}$, $4.4 \times 10^{-4}$ & $8$, $8$ \\
$500$, $500$ & $0.4$ & $1.4 \times 10^{-3}$, $9.9 \times 10^{-4}$ & $\mathbf{0.999}$, $\mathbf{1.000}$ & $0.18$, $0.19$ & $2.1 \times 10^{-3}$, $2.3 \times 10^{-3}$ & $8$, $8$ \\
$500$, $500$ & $0.2$ & NR, NR & NR, NR & NR, NR & NR, NR & NR, NR \\
\hline
$500$, $500$ & $1.0$ & $1.3 \times 10^{-4}$, $2.6 \times 10^{-4}$ & $0.533$, $0.154$ & $0.02$, $0.01$ & $1.7 \times 10^{-5}$, $6.6 \times 10^{-6}$ & $8$, $8$ \\
$500$, $500$ & $0.8$ & $3.3 \times 10^{-4}$, $2.3 \times 10^{-4}$ & $0.564$, $0.527$ & $0.03$, $0.02$ & $6.3 \times 10^{-5}$, $3.5 \times 10^{-5}$ & $8$, $8$ \\
$500$, $500$ & $0.6$ & $2.5 \times 10^{-4}$, $2.0 \times 10^{-4}$ & $\mathbf{0.966}$, $\mathbf{0.955}$ & $0.07$, $0.06$ & $3.3 \times 10^{-4}$, $2.6 \times 10^{-4}$ & $8$, $8$ \\
$500$, $500$ & $0.4$ & $1.7 \times 10^{-3}$, $3.3 \times 10^{-4}$ & $\mathbf{1.000}$, $\mathbf{1.000}$ & $0.21$, $0.21$ & $2.8 \times 10^{-3}$, $2.9 \times 10^{-3}$ & $9$, $9$ \\
$500$, $500$ & $0.2$ & $1.7 \times 10^{-2}$, $4.2 \times 10^{-3}$ & $\mathbf{1.000}$, $\mathbf{1.000}$ & $0.71$, $1.12$ & $3.4 \times 10^{-2}$, $8.8 \times 10^{-2}$ & $8$, $9$ \\
\hline
\end{tabular}
\label{table:performance_combined}
\end{table}

\begin{table}[htb]
\centering
\caption{Under the setting of Table~\ref{tab: 4}, Experiment 4. $R = 1, 000$. Top: $\beta=1.25$ (left entry), $\beta=1.5$ (right entry). Bottom: $\beta=1.75$ (left entry), $\beta=2$ (right entry). NR means that the experiment is not run due to excessive computation time.}
\vspace{1ex}
\begin{tabular}{c|cccccc}
\hline
$m, n$ & $a$ & MSE, $\mu$ & MSE, $\lambda$ & Cov. & T \\
\hline
$1000$, $1000$ & $1.0$ & $1.8 \times 10^{-2}$, $1.1 \times 10^{-2}$ & $2.4 \times 10^{-2}$, $7.6 \times 10^{-3}$ & $\mathbf{1}$, $\mathbf{1}$ & $8.2$, $8.3$ \\
$1000$, $1000$ & $0.8$ & $6.3 \times 10^{-3}$, $6.1 \times 10^{-3}$ & $1.4 \times 10^{-3}$, $1.7 \times 10^{-3}$ & $\mathbf{1}$, $\mathbf{1}$ & $8.1$, $8.5$ \\
$1000$, $1000$ & $0.6$ & $3.98 \times 10^{-1}$, $4.85 \times 10^{-1}$ & $1.55 \times 10^{-1}$, $2.12 \times 10^{-1}$ & $0$, $0$ & $7.5$, $8.4$ \\
$1000$, $1000$ & $0.4$ & $8.7 \times 10^{-4}$, $7.5 \times 10^{-4}$ & $9.6 \times 10^{-3}$, $2.1 \times 10^{-3}$ & $\mathbf{1}$, $\mathbf{1}$ & $7.5$, $8.5$ \\
$1000$, $1000$ & $0.2$ & $3.0 \times 10^{-3}$, NR & $2.3 \times 10^{-3}$, NR & $\mathbf{1}$, NR & $7.7$, NR \\
\hline
$1000$, $1000$ & $1.0$ & $4.87 \times 10^{-1}$, $6.91$ & $5.00 \times 10^{-1}$, $28.88$ & $0$, $0$ & $8.4$, $8.3$ \\
$1000$, $1000$ & $0.8$ & $1.7 \times 10^{-2}$, $2.8 \times 10^{-3}$ & $1.0 \times 10^{-2}$, $2.1 \times 10^{-4}$ & $\mathbf{1}$, $\mathbf{1}$ & $8.5$, $8.1$ \\
$1000$, $1000$ & $0.6$ & $3.28 \times 10^{-1}$, $4.9 \times 10^{-4}$ & $1.36 \times 10^{-1}$, $1.2 \times 10^{-2}$ & $0$, $\mathbf{1}$ & $8.4$, $8.1$ \\
$1000$, $1000$ & $0.4$ & $3.7 \times 10^{-4}$, $9.0 \times 10^{-3}$ & $1.4 \times 10^{-3}$, $1.0 \times 10^{-2}$ & $\mathbf{1}$, $\mathbf{1}$ & $8.4$, $8.1$ \\
$1000$, $1000$ & $0.2$ & NR, $6.9 \times 10^{-4}$ & NR, $3.7 \times 10^{-3}$ & NR, $0$ & NR, $8.3$ \\
\hline
\end{tabular}
\label{table:performance_combined_exp4}
\end{table}

From Table~\ref{table:performance_combined}, it appears that when $\beta$ is closer to 
$1$, the produced confidence set has better coverage. This trend holds in Table~\ref{table:performance_combined_exp4} except for $\alpha=0.6$. Notably, KOSE-Riesz for all the values $\beta=1.25, 1.5, 1.75, 2$ are capable of producing a valid confidence set for $\alpha=0.4$, the most extreme model inexact scenario.

\subsection{Extra Experiments on the Sensitivity of $n$}
\label{sec:sens_n}
We experiment with the choice of $n$ on the convergence and the coverage of the confidence sets for $n=2, 10, 50, 100, 200$. 
The results under the setting of Table~\ref{tab: 4}, Experiment 4 ($p=2$, inexact model), run with an NVIDIA TITAN Xp GPU, are included in Table~\ref{table:performance_sensitivity_combined}.
The results for KOSE-Riesz under the setting of Experiment 1 ($p=1$, exact model), run with an NVIDIA 1080Ti GPU, are included in Table~\ref{table:1-dim}.

\begin{table}[htb]
\centering
\caption{Under the setting of Table~\ref{tab: 4}, Experiment 4. $R = 1, 000$. Sensitivity to $n$ for KOSE-Riesz (left entry) and KOSE-Gaussian (right entry). Iterations is the number of SGD iterations.}
\vspace{1ex}
\begin{tabular}{|c|cccccc|}
\hline
$n, m$ & $a$ & MSE, $\mu$ & MSE, $\lambda$ & Cov. & T & Iterations\\
\hline
$2$, $1000$ & $0.8$ & $8.91$, $10.29$ & $9.92$, $10.95$ & $0$, $0$ & $7.7$, $25.5$ & $800$\\
$10$, $1000$ & $0.8$ & $4.09$, $3.63$ & $3.22$, $2.22$ & $0$, $0$ & $7.7$, $25.3$ & $800$\\
$50$, $1000$ & $0.8$ & $1.21$, $0.57$ & $0.82$, $0.34$ & $0$, $0$ & $7.6$, $25.2$ & $800$\\
$100$, $1000$ & $0.8$ & $0.48$, $0.16$ & $0.37$, $0.13$ & $0$, $0$ & $7.4$, $25.1$ & $800$\\
$500$, $1000$ & $0.8$ & $0.07$, $0.04$ & $0.03$, $0.01$ & $0$, $\mathbf{1}$ & $7.9$, $25.3$ & $800$\\
\hline
$2$, $1000$ & $0.8$ & $9.71$, $6.33$ & $8.40$, $4.89$ & $0$, $0$ & $18.0$, $36.5$ & $2500$\\
$10$, $1000$ & $0.8$ & $1.82$, $1.63$ & $1.25$, $0.93$ & $0$, $0$ & $14.9$, $33.3$ & $2000$\\
$50$, $1,000$ & $0.8$ & $0.44$, $0.12$ & $0.29$, $0.07$ & $0$, $0$ & $12.4$, $30.2$ & $1500$\\
$100$, $1000$ & $0.8$ & $0.35$, $0.12$ & $0.18$, $0.06$ & $0$, $\mathbf{1}$ & $9.2$, $26.9$ & $1000$\\
$500$, $1000$ & $0.8$ & $0.02$, $0.02$ & $0.01$, $0.01$ & $0$, $\mathbf{1}$ & $9.3$, $26.8$ & $1000$\\
\hline
\end{tabular}
\label{table:performance_sensitivity_combined}
\end{table}

\begin{table}[htb]
\centering
\caption{KOSE-Riesz under the setting of Experiment 1. $R=1,000$. Asymp.Var. is the estimated average asymptotic variance.}
\label{table:1-dim}
\vspace{1ex}
\begin{tabular}{|c|cccccc|}
\hline
 & \multicolumn{6}{c|}{KOSE-Riesz, $R=1,000$} \\
$n, m$ & $a$ & MSE & Cov. & Width & Asymp.Var. & T \\
\hline
$2, 500$ & $1$ & $0.015$ & $0.289$ & $0.26$ & $0.185$ & $5$ \\
$10, 500$ & $1$ & $0.003$ & $0.597$ & $0.34$ & $0.035$ & $5$ \\
$50, 500$ & $1$ & $0.001$ & $0.688$ & $0.07$ & $0.000$ & $5$ \\
$100, 500$ & $1$ & $0.001$ & $0.896$ & $0.07$ & $0.000$ & $5$ \\
$200, 500$ & $1$ & $0.000$ & $\mathbf{0.976}$ & $0.07$ & $0.000$ & $5$ \\
\hline
\end{tabular}
\end{table}

From these tables, it is evident that smaller $n$ leads to greater MSE in estimation and worse confidence set coverage if the number of iterations is unchanged. However, performance improves as the number of iterations increases.

\subsection{Extra Experiments on the Bias in Uncertainty Evaluation}
\label{sec:unc_ev}
We conduct additional experiments to investigate the bias in uncertainty evaluation. This bias stems from two primary sources: estimation bias and variance-covariance approximation bias. The estimation bias arises from the learning procedure, where the distribution of the kernel optimum score estimator (KOSE) deviates from the asymptotic normal distribution. The variance-covariance approximation bias is introduced during the estimation of the variance-covariance matrix.

Quantifying the bias in uncertainty evaluation presents challenges. For one-dimensional examples (i.e., $p=1$), the bias is considered small if the mean squared error (MSE) closely approximates the estimated average asymptotic variance. However, for two-dimensional examples (i.e., $p=2$), we lack a definitive measure for this bias.
One approach to assess the magnitude of the bias is to observe whether the asymptotic variance estimate stabilizes. This stabilization can be inferred by examining the average width and height of the produced confidence sets (scaled up by a factor of $\sqrt{m}$).

Table~\ref{tab: 2-asymp} presents the results of Experiment 2, including the estimated average asymptotic variance. Table~\ref{table:kose-riesz-exp4} shows the results of KOSE-Riesz under the conditions of Table~\ref{tab: 4}, Experiment 4.
All experiments, unless otherwise specified, are run with an NVIDIA TITAN Xp GPU.

From Table~\ref{tab: 2-asymp}, it is evident that the MSE closely approximates the asymptotic variance, particularly for KOSE-Riesz. 
Table~\ref{table:kose-riesz-exp4} demonstrates that the width and height (scaled by $\sqrt{m}$) of the confidence set remain stable across $m=1000, 2000$, and $5000$, suggesting a stable asymptotic variance estimate.

\begin{table}[htb]
\centering
\caption{KOSE-Riesz under the setting of Table~\ref{tab: 4}, Experiment 4; $R=100$. Width and Height are the average width and height of the estimated confidence sets, respectively.}
\label{table:kose-riesz-exp4}
\vspace{1ex}
\begin{tabular}{|c|ccccc|}
\hline
 & \multicolumn{5}{c|}{KOSE-Riesz, $R=100$} \\
$n, m$ & $a$ & Cov. & Width & Height & T \\
\hline
$1000, 1000$ & $1.0$ & $\mathbf{1.00}$ & $0.05$ & $2.61$ & $7.9$ \\
$1000, 1000$ & $0.8$ & $0.00$ & $0.05$ & $2.71$ & $7.9$ \\
$1000, 1000$ & $0.6$ & $0.00$ & $0.03$ & $1.28$ & $8.0$ \\
$1000, 1000$ & $0.4$ & $0.00$ & $0.01$ & $1.25$ & $8.1$ \\
$1000, 1000$ & $0.2$ & $0.00$ & $0.00$ & $1.04$ & $8.2$ \\
\hline
$1000, 2000$ & $1.0$ & $\mathbf{1.00}$ & $0.03$ & $1.56$ & $11.4$ \\
$1000, 2000$ & $0.8$ & $0.00$ & $0.04$ & $2.39$ & $11.3$ \\
$1000, 2000$ & $0.6$ & $0.00$ & $0.03$ & $1.37$ & $10.9$ \\
$1000, 2000$ & $0.4$ & $0.00$ & $0.01$ & $1.29$ & $10.7$ \\
$1000, 2000$ & $0.2$ & $0.00$ & $0.00$ & $1.13$ & $10.7$ \\
\hline
$1000, 5000$ & $1.0$ & $\mathbf{0.99}$ & $0.04$ & $1.83$ & $20.4$ \\
$1000, 5000$ & $0.8$ & $\mathbf{1.00}$ & $0.05$ & $2.65$ & $20.2$ \\
$1000, 5000$ & $0.6$ & $0.00$ & $0.02$ & $1.27$ & $19.7$ \\
$1000, 5000$ & $0.4$ & $0.00$ & $0.01$ & $1.27$ & $19.6$ \\
$1000, 5000$ & $0.2$ & $0.00$ & $0.00$ & $1.08$ & $19.5$ \\
\hline
\end{tabular}
\end{table}

\begin{table}[htb]
\centering
\caption{Experiment 2 results; $R=1,000$. Asymp.Var. is the estimated average asymptotic variance.}\label{tab: 2-asymp}
\vspace{1ex}
\begin{tabular}{|c|ccccc|c|}
\hline
& \multicolumn{5}{c|}{KOSE-Riesz, $R=1,000$} & \\
$a$ & MSE & Cov. & Width & Asymp.Var. & T & $\theta_\star^{\operatorname{KS}}$\\
\hline
$1$ & $4.0\cdot 10^{-4}$ & $0.920$ & $0.07$ & $3.1 \cdot 10^{-4}$ & $3$ & $1.2$\\
$0.8$ & $1.6\cdot 10^{-4}$ & $\mathbf{0.996}$ & $0.07$ & $3.6 \cdot 10^{-4}$ & $3$ & $1.5$\\
$0.6$ & $2.6\cdot 10^{-4}$ & $\mathbf{0.999}$ & $0.10$ & $6.7 \cdot 10^{-4}$ & $3$ & $1.9$\\
$0.4$ & $6.7\cdot 10^{-4}$ & $\mathbf{0.999}$ & $0.15$ & $1.5 \cdot 10^{-3}$ & $3$ & $2.5$\\
$0.2$ & $3.0\cdot 10^{-3}$ & $\mathbf{1.000}$ & $0.34$ & $8.0 \cdot 10^{-3}$ & $3$ & $4.1$\\
\hline
\end{tabular}
\vspace{0.5em}
\begin{tabular}{|c|ccccc|c|}
\hline
& \multicolumn{5}{c|}{KOSE-Gaussian, $R=1,000$} & \\
$a$ & MSE & Cov. & Width & Asymp.Var. & T & $\theta_\star^{\operatorname{KS}}$\\
\hline
$1$ & $1.8\cdot 10^{-4}$ & $\mathbf{0.990}$ & $0.06$ & $2.8 \cdot 10^{-4}$ & $4$ & $1.2$\\
$0.8$ & $7.4\cdot 10^{-4}$ & $0.842$ & $0.08$ & $4.0 \cdot 10^{-4}$ & $4$ & $1.5$\\
$0.6$ & $8.0\cdot 10^{-7}$ & $\mathbf{0.942}$ & $0.10$ & $6.6 \cdot 10^{-4}$ & $3$ & $1.9$\\
$0.4$ & $1.0\cdot 10^{-3}$ & $\mathbf{0.997}$ & $0.17$ & $1.8 \cdot 10^{-3}$ & $4$ & $2.5$\\
$0.2$ & $1.9\cdot 10^{-2}$ & $0.833$ & $0.37$ & $9.8 \cdot 10^{-3}$ & $4$ & $4.1$\\
\hline
\end{tabular}
\end{table}

\subsection{Extra Experiments on Contamination Models}
\label{sec:extra_numericals}
We conduct an additional experiment on contamination models under the setting of Experiment 3. We add white noise drawn from a $N(0, 0.01)$ distribution to $(100\epsilon)\%$ of the output data from the target system, where $\epsilon = 0.01, 0.05, 0.1, 0.2, 0.5$. The results are reported in Table~\ref{tab: contamination}. From Table~\ref{tab: contamination}, we observe that our method is robust to noise, as expected from the robustness of the MMD and, consequently, the kernel score \citep{briol2019statistical}.

\begin{table}[htb]
\centering
\caption{Contamination Models under the setting of Experiment 3; $R=100$; $\epsilon$ is the noise level; $m=n=1,000$. MSE, $\mu$ and MSE, $\lambda$ are the average mean-squared errors for the service and arrival rates.}
\label{tab: contamination}
\vspace{1ex}
\begin{tabular}{|c|cc|cc|}
\hline
& \multicolumn{2}{c|}{KOSE-Riesz, $R=100$} & \multicolumn{2}{c|}{KOSE-Gaussian, $R=100$} \\
$\epsilon$ & MSE, $\mu$ & MSE, $\lambda$ & MSE, $\mu$ & MSE, $\lambda$ \\ \hline
$0.01$ & $0.190$ & $0.093$ & $0.071$ & $0.024$ \\ \hline
$0.05$ & $0.222$ & $0.100$ & $0.064$ & $0.023$ \\ \hline
$0.1$ & $0.221$ & $0.099$ & $0.332$ & $0.118$ \\ \hline
$0.2$ & $0.227$ & $0.102$ & $0.270$ & $0.135$ \\ \hline
$0.5$ & $0.244$ & $0.111$ & $0.058$ & $0.029$ \\ \hline
\end{tabular}
\end{table}
\end{proof}
\end{document}